\documentclass{JHEP3}
\usepackage{fullpage}
\usepackage{amssymb,amsmath}
\usepackage{graphicx}
\usepackage{feynmp}
\usepackage{cite}

\def\bra#1{\left\langle #1\right|}
\def\ket#1{\left| #1\right\rangle}
\def\braket#1{\left\langle #1 \right\rangle}

\def\la{\lambda}
\def\lb{\tilde{\lambda}}

\def\Res_#1{\underset{#1}{\text{Res}}}
\title{All one-loop NMHV gluon amplitudes in $\mathcal{N}=1$ SYM}
\author{
Alexander Ochirov\\
Institut de Physique Th\'eorique, CEA-Saclay, F-91191 Gif-sur-Yvette
cedex, France \\
Email: \email{alexander.ochirov@cea.fr}
}

\abstract{
We compute the next-to-maximally-helicity-violating one-loop $n$-gluon amplitudes
in $\mathcal{N}=1$ super-Yang-Mills theory.
These amplitudes contain three negative-helicity gluons
and an arbitrary number of positive-helicity gluons,
and constitute the first infinite series of amplitudes
beyond the simplest, MHV, amplitudes.
We assemble ingredients from the $\mathcal{N}=4$ NMHV tree super-amplitude
into previously unwritten double cuts and use the spinor integration technique
to calculate all bubble coefficients.
We also derive the missing box coefficients from quadruple cuts.
Together with the known formula for three-mass triangles,
this completes the set of NMHV one-loop master integral coefficients
in $\mathcal{N}=1$ SYM.
To facilitate further use of our results,
we provide their Mathematica implementation.
}
\keywords{Scattering amplitudes, Generalized unitarity, Supersymmetry}
\preprint{IPhT-t13/248}

\begin{document}

\section{Introduction}
\label{introduction}

      In the last couple of decades, there have been impressive achievements
in taming gauge theory amplitudes analytically
for increasing and in some cases arbitrary number of particles.
\begin{table}[h]
\centering
    \begin{tabular}{| l | l | l | l |}
    \hline
    &
    $\mathcal{N}=4$ SYM & $\mathcal{N}=1$ SYM & QCD \\ \hline
    MHV &
    $n$-point in 1994 \cite{Bern:1994zx} &
    $n$-point in 1994 \cite{Bern:1994cg} &
    $n$-point in 2006 \cite{Bedford:2004nh,Berger:2006vq} \\ \hline
    NMHV &
    $n$-point in 2004 \cite{Bern:2004bt} &
      6-point in 2005 \cite{Britto:2005ha},
      7-point in 2009 \cite{Dunbar:2009ax}, &
      6-point in 2006 \cite{Britto:2006sj,Xiao:2006vt} \\
    & &
    \textbf{\textit{n}-point} in this work & \\ \hline
    \end{tabular}
\caption{Known analytic results for gluon amplitudes at one loop in gauge theories
         with and without supersymmetry, including the result of the present paper}
\label{known} \end{table}
Table \ref{known} provides a short summary of existing one-loop results
as of November 2013.
In it, ``maximally-helicity-violating'' (MHV) conventionally stands
for amplitudes with two minus-helicity gluons, whereas
the next-to-maximally-helicity-violating (NMHV) case
corresponds to three negative helicities.
In addition to that, general split-helicity color-ordered amplitudes
in $\mathcal{N}=1$ SYM are known as well due to their simple analytic behavior
which permits an elegant one-loop BCFW recursion \cite{Bern:2005hh}.

      The result of this paper is completing the lower middle cell
of table \ref{known} with $n$-point analytic results.
To do that, we use spinor integration \cite{Britto:2005ha,Anastasiou:2006jv}
which provides a sleek way to compute amplitude coefficients
of one-loop master integrals from unitarity cuts in a purely algebraic manner.
We briefly review its idea and recipes in section \ref{method},
and slightly adapt it to make full use of $\mathcal{N}=1$ supersymmetry.

      Intuitively, the main difficulty in finding universal NMHV formulas is that
even at 7 points general patterns are not yet obvious,
because the numbers of minus and plus helicities are still comparable to each other,
whereas MHV amplitudes become ``saturated'' by positive helicities already for 6 external gluons.
So in Section \ref{cutconstruction}, we construct a double cut
for an arbitrary multiplicity from the start,
for which we use the tree input from \cite{Dixon:2010ik}.
Next, in Section \ref{loopmomentum}, we carefully analyze
how the cut depends on loop momentum variables,
which is essential for getting to the master integral coefficients.

      We then obtain the main result of this paper ---
formulas for bubble (Section \ref{nmhvbubbles}) and box (Section \ref{nmhvboxes})
coefficients. In order to facilitate their further use,
we distribute their Mathematica implementation,
briefly described in Appendix \ref{app:implementation}.
To verify our results, we performed a number of non-trivial checks,
summarized in Section \ref{checks}.

      We hope that our all-multiplicity results will provide a helpful testing ground
for further theoretic developments.
For instance, it is an interesting question whether any kind of on-shell recursion
relations can be established between the coefficients we have found.
We only took a quick peek into this, as is mentioned in Section \ref{discussion}.

\section{Set-up: $\mathcal{N}=1$ SYM at one loop}
\label{setup}

\subsection{Supersymmetry expansion}

      There has been remarkable progress
in understanding perturbative expansion of gauge theories
which builds upon the realization that scattering matrix elements
not only constitute basic words of quantum field theoretic language,
but also turn out to be perfect objects to calculate analytically.
This was first seen at tree level \cite{Parke:1986gb},
but was subsequently followed by a number of beautiful insights
at one loop \cite{Bern:1994zx,Britto:2004nc,Drummond:2008vq}
and beyond \cite{Kosower:2011ty,ArkaniHamed:2012nw},
not to mention tree level again \cite{Cachazo:2004kj,Britto:2004ap,Britto:2005fq}.

      There are three most basic tools that came into universal use:
\begin{itemize}
\item decomposing full gauge boson amplitudes into simpler color-ordered components
\cite{Berends:1987cv,Mangano:1988kk,Bern:1990ux};
\item using helicity spinors both for fermions and bosons
\cite{Berends:1981rb,DeCausmaecker:1981bg,Xu:1986xb};
\item supersymmetric Ward identities and superspace coordinates
\cite{Grisaru:1976vm,Grisaru:1977px,Parke:1985pn,Kunszt:1985mg,Nair:1988bq}.
\end{itemize}
These standard techniques are described with great pedagogical skill
in \cite{Dixon:1996wi}.

      Interestingly, supersymmetry proved to be directly useful even for
non-supersymmetric gauge theories.
Whereas gluon tree amplitudes for pure quantum chromodynamics
equal those of supersymmetric Yang-Mills theory,
their one-loop analogues obey a simple expansion \cite{Bern:1993mq,Bern:1994zx}:
\begin{equation}
      A^{\text{1-loop}}_{\text{QCD}} =
      A^{\text{1-loop}}_{\mathcal{N}=4 \text{ SYM}} - 4 \,
      A^{\text{1-loop}}_{\mathcal{N}=1 \text{ chiral}} + 2 \,
      A^{\text{1-loop}}_{\mathcal{N}=0 \text{ scalar}} ,
\label{amplitudeQCD}
\end{equation}
which splits the calculation of direct phenomenological interest
into three problems of increasing difficulty.
Taking into account that
\begin{equation}
      A^{\text{1-loop}}_{\mathcal{N}=1 \text{ SYM}} =
      A^{\text{1-loop}}_{\mathcal{N}=4 \text{ SYM}} - 3 \,
      A^{\text{1-loop}}_{\mathcal{N}=1 \text{ chiral}} ,
\label{amplitudeN1}
\end{equation}
it becomes clear that calculations in $\mathcal{N}=4,1$ SYM are important steps
to full understanding of QCD.

      As stated in the introduction, in this paper we deal with
one-loop NMHV amplitudes in $\mathcal{N}=1$ SYM.
More precisely, we concentrate on $n$-point one-loop contributions
from the $\mathcal{N}=1$ chiral multiplet in the adjoint representation,
which consists of a complex scalar and a Majorana fermion.
In fact, its effective number of supersymmetries is two,
which is reflected in its alternative name, $\mathcal{N}=2$ hyper multiplet,
and can be easily seen from its relation to $\mathcal{N}=2$ SYM:
\begin{equation}
      A^{\text{1-loop}}_{\mathcal{N}=2 \text{ SYM}} =
      A^{\text{1-loop}}_{\mathcal{N}=4 \text{ SYM}} - 2 \,
      A^{\text{1-loop}}_{\mathcal{N}=1 \text{ chiral}} .
\label{amplitudeN2}
\end{equation}

      Moreover, amplitudes in four dimensions are known to be reducible
\cite{Brown:1952eu,Melrose:1965kb,'tHooft:1978xw,Passarino:1978jh,vanNeerven:1983vr}
to the following basis of master integrals:
\begin{equation}
      A^{\text{1-loop}} = \mu^{2\epsilon} \Big(
                          \sum C^{\text{box}} \, I_4
                        + \sum C^{\text{tri}} \, I_3
                        + \sum C^{\text{bub}} \, I_2 + R \Big) ,
\label{amplitude}
\end{equation}
where the sums go over all distinct scalar integrals
and $R$ is the purely rational part.
However, we know supersymmetry can constrain the general expansion (\ref{amplitude}):
the strongest, $\mathcal{N}=4$, supersymmetry leaves nothing but boxes $\{I_4\}$
\cite{Bern:1994zx},
while $\mathcal{N}=1,2$ supersymmetries eliminate the rational part $R$
\cite{Bern:1994cg}.
Since $R$ is the only term in (\ref{amplitude}) invisible to four-dimensional cuts,
supersymmetric amplitudes can be characterized as cut-constructible.

\subsection{UV and IR behavior}
\label{singular}

      In this paper, we adopt the conventional definition
\cite{Bern:1992em,Bern:1993kr}
for dimensionally-regularized massless scalar integrals:
\begin{equation}
      I_n = (-1)^{n+1}(4\pi)^{\frac{d}{2}}i
            \int \frac{d^d l_1}{(2\pi)^d}
            \frac{ 1 }
                 { l_1^2 (l_1-K_1)^2 \dots (l_1-K_{n-1})^2 } ,
\label{integrals}
\end{equation}
where $d = 4-2\epsilon$.
Due to the normalization, all the coefficient formulas we provide further
contain trivial pre-factors $(4\pi)^{-d/2}$.
Analytic expressions for these integrals are well-known
and are given in \cite{Bern:1993kr}.


      Now we review a useful result from \cite{Britto:2005ha},
where it was derived that
one can include all the infrared divergent one-mass and two-mass triangles
into the definition of new, finite, boxes
and thus leave only three-mass triangles in expansion (\ref{amplitude}).
Moreover, the only remaining divergent integrals are the bubbles:
\begin{equation}
      I_2 = \frac{1}{\epsilon} + O(1) ,
\label{bubdiv}
\end{equation}
so they alone must produce the remaining singular behavior of the amplitude.
As the latter is proportional to the tree amplitude
\begin{equation}
      A^{\text{1-loop}}_{\mathcal{N}=1 \text{ chiral}}
    = \frac{1}{\epsilon} \sum C^{\text{bub}} + O(1)
    = \frac{1}{(4\pi)^\frac{d}{2} \epsilon} A^{\text{tree}} + O(1) ,
\label{amplitudeUV}
\end{equation}
we retrieve a non-trivial relation among bubble coefficients:
\begin{equation}
      \sum C^{\text{bub}} = \frac{1}{(4\pi)^\frac{d}{2}} A^{\text{tree}} ,
\label{bubblesum}
\end{equation}
which we use as the first consistency check for our analytic results.

      Having considerably reduced our problem,
we now summarize how we deal with the rest.
The best and immediately algebraic method to compute box coefficients
is from quadruple cuts, first introduced in \cite{Britto:2004nc}.
Three-mass triangle coefficients can be found from triple cuts
\cite{Bern:1997sc,Forde:2007mi},
and it was done in full generality in \cite{BjerrumBohr:2007vu,Dunbar:2009ax}.
In the following, we will thus concentrate mostly on bubbles,
for which we use the spinor integration technique \cite{Britto:2005ha},
described in Section \ref{method}.

      In the following, we extensively use the following notation for momentum sums:
\begin{equation}
      P_{i,j} \equiv p_i + p_{i+1} + \dots + p_{j-1} + p_j ,
\label{momentumsum}
\end{equation}
where indices are taken modulo the number of legs $n$.
For all calculations in this work, we pick the cut-channel momentum to be $P_{1,k}$.
(For brevity, we will spell it simply as $P_{1k}$.)
If one wishes to compute another channel cut $P_{r,s}$,
one should simply cyclically relabel the legs $i \rightarrow (i\!-\!r\!+\!1)$
and set $k=s-r+1$.
As described in Appendix \ref{app:implementation},
the functions provided in the attached Mathematica notebook
have input arguments that are adapted for such relabeling.


\section{Method: spinor integration}
\label{method}

      In this section, we go through the spinor integration method in four dimensions \cite{Britto:2005ha,Anastasiou:2006jv,Britto:2006fc,Britto:2007tt,Britto:2008sw,
Mastrolia:2009dr}
and write down the formulas that we use to find
the coefficients of the master integrals.

\subsection{General coefficient formulas}
\label{generalformulas}

      We start by constructing the standard unitarity cut, the double cut,
from two tree amplitudes.
For simplicity, we define the four-dimensional $K$-channel cut without any prefactors:
      \begin{equation}
            \text{Cut} = \sum_{h_1,h_2}
                         \int \! d^4l_1 \, \delta(l_1^2) \delta(l_2^2)
                         A(-l_1^{\bar{h}_1},\dots,l_2^{h_2})
                         A(-l_2^{\bar{h}_2},\dots,l_1^{h_1}) .
      \label{cut}
	\end{equation}
The most important step is then
to trade the constrained loop variables $l_1$ and $l_2=l_1-K$
for homogeneous spinor variables $\la$ and $\lb$ such that
      \begin{subequations} \begin{align}
            l_1^{\mu} & =  \frac{K^2}{2}
                           \frac{ \bra{\la} \gamma^{\mu}|\lb] }
                                { \bra{\la}\!K|\lb] } , \\
            l_2^{\mu} & = -\frac{1}{2}
                           \frac{ \bra{\la}\!K|\gamma^{\mu}|K|\lb] }
                                { \bra{\la}\!K|\lb] } .
	\end{align} \label{lvariables} \end{subequations}
The integration measure transforms as follows:
      \begin{equation}
            \int \! d^4l_1 \, \delta(l_1^2) \delta(l_2^2)
                      = -\frac{K^2}{4} \int_{\tilde{\la}=\bar{\la}} \!
                         \frac{ \braket{\la d\la} [\lb d\lb] }
                              { \bra{\la}\!K|\lb]^2 } .
      \label{dLIPS4}
	\end{equation}

      If one then expands these homogeneous variables in arbitrary basis spinors:
      \begin{equation}
            \la = \la_{p} + z \la_{q} , \quad
            \lb = \lb_{p} + \bar{z} \lb_{q} ,
      \label{complexvariables}
	\end{equation}
then the connection to the integral over the complex plane becomes evident:
      \begin{equation}
            \int_{\tilde{\la}=\bar{\la}} \! \braket{\la d\la} [\lb d\lb]
                      = -(p+q)^2 \! \int \! dz \wedge d\bar{z} .
      \label{complexmeasure}
	\end{equation}
So the phase space spinor integration can be treated as
a complex plane integration in disguise. 
In this spinorial language, it is possible to define
simple and self-consistent rules for taking residues.
For instance, we calculate the residue of simple pole $\braket{\zeta|\la}$ as follows:
      \begin{equation}
            \Res_{\lambda=\zeta} \; \frac{F(\la,\lb)}{\braket{\zeta|\la}}
                  = F(\zeta,\tilde{\zeta}) .
      \label{simplepolerule}
	\end{equation}
The full set of rules is given in detail in Appendix \ref{app:residues}.

      In essence, the method of \emph{spinor integration} uses
a spinorial version of Cauchy's integral theorem \cite{Mastrolia:2009dr}
to actually perform that complex plane integration
in a manner which exposes coefficients of different scalar integrals.

      In short, once we rewrite the cut (\ref{cut}) using homogeneous spinor variables
      \begin{equation}
            \text{Cut} = \int_{\tilde{\la}=\bar{\la}} \!
                         \braket{\la d\la} [\lb d\lb] \;
                         \mathcal{I}_{\text{spinor}} ,
      \label{cutspinor}
	\end{equation}
integral coefficients are given by \emph{general algebraic formulas}
which are given below.
To write them we only need to introduce a short notation for the following vectors:
      \begin{equation}
            Q_i^{\mu}(K_i,K) = -K_i^{\mu}+ \frac{K_i^2}{K^2} K^{\mu} .
      \label{Qi4dim}
	\end{equation}
These arise naturally because
all loop-dependent physical poles come from propagators
which can be rewritten in homogeneous variables as
      \begin{equation} \begin{aligned}
            (l_1 - K_i)^2 = K^2 \frac{ \bra{\la}Q_i|\lb] }
                                     { \bra{\la}\!K|\lb] } .
      \label{propagators4dim}
      \end{aligned} \end{equation}

\subsubsection{Box coefficient}
\label{box}

      The coefficient of the scalar box labeled by two uncut propagators $i$ and $j$
can be expressed as
      \begin{equation}
            C_{ij}^{\text{box}} =-\frac{2 K^2}{(4\pi)^{\frac{d}{2}}i}
                                  \mathcal{I}_{\text{spinor}}
                                  \bra{\la}Q_i|\lb] \bra{\la}Q_j|\lb]
                                  \bigg\{ \bigg|_{\substack{
                                             \la = \la^{ij}_+ \\
                                             \lb = \lb^{ij}_-
                                             }}
                                          +
                                          \bigg|_{\substack{
                                             \la = \la^{ij}_- \\
                                             \lb = \lb^{ij}_+
                                             }} 
                                  \bigg\} ,
      \label{Cbox}
	\end{equation}
where spinors $\la = \la^{ij}_\pm$ and $\lb = \lb^{ij}_\pm$ correspond to
on-shell combinations of propagator momenta:
      \begin{equation}
            P^{ij}_{\pm}(K_i,K_j,K) = Q_i + x^{ij}_{\pm} Q_j ,
      \label{Pbox}
	\end{equation}
      \begin{equation}
            x^{ij}_{\pm} = \frac{ -Q_i \cdot Q_j
                    \pm \sqrt{ (Q_i \cdot Q_j)^2 - Q_i^2 Q_j^2 } }{Q_j^2} .
      \label{Xbox}
	\end{equation}

      It is easy to see that these formulae are equivalent
to the well-understood quadruple cut method \cite{Britto:2004nc,Kosower:2011ty}.
Indeed, the sole purpose of factors $\bra{\la}Q_i|\lb]$ and $\bra{\la}Q_j|\lb]$
in (\ref{Cbox}) is just to cancel the corresponding propagator factors
in the denominator of $\mathcal{I}_{\text{spinor}}$.
Now, by definition
      \begin{equation}
            \bra{\la^{ij}_\pm}Q_i|\lb^{ij}_\mp] =
            \bra{\la^{ij}_\pm}Q_j|\lb^{ij}_\mp] = 0 ,
      \label{quadruplecut}
	\end{equation}
so formula (\ref{Cbox}) effectively puts propagators $i$ and $j$ on shell,
thus converting the original double cut into a quadruple cut
and summing over the two solutions.

\subsubsection{Triangle coefficient}
\label{triangle}

      The coefficient of the scalar triangle labeled by one uncut propagator $i$
can be found to be equal to
      \begin{equation} \begin{aligned}
          C_i^{\text{tri}} = & \frac{2}{(4\pi)^{\frac{d}{2}}i}
                               \frac{1}{(K^2 (x^i_+-x^i_-)^2)^{n-k+1}} \\ \times
                             & \frac{1}{(n\!-\!k\!+\!1)!}
                               \frac{\mathrm{d}^{(n-k+1)}}{\mathrm{d}t^{(n-k+1)}}
                               \mathcal{I}_{\text{spinor}}
                               \bra{\la}Q_i|\lb] \bra{\la}\!K|\lb]^{n-k+2}
                               \bigg\{ \bigg|_{\substack{
                                          \la = \la^i_+ - t \la^i_- \\
                                          \lb = x^i_+ \lb^i_- - t x^i_- \lb^i_+
                                          }}
                                          +
                                          \bigg|_{\substack{
                                          \la = \la^i_- - t \la^i_+ \\
                                          \lb = x^i_- \lb^i_+ - t x^i_+ \lb^i_-
                                          }} 
                               \bigg\} \bigg|_{t=0} ,
      \label{Ctri}
	\end{aligned} \end{equation}
where spinors $\la = \la^{i}_\pm$ and $\lb = \lb^{i}_\pm$ correspond to
the following on-shell momenta:
      \begin{equation}
            P^{i}_{\pm}(K_i,K) = Q_i + x^{i}_{\pm} K ,
      \label{Ptri}
	\end{equation}
      \begin{equation}
            x^{i}_{\pm} = \frac{ - K \cdot Q_i
                       \pm \sqrt{ (K \cdot Q_i)^2 - K^2 Q_i^2 } }{K^2} .
      \label{Xtri}
	\end{equation}
Here and below in this section,
$(n-k)$ is the difference between the numbers of $\la$-factors
in the numerator and the denominator of $\mathcal{I}_{\text{spinor}}$,
excluding the homogeneity-restoring factor $\bra{\la}\!K|\lb]^{n-k+2}$.

\subsubsection{Bubble coefficient}
\label{bubble}

      Finally, we find the coefficient of the $K$-channel scalar bubble
through the following general formula:
      \begin{equation}
          C^{\text{bub}}\! = \frac{4}{(4\pi)^{\frac{d}{2}}i}
                             \sum_{\text{residues}}
                             \frac{1}{(n\!-\!k)!}
                             \frac{\mathrm{d}^{(n-k)}}{\mathrm{d}s^{(n-k)}}
                             \frac{1}{s}
                             \ln \left( 1 + s \frac{\bra{\la}\!q|\lb]}
                                                   {\bra{\la}\!K|\lb]}
                                 \right)
                             \bigg[ \mathcal{I}_{\text{spinor}}
                                    \frac{\bra{\la}\!K|\lb]^{n-k+2}}
                                         {\bra{\la}\!K|q\!\ket{\la}}
                             \bigg|_{|\lb]=|K+s\,q\ket{\la}}
                             \bigg] \bigg|_{s=0} ,
      \label{Cbub}
	\end{equation}
where the derivative in $s$ is just a way to encode the extraction of the
$(n\!-\!k)$-th Taylor coefficient around $s = 0$.
Note that the formula contains an arbitrary light-like vector $q$.
Nonetheless, the answer does not depend on it
and thus can be simplified by an appropriate choice of $q$.

      We point out the fact that (\ref{Cbub}) looks different
from equivalent spinor integration formulas given earlier in
\cite{Britto:2006fc,Britto:2007tt,Britto:2008sw,Britto:2010xq},
because here we chose to write it using as a sum over spinor residues
thus leaving the next step to be carried out afterwards
according to the conventions given in Appendix \ref{app:residues}.
So in fact, (\ref{Cbub}) can be considered as an intermediate step
in derivation of more involved formulas with all pole residues
already taken explicitly in full generality with the price of generating
extra sums and derivatives in another auxiliary parameter.
Further in Section \ref{simplifiedbubble}, we provide another formula
which is even better suited for calculations with $\mathcal{N}=1$ supersymmetry.

\subsection{Example: $\overline{\text{MHV}}$-MHV bubbles in $\mathcal{N}=1$ SYM}
\label{simplestbubbles}

      In this section, we employ the spinor integration technique
to derive explicitly a simple but non-trivial family of bubble coefficients
in $\mathcal{N}=1$ SYM.
To be more precise, we consider the contribution of
the $\mathcal{N}=1$ chiral multiplet in the loop.
To get bubble coefficients in pure $\mathcal{N}=1$ SYM
one just needs to multiply our results by $-3$.

      \begin{figure}[h]
      \centering
      \parbox{127pt}{ \begin{fmffile}{graph1}
      \fmfframe(10,10)(10,10){ \begin{fmfgraph*}(100,60)
            \fmflabel{$l_1$}{bottom}
            \fmflabel{$1^-$}{g1}
            \fmflabel{$\dots$}{dots1}
            \fmflabel{$p^+$}{gp}
            \fmflabel{$\dots$}{dots2}
            \fmflabel{$k^-$}{gk}
            \fmflabel{$l_2$}{top}
            \fmflabel{$(k\!+\!1)^+$}{gk1}
            \fmflabel{$\dots$}{dots3}
            \fmflabel{$m^-$}{gm}
            \fmflabel{$\dots$}{dots4}
            \fmflabel{$n^+$}{gn}
            \fmfleft{g1,dots1,gp,dots2,gk}
            \fmfright{gn,dots4,gm,dots3,gk1}
            \fmftop{top}
            \fmfbottom{bottom}
            \fmf{wiggly}{g1,vleft}
            \fmf{wiggly}{gp,vleft}
            \fmf{wiggly}{gk,vleft}
            \fmf{wiggly}{gk1,vright}
            \fmf{wiggly}{gm,vright}
            \fmf{wiggly}{gn,vright}
            \fmf{dashes}{top,bottom}
            \fmf{plain,left,tension=1.0}{vleft,vright}
            \fmf{plain,right,tension=1.0}{vleft,vright}
            \fmfblob{0.17w}{vleft}
            \fmfblob{0.17w}{vright}
      \end{fmfgraph*} }
      \end{fmffile} }
      \caption{$P_{1k}$-channel cut for
               $ A_{\mathcal{N}=1 \text{ chiral}}^{\text{1-loop}}
                   (p^+\!\in\!\{1,\dots,k\}, m^-\!\in\!\{k\!+\!1,\dots,n\}) $
               \label{cut1ksimplest}}
      \end{figure}
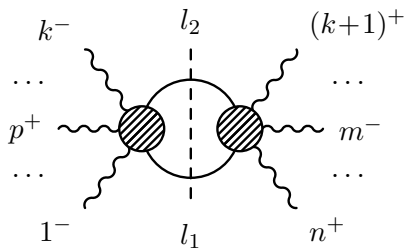
      Consider the $P_{1k}$-channel cut of $\mathcal{N}=1$ chiral one-loop amplitude
with one plus-helicity gluon $p^+$ to the left of the cut
and one minus-helicity gluon $m^-$ to the right, see fig. \ref{cut1ksimplest}.
The amplitude to the right of the cut is then just MHV,
whereas the one to the left is then
$(k+2)$-point $\text{N}^{k-2}$MHV=$\overline{\text{MHV}}$.
A nice property of this cut is that it is omnipresent
as a two-particle cut in MHV amplitudes,
a three-particle cut in NMHV amplitudes,
a four-particle cut in NNMHV amplitudes and so on.
At the same time, it is very simple to write down:
      \begin{equation} \begin{aligned}
            \sum_{h_1,h_2} A(-l_1^{\bar{h}_1},\dots,l_2^{h_2})
                           A(-l_2^{\bar{h}_2},\dots,l_1^{h_1})
                = & \bigg(
                         - \frac{ [l_1 p] \braket{l_1 m} }
                                { [l_2 p] \braket{l_2 m} }
                         + 2
                         - \frac{ [l_2 p] \braket{l_2 m} }
                                { [l_1 p] \braket{l_1 m} }
                    \bigg) \\ \times
                    \frac{ (-1)^k i [l_1 p]^2 [l_2 p]^2 }
                         { [l_1 1] [12] \dots [k\!-\!1|k] [k l_2] [l_2 l_1] } &
                    \frac{ i \braket{l_1 m}^2 \braket{l_2 m}^2 }
                         { \braket{l_2|k\!+\!1} \braket{k\!+\!1|k\!+\!2}
                                    \dots \braket{n\!-\!1|n}
                           \braket{n l_1} \braket{l_1 l_2} } .
      \label{simplestcut}
      \end{aligned} \end{equation}
The second line is just a product of tree amplitudes with two scalar legs
and the factor in the first line sums supersymmetric Ward identity
(SWI) factors \cite{Grisaru:1976vm,Grisaru:1977px,Parke:1985pn}
due to two scalars and two helicities of the Majorana fermion circulating in the loop.
Due to supersymmetry, instead of complicating the cut integrand,
that sum helps to simplify it:
      \begin{equation} \begin{aligned}
            \sum_{h_1,h_2} \! A_1 A_2
                & = \frac{ (-1)^{k} \bra{m}\!P_{1k}|p]^2 }
                         { P_{1k}^2 [12] \dots [k\!-\!1|k]
                           \braket{k\!+\!1|k\!+\!2} \dots \braket{n\!-\!1|n} }
                    \frac{ \braket{l_1 m} \braket{l_2 m} [l_1 p] [l_2 p] }
                         { \braket{l_1 n} \braket{l_2|k\!+\!1} [l_1 1] [l_2 k] } \\
                  & \equiv \frac{F}{P_{1k}^2}
                    \frac{ \braket{l_1 m} \braket{l_2 m} [l_1 p] [l_2 p] }
                         { \braket{l_1 n} \braket{l_2|k\!+\!1} [l_1 1] [l_2 k] } \\
                & = \frac{F}{P_{1k}^2}
                    \frac{ \braket{l_1 m} \bra{l_1}\!P_{1k}|p]
                           \bra{m}\!P_{1k}|l_1] [p l_1] }
                         { \braket{l_1 n} \bra{l_1}\!P_{1k}|k]
                           \bra{k\!+\!1}\!P_{1k}|l_1] [1 l_1] } ,
      \label{simplestcut2}
      \end{aligned} \end{equation}
where in the last line by $F$ we denoted a kinematic factor
independent of loop momenta
and then we eliminated $l_2$ in favor of $l_1$.
Now the introduction of the homogeneous variables is trivial,
so after restoring the integration measure (\ref{dLIPS4}) we get:
      \begin{equation} \begin{aligned}
            \text{Cut}(P_{1k}) = &
                  - \frac{F}{4}
                    \int_{\tilde{\la}=\bar{\la}} \!
                    \frac{ \braket{\la d\la} [\lb d\lb] }
                         { \bra{\la}\!P_{1k}|\lb]^2 }
                    \frac{ \braket{\la|m} \bra{\la}\!P_{1k}|p]
                                          \bra{m}\!P_{1k}|\lb] [p|\lb] }
                         { \braket{\la|n} \bra{\la}\!P_{1k}|k]
                                          \bra{k\!+\!1}\!P_{1k}|\lb] [1|\lb] } .
      \label{simplestcut3}
      \end{aligned} \end{equation}
We then plug the spinorial integrand into (\ref{Cbub})
to obtain the bubble coefficient:
      \begin{equation} \begin{aligned}
            C^{\text{bub},P_{1k}}_{\mathcal{N}=1 \text{ chiral}} = & -
                    \frac{F}{(4\pi)^{\frac{d}{2}}i}
                    \sum_{\text{residues}}
                    \frac{1}{s}
                    \ln \left( 1 + s \frac{\bra{\la}\!q|\lb]}
                                          {\bra{\la}\!P_{1k}|\lb]}
                        \right) \\ & \times
                    \bigg\{
                        \frac{ \braket{\la|m} \bra{\la}\!P_{1k}|p]
                                              \bra{m}\!P_{1k}|\lb] [p|\lb] }
                             { \braket{\la|n} \bra{\la}\!P_{1k}|k]
                                              \bra{k\!+\!1}\!P_{1k}|\lb] [1|\lb] }
                        \frac{1}{\bra{\la}\!P_{1k}|q\!\ket{\la}}
                        \bigg|_{|\lb]=|P_{1k}+s\,q\ket{\la}}
                    \bigg\} \bigg|_{s=0} .
      \label{simplestbubble}
      \end{aligned} \end{equation}
Here we used the fact that the integrand (\ref{simplestcut2})
was homogeneous in $l_1$,
so the power of the derivative in $s$ is zero.
Therefore, only the first term in the expansion of
the logarithm $ \frac{1}{s} \ln(1+st) = t + O(s) $
survives in the limit $s \rightarrow 0$:
      \begin{equation} \begin{aligned}
            C^{\text{bub},P_{1k}}_{\mathcal{N}=1 \text{ chiral}} =
                    \frac{F}{(4\pi)^{\frac{d}{2}}i}
                    \sum_{\text{residues}}
                    \frac{[q|\lb]}
                         {\bra{\la}\!P_{1k}|\lb]}
                    \frac{ \braket{\la|m}^2 \bra{\la}\!P_{1k}|p]^2 }
                         { \braket{\la|k\!+\!1} \braket{\la|n} 
                           \bra{\la}\!P_{1k}|1] \bra{\la}\!P_{1k}|k]
                           \bra{\la}\!P_{1k}|q] } .
      \label{simplestbubble2}
      \end{aligned} \end{equation}
We see 5 poles in the denominator: $\ket{\la}=\ket{k\!+\!1}$, $\ket{\la}=\ket{n}$,
$\ket{\la}=|P_{1k}|1]$, $\ket{\la}=|P_{1k}|k]$ and $\ket{\la}=|P_{1k}|q]$.
Note that the factor $\bra{\la}\!K|\lb]$ never contains any poles
because in complex variable representation (\ref{complexvariables})
it becomes proportional to $(1+z\bar{z})$.
The sum of the residues produces the final answer:
      \begin{equation} \begin{aligned}
            C^{\text{bub},P_{1k}}_{\mathcal{N}=1 \text{ chiral}}
              (p^+\!\in\!\{1,\dots,k\}, m^-\!\in\!\{k\!+\!1,\dots,n\}) =
                    \frac{ (-1)^k }{ (4\pi)^{\frac{d}{2}}i }
                    \frac{ \bra{m}\!P_{1k}|p]^2 }
                         { [12] \dots [k\!-\!1|k]
                           \braket{k\!+\!1|k\!+\!2} \dots \braket{n\!-\!1|n} } \\
             \times \bigg\{
                    \frac{ \braket{m|k\!+\!1}^2
                           \bra{k\!+\!1}\!P_{1k}|p]^2 [k\!+\!1|q] }
                         { \braket{k\!+\!1|n}
                           \bra{k\!+\!1}\!P_{1k}|1] \bra{k\!+\!1}\!P_{1k}|k] 
                           \bra{k\!+\!1}\!P_{1k}|k\!+\!1] \bra{k\!+\!1}\!P_{1k}|q] }
                    & \\ +
                    \frac{ \braket{mn}^2 \bra{n}\!P_{1k}|p]^2 [nq] }
                         { \braket{n|k\!+\!1}
                           \bra{n}\!P_{1k}|1] \bra{n}\!P_{1k}|k] 
                           \bra{n}\!P_{1k}|n] \bra{n}\!P_{1k}|q] }
                    & \\ +
                    \frac{1}{P_{1k}^2}
                    \bigg(
                    \frac{[1p]^2}{[1k][1q]}
                    \frac{ \bra{m}\!P_{1k}|1]^2 \bra{1}\!P_{1k}|q] }
                         { \bra{1}\!P_{1k}|1]
                           \bra{k\!+\!1}\!P_{1k}|1] \bra{n}\!P_{1k}|1] }
                    & \\ +
                    \frac{[kp]^2}{[k1][kq]}
                    \frac{ \bra{m}\!P_{1k}|k]^2 \bra{k}\!P_{1k}|q] }
                         { \bra{k}\!P_{1k}|k]
                           \bra{k\!+\!1}\!P_{1k}|k] \bra{n}\!P_{1k}|k] }
                    & \\ +
                    \frac{[pq]^2}{[1q][kq]}
                    \frac{ \bra{m}\!P_{1k}|q]^2 }
                         { \bra{k\!+\!1}\!P_{1k}|q] \bra{n}\!P_{1k}|q] } &
                    \bigg)
                    \bigg\} .
      \label{simplestbubbleq}
      \end{aligned} \end{equation}

      Each term in (\ref{simplestbubble}) can be generically eliminated
by an appropriate choice of reference spinor $|q]$.
Moreover, specific helicity configurations can further simplify the formula.
For instance, if in a $P_{1,3}$-channel NMHV bubble
the plus-helicity leg gluon $p^+$ is $3^+$
followed by the minus-helicity gluon $m^- = 4^-$ and we pick $|q]=|3]$,
then only two terms survive:
      \begin{equation} \begin{aligned}
            C^{\text{bub},P_{1,3}}_{\mathcal{N}=1 \text{ chiral}}
              (1^-,2^-,3^+,4^-,5^+,\dots,n^+) = & \\
              \frac{1}{(4\pi)^{\frac{d}{2}} i}
              \frac{ \bra{4}\!P_{1,3}|3]^2 }
                   { [12] \braket{45} \dots \braket{n\!-\!1|n} \bra{n}\!P_{1,3}|1] } &
              \bigg\{
                \frac{ \bra{4}\!n|3] }
                     { [23] \bra{n}\!P_{1,3}|n] }
              + \frac{ \bra{2}\! 1|P_{1,3}\!\ket{4} }
                     { P_{1,3}^2 \bra{1}\!P_{1,3}|1] }
              \bigg\} .
      \label{N1chBubP13}
	\end{aligned} \end{equation}

      We checked on various examples that our result numerically coincides
with the equivalent all-$n$ formula found earlier in \cite{Dunbar:2009ax}.
More than that, we found that we can reproduce their formula term-by-term
by choosing in (\ref{simplestbubbleq}) $|q]=|P_{1k}\!\ket{m}$:
      \begin{equation} \begin{aligned}
            C^{\text{bub},P_{1k}}_{\mathcal{N}=1 \text{ chiral}}
              (p^+\!\in\!\{1,\dots,k\}, m^-\!\in\!\{k\!+\!1,\dots,n\}) =
                    \frac{ (-1)^k }{ (4\pi)^{\frac{d}{2}}i }
                    \frac{ \bra{m}\!P_{1k}|p]^2 }
                         { [12] \dots [k\!-\!1|k]
                           \braket{k\!+\!1|k\!+\!2} \dots \braket{n\!-\!1|n} }
             \! & \\ 
             \times \bigg\{
                    \frac{1}{P_{1k}^2 \braket{k\!+\!1|n}}
                    \bigg(
                    \frac{ \bra{k\!+\!1}\!P_{1k}|p]^2 \braket{m|P_{1k}|k\!+\!1|m} }
                         { \bra{k\!+\!1}\!P_{1k}|1] \bra{k\!+\!1}\!P_{1k}|k] 
                           \bra{k\!+\!1}\!P_{1k}|k\!+\!1] }
                    -
                    \frac{ \bra{n}\!P_{1k}|p]^2 \braket{m|P_{1k}|n|m} }
                         { \bra{n}\!P_{1k}|1] \bra{n}\!P_{1k}|k] 
                           \bra{n}\!P_{1k}|n] }
                    \bigg) & \\
                    +
                    \frac{1}{[1k]}
                    \bigg(
                    \frac{ [1p]^2 \braket{m|P_{1k}|1|m} }
                         { \bra{1}\!P_{1k}|1] \bra{k\!+\!1}\!P_{1k}|1]
                           \bra{n}\!P_{1k}|1] }
                    -
                    \frac{ [kp]^2 \braket{m|P_{1k}|k|m} }
                         { \bra{k}\!P_{1k}|k] \bra{k\!+\!1}\!P_{1k}|k]
                           \bra{n}\!P_{1k}|k] }
                    \bigg) &
                    \bigg\} .
      \label{simplestbubbledpw}
      \end{aligned} \end{equation}

      Needless to say, any bubble with an $\overline{\text{MHV}}$-MHV cut
can be obtained from the $P_{1k}$-channel bubble (\ref{simplestbubbleq}) by appropriate relabeling.

\subsection{Modified bubble formula}
\label{simplifiedbubble}

      We can already learn a more general lesson from the calculation
in Section \ref{simplestbubbles}.
The supersymmetric helicity sum is well known \cite{Bern:1994cg}
to simplify cut integrands instead of complicating them.

      We consider the $\mathcal{N}=1$ chiral multiplet
in the adjoint representation of the gauge group
which in fact has a effective $\mathcal{N}=2$ supersymmetry.
Thanks to that,
as we will show later in Section \ref{nmhvbubbles},
for all $\mathcal{N}=1$ chiral double cuts
the numerator and the denominator
have the same number of loop-momentum-dependent factors
and after introducing homogeneous variables $\la,\lb$
the only overall factor $\bra{\la}\!K|\lb]^{-2}$ comes
from the cut measure (\ref{dLIPS4}).
This means that when plugging the cut integrand
into the general bubble formula (\ref{Cbub})
we will always have a zero power of the derivative in $s$,
so we can set $s$ to zero from the start:
      \begin{equation}
            C^{\text{bub}}_{\mathcal{N}=1 \text{ chiral}} =
                  \frac{4}{(4\pi)^{\frac{d}{2}}i} \!
                  \sum_{\text{residues}}
                  \frac{\bra{\la}\!q|\lb]}
                       {\bra{\la}\!K|\lb]}
                  \bigg[ \mathcal{I}_{\text{spinor}}
                         \frac{\bra{\la}\!K|\lb]^2}
                              {\bra{\la}\!K|q\!\ket{\la}}
                  \bigg|_{|\lb]=|K\ket{\la}}
                  \bigg] .
      \label{CbubN1}
	\end{equation}
Taking into account that
      \begin{equation}
            \mathcal{I}_{\text{spinor}} =
                - \frac{K^2}{4} \frac{1}{\bra{\la}\!K|\lb]^2}
                  \sum_{h_1,h_2} \! A_1 A_2 ,
      \label{IspinorN1}
	\end{equation}
where by $\sum_{h_1,h_2} \! A_1 A_2$ we just mean the double cut
after loop-variable change, 
we retrieve a more direct formula for $\mathcal{N}=1$ chiral bubble coefficients:
      \begin{equation}
            C^{\text{bub}}_{\mathcal{N}=1 \text{ chiral}} =
                - \frac{K^2}{(4\pi)^{\frac{d}{2}}i} \!
                  \sum_{\text{residues}}
                  \frac{ [\lb|q] }
                       { \bra{\la}\!K|\lb] \bra{\la}\!K|q] }
                  \bigg[ \sum_{h_1,h_2} \! A_1 A_2
                  \bigg|_{|\lb]=|K\ket{\la}}
                  \bigg] .
      \label{CbubN2}
	\end{equation}

      Incidentally, a close analogue of (\ref{CbubN2}) has already been discovered
in \cite{Elvang:2011fx} with the help of $\mathcal{N}=1$ superspace.

\section{Cut integrand construction}
\label{cutconstruction}

      \begin{figure}[h]
      \centering
      \parbox{127pt}{ \begin{fmffile}{graph2}
      \fmfframe(10,10)(10,10){ \begin{fmfgraph*}(100,60)
            \fmflabel{$l_1$}{bottom}
            \fmflabel{$1^+$}{g1}
            \fmflabel{$\dots$}{dots1}
            \fmflabel{$m_1^-$}{gm1}
            \fmflabel{$\dots$}{dots2}
            \fmflabel{$m_2^-$}{gm2}
            \fmflabel{$\dots$}{dots3}
            \fmflabel{$k^+$}{gk}
            \fmflabel{$l_2$}{top}
            \fmflabel{$(k\!+\!1)^+$}{gk1}
            \fmflabel{$\dots$}{dots4}
            \fmflabel{$m_3^-$}{gm3}
            \fmflabel{$\dots$}{dots5}
            \fmflabel{$n^+$}{gn}
            \fmfleft{,dots1,,gm1,,dots2,,gm2,,dots3,}
            \fmfright{gn,dots5,gm3,dots4,gk1}
            \fmftop{,gk,top,,}
            \fmfbottom{,g1,bottom,,}
            \fmf{wiggly}{g1,vleft}
            \fmf{wiggly}{gm1,vleft}
            \fmf{wiggly}{gm2,vleft}
            \fmf{wiggly}{gk,vleft}
            \fmf{wiggly}{gk1,vright}
            \fmf{wiggly}{gm3,vright}
            \fmf{wiggly}{gn,vright}
            \fmf{dashes}{top,bottom}
            \fmf{plain,left,tension=1.0}{vleft,vright}
            \fmf{plain,right,tension=1.0}{vleft,vright}
            \fmfblob{0.17w}{vleft}
            \fmfblob{0.17w}{vright}
      \end{fmfgraph*} }
      \end{fmffile} }
      \caption{$P_{1k}$-channel cut for
               $ A_{\mathcal{N}=1 \text{ chiral}}^{\text{1-loop,NMHV}}
                   (m_1^-,m_2^-\!\in\!\{1,\dots,k\},
                          m_3^-\!\in\!\{k\!+\!1,\dots,n\}) $
               \label{cut1k}}
      \end{figure}
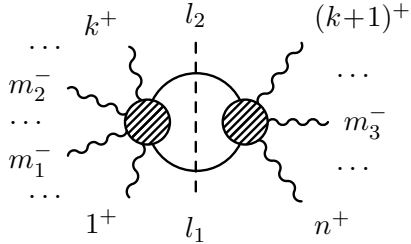

      Constructing appropriate cut integrands is crucial for using spinor integration
and getting clean analytic expressions.
In short, what we do is we sew tree amplitudes
and sum over the $\mathcal{N}=1$ chiral multiplet circulating in the cut.
An NMHV amplitude has 3 minus-helicity gluons,
so all its non-zero double cuts have an MHV amplitude on one side of the cut
and an NMHV one on the other side, as shown in fig. \ref{cut1k}.
For some 3-particle cuts the NMHV amplitude happens to be $\overline{\text{MHV}}$
and the computation is greatly simplified which we exploited
in Section \ref{simplestbubbles}.
But for all other integrands one needs to sew NMHV tree amplitudes,
which we describe in detail in the following section.

\subsection{NMHV tree amplitudes}
\label{nmhvtrees}

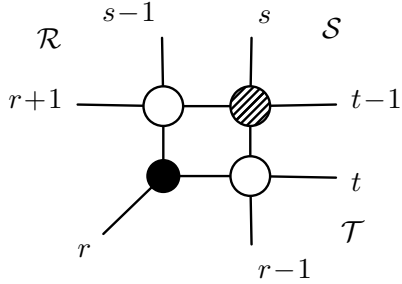
\begin{figure}[h]
      \centering
      \parbox{127pt}{ \begin{fmffile}{graph4}
      \fmfframe(10,10)(10,10){ \begin{fmfgraph*}(100,80)
            \fmflabel{$\mathcal{R}$}{R}
            \fmflabel{$\mathcal{S}$}{S}
            \fmflabel{$\mathcal{T}$}{T}
            \fmflabel{$r$}{r}
            \fmflabel{$s$}{s}
            \fmflabel{$t$}{t}
            \fmflabel{$r\!+\!1$}{rp1}
            \fmflabel{$r\!-\!1$}{rm1}
            \fmflabel{$s\!-\!1$}{sm1}
            \fmflabel{$t\!-\!1$}{tm1}
            \fmfbottom{,r,,ph1,,,rm1,,,}
            \fmftop{R,,,sm1,,,s,,S,}
            \fmfleft{,,,,ph2,,,,rp1,,,,}
            \fmfright{,,T,,t,,,,tm1,,,,}
            \fmf{plain,tension=0}{r,rv}
            \fmf{plain,tension=5.0}{rp1,Rv,Sv,tm1}
            \fmf{plain,tension=5.0}{s,Sv,Tv,rm1}
            \fmf{phantom,tension=5.0}{ph1,rv,Rv,sm1}
            \fmf{phantom,tension=5.0}{ph2,rv,Tv,t}
            \fmf{plain,tension=0}{rv,Rv,sm1}
            \fmf{plain,tension=0}{rv,Tv,t}
            \fmfv{decor.shape=circle,decor.size=0.12w}{rv}
            \fmfv{decor.shape=circle,decor.size=0.15w,decor.filled=empty}{Rv}
            \fmfblob{0.15w}{Sv}
            \fmfv{decor.shape=circle,decor.size=0.15w,decor.filled=empty}{Tv}
      \end{fmfgraph*} }
      \end{fmffile} }
      \caption{Cut box diagram for determining the values of $r$, $s$ and $t$
               \label{RST}}
      \end{figure}

      NMHV tree amplitudes are known to be encoded in the $\mathcal{N}=4$ SYM
$n$-point superamplitude \cite{Drummond:2008bq}:
      \begin{equation}
            \mathcal{A}_n^{\text{NMHV}} = \mathcal{A}_n^{\text{MHV}}
                                          \sum_{s=r+2}^{r+n-3} \, \sum_{t=s+2}^{r+n-1}
                                          \mathcal{R}_{rst} ,
      \label{NMHVtree}
      \end{equation}
where $r$ can be chosen arbitrarily.
The possible values of $s$ and $t$ (mod $n$) are already given
in the explicit double sum in \ref{NMHVtree},
but we also find insightful the following graphic approach
from \cite{Drummond:2008bq}.
After picking the $r$, one draws all cut-box-like diagrams
with one vertex having only one external leg $r$,
the opposite vertex with at least two external legs $s,\dots,t-1$ and
the other two vertices having at least one external legs, see fig. \ref{RST}.
For brevity, we denote the three collections of external legs as
$\mathcal{R} = \{r+1,\dots,s-1\}$,
$\mathcal{S} = \{s,\dots,t-1\}$ and
$\mathcal{T} = \{t,\dots,r-1\}$.

      Component amplitudes can then be extracted
from the super-amplitude (\ref{NMHVtree}) using, for example,
the package GGT \cite{Dixon:2010ik} in the following representation:
      \begin{subequations} \begin{align}
            A^{\text{tree}}(1_g^+, \overset{+}{\dots},
                            a_g^-, \overset{+}{\dots},
                            b_g^-, \overset{+}{\dots}, n_g^-) &
            = \frac{i}{ \braket{12} \braket{23} \dots \braket{n1} }
              \sum_{s=2}^{n-3} \sum_{t=s+2}^{n-1} R_{nst} D_{nst;ab}^4 , \\
            A^{\text{tree}}(1_g^+, \overset{+}{\dots},
                   a_\Lambda^{-A}, \overset{+}{\dots},
                 b_\Lambda^{+BCD}, \overset{+}{\dots},
                            c_g^-, \overset{+}{\dots}, n_g^-) &
            = \frac{i \epsilon^{ABCD} }{ \braket{12} \braket{23} \dots \braket{n1} }
              \sum_{s=2}^{n-3} \sum_{t=s+2}^{n-1}
              R_{nst} D_{nst;ac}^3 D_{nst;bc} , \\
            A^{\text{tree}}(1_g^+, \overset{+}{\dots},
                         a_S^{AB}, \overset{+}{\dots},
                         b_S^{CD}, \overset{+}{\dots},
                            c_g^-, \overset{+}{\dots}, n_g^-) &
            = \frac{i \epsilon^{ABCD} }{ \braket{12} \braket{23} \dots \braket{n1} }
              \sum_{s=2}^{n-3} \sum_{t=s+2}^{n-1}
              R_{nst} D_{nst;ac}^2 D_{nst;bc}^2 ,
	\end{align} \label{NMHVtrees} \end{subequations}
where $R_{rst}$ is just the bosonic part of $\mathcal{R}_{rst}$:
      \begin{equation}
            R_{rst} = \frac{-\braket{s\!-\!1|s} \braket{t\!-\!1|t} }
                           { P_{\mathcal{S}}^2
                             \braket{s\!-\!1|P_{\mathcal{S}}|P_{\mathcal{T}}|r}
                             \braket{s|P_{\mathcal{S}}|P_{\mathcal{T}}|r}
                             \braket{t\!-\!1|P_{\mathcal{S}}|P_{\mathcal{R}}|r}
                             \braket{t|P_{\mathcal{S}}|P_{\mathcal{R}}|r} } ,
      \label{Rrst}
      \end{equation}
whereas $D_{rst;ab}$ arise from differentiating the product of super-delta
functions inside $\mathcal{A}_n^{\text{MHV}}$ and $\mathcal{R}_{rst}$ :
      \begin{equation} \begin{aligned}
            D_{rst;ab} =
            \begin{cases}
                  \braket{ab} \braket{r|P_{\mathcal{S}}|P_{\mathcal{T}}|r} &
                  \text{if } a,b \in \mathcal{S} \\
                - \braket{br} \braket{a|P_{\mathcal{S}}|P_{\mathcal{T}}|r} &
                  \text{if } a \in \mathcal{S}, b \in \mathcal{R} \\
                  \braket{ar} \braket{b|P_{\mathcal{S}}|P_{\mathcal{T}}|r} &
                  \text{if } a \in \mathcal{R}, b \in \mathcal{S} \\
                  \braket{br} \braket{a|P_{\mathcal{S}}|P_{\mathcal{R}}|r} &
                  \text{if } a \in \mathcal{S}, b \in \mathcal{T} \\
                - \braket{ar} \braket{b|P_{\mathcal{S}}|P_{\mathcal{R}}|r} &
                  \text{if } a \in \mathcal{T}, b \in \mathcal{S} \\
                - P_{\mathcal{S}}^2 \braket{ar} \braket{br} &
                  \text{if } a \in \mathcal{R}, b \in \mathcal{T} \\
                  P_{\mathcal{S}}^2 \braket{ar} \braket{br} &
                  \text{if } a \in \mathcal{T}, b \in \mathcal{R} \\
                  0 & \text{otherwise} .
            \end{cases}
      \label{Drstab}
      \end{aligned} \end{equation}
By the derivation in Grassmann variables,
$D_{rst;ab}$ is antisymmetric in $a$ and $b$.

      From (\ref{NMHVtrees}) it is clear that,
much like ratios of spinor products relate
MHV amplitudes with fermions and scalars to purely gluonic ones
through standard supersymmetric Ward identities (SWI),
ratios of different $D_{rst;ab}$ do the same job for NMHV amplitude contributions.
We note here that one could in principle try to encode this information
using $\mathcal{N}=1$ superfields \cite{Bern:2009xq,Lal:2009gn,Elvang:2011fx}.
More than that, as we have already noted, the effective number of supersymmetries of
the $\mathcal{N}=1$ chiral multiplet in the adjoint representation is two,
so one can imagine even defining $\mathcal{N}=2$ hyper superspace.
However, even if Grassmann variables are undoubtedly
an indispensable tool for describing the theory in general,
sometimes they seem to put us farther away from calculating explicit formulas.
In this paper, we find it direct enough
to assemble the cut without introducing a superspace.

\subsection{Cut integrand}
\label{cutintegrand}

      Now we are ready to write down the cut integrand in full generality.
Consider the $P_{1k}$-channel cut shown on fig. \ref{cut1k}.
It has two minus-helicity gluons labeled $m_1^-$ and $m_2^-$ on the left of the cut
and one such gluon $m_3^-$ on the right.
Evidently, all other cuts can be obtained from this one by appropriate relabeling.

      A scalar cut would be just a product of
the right-hand side scalar MHV amplitude
and left-hand side scalar NMHV amplitude.
As explained above, to account for the fact that there are two scalars
and two helicities of the Majorana fermion circulating in the loop,
we multiply it further by a sum of SWI factors:
      \begin{equation} \begin{aligned}
            \sum_{h_1,h_2} A(-l_1^{\bar{h}_1},\dots,l_2^{h_2})
                           A(-l_2^{\bar{h}_2},\dots,l_1^{h_1})
                  = \frac{ i \braket{-l_2|m_3}^2 \braket{l_1|m_3}^2 }
                         { \braket{-l_2|k\!+\!1}
                           \braket{k\!+\!1|k\!+\!2} \dots \braket{n\!-\!1|n}
                           \braket{n|l_1} \braket{l_1|\!-\!l_2} } &\\\times
                    \frac{ i }
                         { \braket{-l_1|1} \braket{12} \dots \braket{k\!-\!1|k}
                           \braket{k|l_2} \braket{l_2|\!-\!l_1} }
                    \sum_{s=m_1+2}^{m_1-3} \, \sum_{t=s+2}^{m_1-1}
                    R_{m_1 st} D_{m_1 st; m_2(-l_1)}^2 D_{m_1 st; m_2 l_2}^2 &\\\times
                    \bigg(
                           \frac{ \braket{-l_2|m_3} D_{m_1 st; m_2(-l_1)} }
                                { \braket{ l_1|m_3} D_{m_1 st; m_2  l_2 } }
                         + 2
                         + \frac{ \braket{ l_1|m_3} D_{m_1 st; m_2  l_2 } }
                                { \braket{-l_2|m_3} D_{m_1 st; m_2(-l_1)} }
                    \bigg) & ,
      \label{genericcuta}
      \end{aligned} \end{equation}
where both sums in the second line go cyclically
over labels $\{-l_1,1,,\dots,k,l_2\}$.

      Note that in (\ref{genericcuta}) we picked $m_1$ to be the first argument
of $R_{rst}$ and $D_{rst;ab}$ and $m_2$ to be the last,
but in principle $m_1$ and $m_2$ can be interchanged
due to the arbitrariness of the choice of $r$ in the NMHV expansion (\ref{NMHVtree}),
which is a non-trivial property of tree amplitudes.
It comes from the BCFW recursion that underlies formulas
(\ref{NMHVtree})-(\ref{Drstab}) \cite{Drummond:2008cr}
and is related to the freedom of choosing BCFW shifts.
Anyway, the roles of $m_1$ and $m_2$ can also be interchanged
by a vertical flip of the amplitude.

      To make full use of the effective $\mathcal{N}=2$ supersymmetry
of the $\mathcal{N}=1$ chiral multiplet
in the adjoint representation of the gauge group
we rewrite it as follows:
      \begin{equation} \begin{aligned}
            \sum_{h_1,h_2} \! A_1 & A_2
                 = -\frac{ 1 }
                         { \braket{12} \dots \braket{k\!-\!1|k}
                           \braket{k\!+\!1|k\!+\!2} \dots \braket{n\!-\!1|n} }
                    \frac{ \braket{l_1|m_3} \braket{m_3|-\!l_2} }
                         { \braket{l_1|1} \braket{l_1|n} \braket{l_1|l_2}^2
                           \braket{k|l_2} \braket{k\!+\!1|l_2} } \\ \times & \!\!\!
                    \sum_{s=m_1+2}^{m_1-3} \, \sum_{t=s+2}^{m_1-1}
                    R_{m_1 st} D_{m_1 st; m_2(-l_1)} D_{m_1 st; m_2 l_2}
                    \big(
                      \!\braket{-l_2|m_3} D_{m_1 st; m_2(-l_1)}
                      + \braket{ l_1|m_3} D_{m_1 st; m_2  l_2 }
                    \big)^2 ,
      \label{genericcut}
      \end{aligned} \end{equation}
where the last factor squared is typically subject to non-trivial simplifications
involving the Schouten identity.

      The most important thing for applying spinor integration
is the dependence of the integrand on the loop variables.
Thus we need to do a case-by-case analysis of (\ref{genericcut}) to expose them.
But first we consider a helicity configuration
for which there is only one case that contributes.

\subsection{Simpler bubble coefficients}
\label{simplebubbles}

      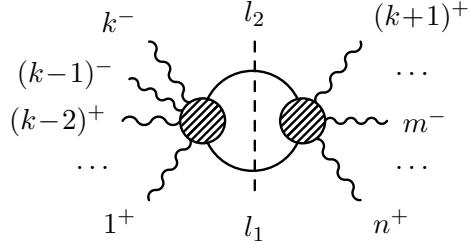
\begin{figure}[h]
      \centering
      \parbox{127pt}{ \begin{fmffile}{graph3}
      \fmfframe(10,10)(10,10){ \begin{fmfgraph*}(100,60)
            \fmflabel{$l_1$}{bottom}
            \fmflabel{$1^+$}{g1}
            \fmflabel{$\dots$}{dots1}
            \fmflabel{$(k\!-\!2)^+$}{gk2}
            \fmflabel{$(k\!-\!1)^-$}{gk1}
            \fmflabel{$k^-$}{gk}
            \fmflabel{$l_2$}{top}
            \fmflabel{$(k\!+\!1)^+$}{gk3}
            \fmflabel{$\dots$}{dots2}
            \fmflabel{$m^-$}{gm}
            \fmflabel{$\dots$}{dots3}
            \fmflabel{$n^+$}{gn}
            \fmfleft{g1,dots1,gk2,gk1,gk}
            \fmfright{gn,dots3,gm,dots2,gk3}
            \fmftop{top}
            \fmfbottom{bottom}
            \fmf{wiggly,tension=0.60}{g1,vleft}
            \fmf{phantom,tension=0.60}{dots1,vleft}
            \fmf{wiggly,tension=0.60}{gk2,vleft}
            \fmf{wiggly,tension=0.60}{gk1,vleft}
            \fmf{wiggly,tension=0.60}{gk,vleft}
            \fmf{wiggly}{gk3,vright}
            \fmf{wiggly}{gm,vright}
            \fmf{wiggly}{gn,vright}
            \fmf{dashes}{top,bottom}
            \fmf{plain,left,tension=1.0}{vleft,vright}
            \fmf{plain,right,tension=1.0}{vleft,vright}
            \fmfblob{0.17w}{vleft}
            \fmfblob{0.17w}{vright}
      \end{fmfgraph*} }
      \end{fmffile} }
      \caption{$P_{1k}$-channel cut for
               $ A_{\mathcal{N}=1 \text{ chiral}}^{\text{1-loop,NMHV}}
                   ((k\!-\!1)^-,k^-\!\in\!\{1,\dots,k\},
                          m^-\!\in\!\{k\!+\!1,\dots,n\}) $
               \label{cut1ksimple}}
      \end{figure}

      Consider a $P_{1k}$-channel cut
with minus-helicity gluons $k^-$ and $(k\!-\!1)^-$ adjacent to the cut.
The other negative helicity leg $m^-$ is at an arbitrary position
on the other side of the cut, see fig. \ref{cut1ksimple}.
It turns out that the general formula (\ref{genericcut})
simplifies greatly in this case.
Indeed, if we take $r=m_1=(k\!-\!1)$, $m_2=k$ and $m_3=m$,
we can see from definition (\ref{Drstab}) that
$ D_{(k-1)st; k l_2} $ is non-zero only for $s=l_2$,
because for all subsequent values of $s \in \{-l_1,\dots,k\!-\!4\} $
both $a=k$ and $b=l_2$ will belong to $\mathcal{R} = \{k\!-\!1,k,\dots,s\!-\!1\} $.
So the double sum in $s \in \{l_2,-l_1,\dots,k\!-\!4\} $
and $t \in \{s\!+\!2,\dots,k\!-\!2\} \subset \{1,\dots,k\!-\!2\} $
collapses to a single sum in $t \in \{1,\dots,k\!-\!2\} $:
      \begin{equation} \begin{aligned}
            \sum_{h_1,h_2} \! A_1 A_2
                 = &\frac{ 1 }
                         { \braket{12} \dots \braket{k\!-\!1|k}
                           \braket{k\!+\!1|k\!+\!2} \dots \braket{n\!-\!1|n} }
                    \frac{ \braket{l_1|m} \braket{-l_2|m} }
                         { \braket{n|l_1} \braket{l_1|1} \braket{l_1|l_2}^2
                           \braket{k|l_2} \braket{l_2|k\!+\!1} } \\ \times
                 &\!\sum_{t=1}^{k-2}
                    R_{(k-1) l_2 t} D_{(k-1) l_2 t; k(-l_1)} D_{(k-1) l_2 t; k l_2}
                    \big(
                      \!\braket{-l_2|m} D_{(k-1) l_2 t; k(-l_1)}
                      + \braket{ l_1|m} D_{(k-1) l_2 t; k l_2}
                    \big)^2 ,
      \label{simplecut}
      \end{aligned} \end{equation}
where we compute
      \begin{equation}
            R_{(k-1) l_2 t}
                  = \frac{ \braket{k|l_2} \braket{t\!-\!1|t} }
                         { P_{t,k-1}^2 P_{t,k}^2 \braket{k\!-\!1|k}^3
                           \braket{l_2|P_{t,k}|P_{t,k-1}|k\!-\!1}
                           \bra{t\!-\!1}\!P_{t,k}|k] \bra{t}\!P_{t,k}|k] } ,
      \label{Rsimple}
      \end{equation}
      \begin{subequations} \begin{align}
            D_{(k-1) l_2 t; k(-l_1)} & = \braket{k\!-\!1|k}
                                         \braket{-l_1|P_{t,k}|P_{t,k-1}|k\!-\!1} , \\
            D_{(k-1) l_2 t; k l_2}   & = \braket{k\!-\!1|k}
                                         \braket{ l_2|P_{t,k}|P_{t,k-1}|k\!-\!1} ,
      \end{align} \label{Dsimple} \end{subequations}
and the chiral sum is simplified by a Schouten identity:
      \begin{equation}
                   \braket{-l_2|m} D_{(k-1) l_2 t; k(-l_1)}
                 + \braket{ l_1|m} D_{(k-1) l_2 t; k l_2}
                 = \braket{k\!-\!1|k} \braket{l_1|l_2}
                   \braket{m|P_{t,k}|P_{t,k-1}|k\!-\!1} .
      \label{chsumsimple}
      \end{equation}
Putting all these ingredients together, we observe numerous cancellations and find 
      \begin{equation} \begin{aligned}
            \sum_{h_1,h_2} \! A_1 A_2 &
                  = \frac{ 1 }
                         { \braket{12} \dots \braket{k\!-\!2|k\!-\!1}
                           \braket{k\!+\!1|k\!+\!2} \dots \braket{n\!-\!1|n} } \\ &
             \times \sum_{t=1}^{k-2}
                    \frac{ \braket{m|P_{t,k}|P_{t,k-1}|k\!-\!1}^2
                           \braket{t\!-\!1|t} }
                         { P_{t,k-1}^2 P_{t,k}^2
                           \bra{t}\!P_{t,k}|k] \bra{t\!-\!1}\!P_{t,k}|k] }
                    \frac{ \braket{l_1|m}
                           \braket{l_1|P_{t,k}|P_{t,k-1}|k\!-\!1}
                           \braket{m|l_2} }
                         { \braket{l_1|1} \braket{l_1|n}
                           \braket{k\!+\!1|l_2} } \\ &
             \equiv \sum_{t=1}^{k-2}
                    \frac{ F_t \braket{t\!-\!1|t} }
                         { \bra{t\!-\!1}\!P_{t,k}|k] }
                    \frac{ \braket{l_1|m}
                           \braket{l_1|P_{t,k}|P_{t,k-1}|k\!-\!1}
                           \braket{m|l_2} }
                         { \braket{l_1|1} \braket{l_1|n}
                           \braket{k\!+\!1|l_2} } ,
      \label{simplecut2}
      \end{aligned} \end{equation}
where in the last line for brevity
we denoted the common factor independent of the loop momenta by $F_t$.
Note that, as expected, the number of loop momentum spinors is the same
for the numerator and the denominator.
Moreover, one should not miss the fact
that the $(t\!-\!1)$-th leg can become $(-l_1)$,
so poles are different for $t = 1$ and $t \neq 1$.
We then trade $l_2$ for $l_1$,
introduce the homogeneous variables
to find the following expression for the cut:
      \begin{equation}
            \sum_{h_1,h_2} \! A_1 A_2 = \! F_1
                    \frac{ \braket{\la|m}
                           \braket{\la|P_{1k}|P_{1,k-1}|k\!-\!1}
                           \bra{m}\!P_{1k}|\lb] }
                         { \bra{\la}\!P_{1k}|k] \braket{\la|n}
                           \bra{k\!+\!1}\!P_{1k}|\lb] }
                    + \!
                    \sum_{t=2}^{k-2}
                    \frac{ F_t \braket{t\!-\!1|t} }
                         { \bra{t\!-\!1}\!P_{t,k}|k] }
                    \frac{ \braket{\la|m}
                           \braket{\la|P_{t,k}|P_{t,k-1}|k\!-\!1}
                           \bra{m}\!P_{1k}|\lb] }
                         { \braket{\la|1} \braket{\la|n}
                           \bra{k\!+\!1}\!P_{1k}|\lb] } .
      \label{simplecut3}
      \end{equation}
To obtain the bubble coefficient, we plug this expression directly into
our simplified formula (\ref{CbubN2}):
      \begin{equation} \begin{aligned}
            C^{\text{bub},P_{1k}}_{\mathcal{N}=1 \text{ chiral}} =
                    \frac{P_{1k}^2}{(4\pi)^{\frac{d}{2}}i}
                    \sum_{\text{residues}}
                    \frac{[q|\lb]}
                         {\bra{\la}\!P_{1k}|\lb]}
                    \bigg\{
                    F_1
                    \frac{ \braket{\la|m}^2
                           \braket{\la|P_{1k}|P_{1,k-1}|k\!-\!1} }
                         { \braket{\la|k\!+\!1} \braket{\la|n} 
                           \bra{\la}\!P_{1k}|k] \bra{\la}\!P_{1k}|q] } & \\
                    + \sum_{t=2}^{k-2}
                    \frac{ F_t \braket{t\!-\!1|t} }
                         { \bra{t\!-\!1}\!P_{t,k}|k] }
                    \frac{ \braket{\la|m}^2
                           \braket{\la|P_{t,k}|P_{t,k-1}|k\!-\!1} }
                         { \braket{\la|1} \braket{\la|k\!+\!1}
                           \braket{\la|n} \bra{\la}\!P_{1k}|q] } &
                    \bigg\} .
      \label{simplebubble2}
      \end{aligned} \end{equation}
      We see 5 poles in the denominators:
$\ket{\la}=\ket{1}$, $\ket{\la}=\ket{k\!+\!1}$, $\ket{\la}=\ket{n}$,
$\ket{\la}=|P_{1k}|k]$ and $\ket{\la}=|P_{1k}|q]$.
The answer is then given by the sum of their residues:
      \begin{equation} \begin{aligned}
            C^{\text{bub},P_{1k}}_{\mathcal{N}=1 \text{ chiral}} =
                  \frac{1}{(4\pi)^{\frac{d}{2}}i}
                  \frac{ 1 }
                       { \braket{12} \dots \braket{k\!-\!2|k\!-\!1}
                         \braket{k\!+\!1|k\!+\!2} \dots \braket{n\!-\!1|n} }
                  \hspace{195pt} \\ \times
                  \bigg\{
                  \sum_{t=2}^{k-2} \!
                  \frac{ P_{1k}^2 \! \braket{m|P_{t,k}|P_{t,k-1}|k\!-\!1}^2 \!
                         \braket{t\!-\!1|t} \! }
                       { P_{t,k}^2 P_{t,k-1}^2 \!
                         \bra{t\!-\!1}\!P_{t,k}|k] \bra{t}\!P_{t,k}|k] }
                  \bigg(
                  \frac{ \braket{1m}^2 \!
                         \braket{1|P_{t,k}|P_{t,k-1}|k\!-\!1} [1q] }
                       { \braket{1n} \! \braket{1|k\!+\!1} \!
                         \bra{1}\!P_{1k}|1] \! \bra{1}\!P_{1k}|q] }
             \!+\!\frac{ \braket{nm}^2 \!
                         \braket{n|P_{t,k}|P_{t,k-1}|k\!-\!1} [nq] }
                       { \braket{n1} \! \braket{n|k\!+\!1} \!
                         \bra{n}\!P_{1k}|n] \! \bra{n}\!P_{1k}|q] } & \\
                + \frac{ \braket{k\!+\!1|m}^2
                         \braket{k\!+\!1|P_{t,k}|P_{t,k-1}|k\!-\!1} [k\!+\!1|q] }
                       { \braket{k\!+\!1|1} \! \braket{k\!+\!1|n} \!
                         \bra{k\!+\!1}\!P_{1k}|k\!+\!1] \!
                         \bra{k\!+\!1}\!P_{1k}|q] }
             \!+\!\frac{ \bra{m}\!P_{1k}|q]^2
                         \bra{k\!-\!1}\!P_{t,k-1}|P_{t,k}|P_{1,k}|q] }
                       { P_{1k}^2 \bra{1}\!P_{1k}|q] \!
                         \bra{k\!+\!1}\!P_{1k}|q] \! \bra{n}\!P_{1k}|q] } &
                  \bigg) \\
                + \frac{ \braket{m|P_{1k}|P_{1,k-1}|k\!-\!1}^2 }
                       { P_{1,k-1}^2 \! \bra{1}\!P_{1k}|k] }
                  \bigg(
                  \frac{1}{P_{1k}^2}
                  \bigg(
                  \frac{ \bra{m}\!P_{1k}|k]^2 \bra{k\!-\!1}\!P_{1k}|k]
                         \bra{k}\!P_{1k}|q] }
                       { \bra{n}\!P_{1k}|k] \bra{k\!+\!1}\!P_{1k}|k]
                         \bra{k}\!P_{1k}|k] [kq] }
             \!-\!\frac{ \bra{m}\!P_{1k}|q]^2 \bra{k\!-\!1}\!P_{1k}|q] }
                       { \bra{n}\!P_{1k}|q] \! \bra{k\!+\!1}\!P_{1k}|q] [kq] }
                  \bigg) & \\
                + \frac{ \braket{k\!+\!1|m}^2
                         \braket{k\!+\!1|P_{1k}|P_{1,k-1}|k\!-\!1} [k\!+\!1|q] }
                       { \braket{k\!+\!1|n} \! \bra{k\!+\!1}\!P_{1k}|k] \!
                         \bra{k\!+\!1}\!P_{1k}|k\!+\!1] \! \bra{k\!+\!1}\!P_{1k}|q] }
             \!+\!\frac{ \braket{nm}^2
                         \braket{n|P_{1k}|P_{1,k-1}|k\!-\!1} [nq] }
                       { \braket{n|k\!+\!1} \! \bra{n}\!P_{1k}|k] \!
                         \bra{n}\!P_{1k}|n] \! \bra{n}\!P_{1k}|q] } &
                  \bigg)
                  \bigg\} .
      \label{simplebubbleq}
      \end{aligned} \end{equation}
This expression can be further simplified by an appropriate choice
of the arbitrary spinor $|q]$.
For example, setting it equal to $|P_{1k}\!\ket{m}$ gives
the following formula:
      \begin{equation} \begin{aligned}
            C^{\text{bub},P_{1k}}_{\mathcal{N}=1 \text{ chiral}} =
                  \frac{1}{(4\pi)^{\frac{d}{2}}i}
                  \frac{ 1 }
                       { \braket{12} \dots \braket{k\!-\!2|k\!-\!1}
                         \braket{k\!+\!1|k\!+\!2} \dots \braket{n\!-\!1|n} }
                  \hspace{77pt} \\ \times
                  \bigg\{
                  \sum_{t=2}^{k-2}
                  \frac{ \braket{m|P_{t,k}|P_{t,k-1}|k\!-\!1}^2
                         \braket{t\!-\!1|t} }
                       { P_{t,k}^2 P_{t,k-1}^2
                         \bra{t\!-\!1}\!P_{t,k}|k] \bra{t}\!P_{t,k}|k] }
                  \bigg(
                  \frac{ \braket{m|P_{1k}|k\!+\!1|m}
                         \braket{k\!+\!1|P_{t,k}|P_{t,k-1}|k\!-\!1} }
                       { \braket{k\!+\!1|1} \braket{k\!+\!1|n}
                         \bra{k\!+\!1}\!P_{1k}|k\!+\!1] } & \\ +
                  \frac{ \braket{m|P_{1k}|1|m}
                         \braket{1|P_{t,k}|P_{t,k-1}|k\!-\!1} }
                       { \braket{1n} \braket{1|k\!+\!1}
                         \bra{1}\!P_{1k}|1] } +
                  \frac{ \braket{m|P_{1k}|n|m}
                         \braket{n|P_{t,k}|P_{t,k-1}|k\!-\!1} }
                       { \braket{n1} \braket{n|k\!+\!1}
                         \bra{n}\!P_{1k}|n] } &
                  \bigg) \\ +
                  \frac{ \braket{m|P_{1k}|P_{1,k-1}|k\!-\!1}^2 }
                       { P_{1k}^2 P_{1,k-1}^2
                         \bra{1}\!P_{1k}|k] }
                  \bigg(
                  \frac{ \braket{m|P_{1k}|k\!+\!1|m}
                         \braket{k\!+\!1|P_{1k}|P_{1,k-1}|k\!-\!1} }
                       { \braket{k\!+\!1|n}
                         \bra{k\!+\!1}\!P_{1k}|k]
                         \bra{k\!+\!1}\!P_{1k}|k\!+\!1] } \hspace{49pt} & \\ + \,
                  \frac{ \braket{m|P_{1k}|n|m}
                         \braket{n|P_{1k}|P_{1,k-1}|k\!-\!1} }
                       { \braket{n|k\!+\!1}
                         \bra{n}\!P_{1k}|k] \bra{n}\!P_{1k}|n] } \, + \,
                  \frac{ P_{1k}^2 \braket{m|P_{1k}|k|m}
                         \bra{k\!-\!1}\!P_{1k}|k] }
                       { \bra{n}\!P_{1k}|k] \bra{k\!+\!1}\!P_{1k}|k]
                         \bra{k}\!P_{1k}|k] } &
                  \bigg)
                  \bigg\} .
      \label{simplebubble}
      \end{aligned} \end{equation}
In the following sections, we choose to provide only formulas with $q$ left arbitrary.

\section{Loop momentum dependence}
\label{loopmomentum}

In this section, we carefully study the dependence
of the cut expression \eqref{genericcut} on the cut loop momenta $l_1$ and $l_2$.
Later in Section \ref{nmhvbubbles}, we change them in favor of
homogeneous variables $\la$, $\lb$ to find the bubble coefficient
corresponding to that cut.

\subsection{Case-by-case analysis}
\label{caseanalysis}

\begin{figure}[h]
      \centering
            \includegraphics[scale=0.67]{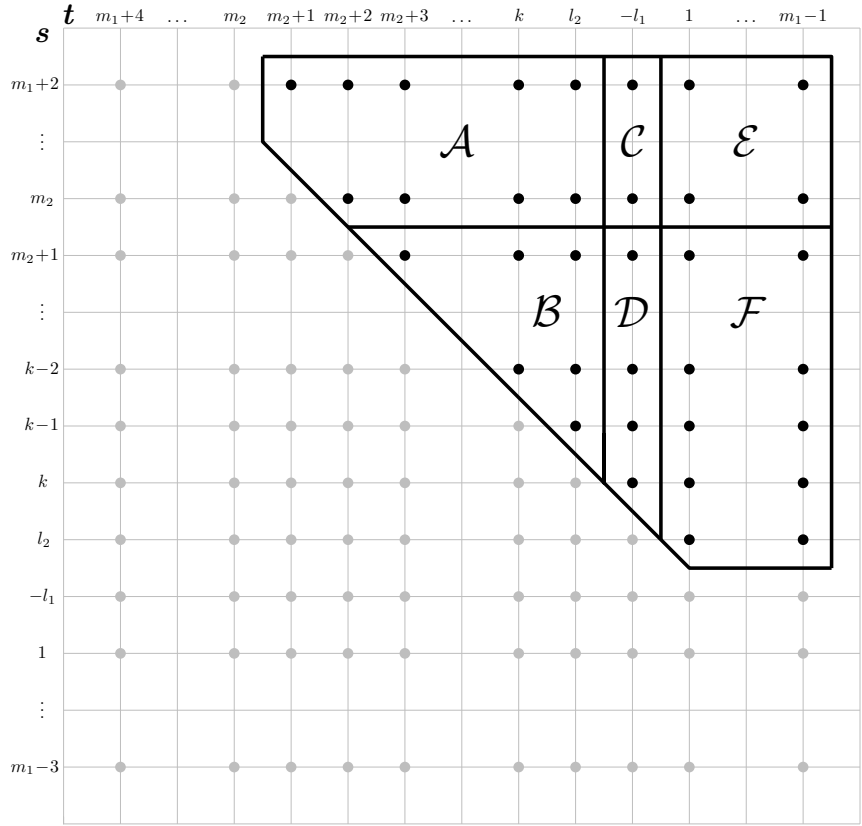}
      \caption{Values of $s$ and $t$ corresponding to non-zero contributions to the
               $P_{1k}$-channel cut.
               \label{stmesh}}
      \end{figure}

First of all, we find that in \eqref{genericcut}
the factor most frequently equal to zero is
      \begin{equation} \begin{aligned}
            D_{m_1 st; m_2 l_2} \! = \!
            \begin{cases}
            \quad \braket{m_1|l_2} \braket{m_1|P_{m_1+1,s-1}|P_{s,t-1}|m_2} &
                  \text{if } \{s,t\} \in \mathcal{A} \\
                - P_{s,t-1}^2 \braket{m_1|m_2} \braket{m_1|l_2} &
                  \text{if } \{s,t\} \in \mathcal{B} \\
            \quad \braket{m_2|l_2}
                  \big( \bra{m_1}\!P_{m_1+1,s-1}|l_1] \braket{l_1|m_1}
                      - \braket{m_1|P_{m_1+1,s-1}|P_{1,m_1-1}|m_1} \big) &
                  \text{if } \{s,t\} \in \mathcal{C} \\
                - \braket{m_1|m_2}
                  \big( \braket{m_1|l_1} [l_1|P_{s,k}\ket{l_2}
                      - \braket{m_1|P_{1,m_1-1}|P_{s,k}|l_2} \big) &
                  \text{if } \{s,t\} \in \mathcal{D} \\
            \quad \braket{m_2|l_2} \braket{m_1|P_{t,m_1-1}|P_{m_1+1,s-1}|m_1} &
                  \text{if } \{s,t\} \in \mathcal{E} \\
                - \braket{m_1|m_2} \braket{m_1|P_{t,m_1-1}|P_{t,s-1}|l_2} &
                  \text{if } \{s,t\} \in \mathcal{F} \\
            \quad 0 & \text{otherwise} ,
            \end{cases}
      \label{D2cases}
      \end{aligned} \end{equation}
where we define the non-zero cases:
      \begin{subequations} \begin{align}
                  \mathcal{A}: \; & s \in \{m_1\!+\!2,\dots,m_2\},
                             \;\;\; t \in \{m_2\!+\!1,\dots,l_2\} \\
                  \mathcal{B}: \; & s \in \{m_2\!+\!1,\dots,k\!-\!1\},
                               \,   t \in \{m_2\!+\!3,\dots,l_2\} \\
                  \mathcal{C}: \; & s \in \{m_1\!+\!2,\dots,m_2\},
                               \;\; t = -l_1 \\
                  \mathcal{D}: \; & s \in \{m_2\!+\!1,\dots,k\},
                            \quad\; t = -l_1 \\
                  \mathcal{E}: \; & s \in \{m_1\!+\!2,\dots,m_2\},
                               \;   t \in \{1,\dots,m_1\!-\!1\} \\
                  \mathcal{F}: \; & s \in \{m_2\!+\!1,\dots,l_2\},
                             \;\;\; t \in \{1,\dots,m_1\!-\!1\} .
      \label{cases}
      \end{align} \end{subequations}
Thus, we need to consider all other factors solely in these six cases.
For clearness, we depict them on a two-dimensional mesh in fig. \ref{stmesh}.

      Next, we expose the loop-momentum dependence of $D_{m_1 st; m_2(-l_1)}$:
      \begin{equation} \begin{aligned}
            D_{m_1 st; m_2(-l_1)} \! = \!
            \begin{cases}
                - \braket{-l_1|m_1} \braket{m_1|P_{m_1+1,s-1}|P_{s,t-1}|m_2} &
                  \text{if } \{s,t\} \in \mathcal{A} \\
                  P_{s,t-1}^2 \braket{-l_1|m_1} \braket{m_1|m_2} &
                  \text{if } \{s,t\} \in \mathcal{B} \\
                - \braket{-l_1|m_1}
                  \big( \bra{m_1}\!P_{m_1+1,s-1}|l_1] \braket{l_1|m_2}
                      - \braket{m_2|P_{1,s-1}|P_{s,m_1-1}|m_1} \big) &
                  \text{if } \{s,t\} \in \mathcal{C} \\
                  P_{-l_1,s-1}^2 \braket{-l_1|m_1} \braket{m_1|m_2} &
                  \text{if } \{s,t\} \in \mathcal{D} \\
                - \braket{-l_1|m_2} \braket{m_1|P_{t,m_1-1}|P_{m_1+1,s-1}|m_1} &
                  \text{if } \{s,t\} \in \mathcal{E} \\
                  \braket{m_1|m_2} \braket{-l_1|P_{t,s-1}|P_{t,m_1-1}|m_1} &
                  \text{if } \{s,t\} \in \mathcal{F} .
           \end{cases}
      \label{D1cases}
      \end{aligned} \end{equation}
Then we combine the $D$-terms together,
and, after applying Schouten identities where necessary, we find:
      \begin{equation} \begin{aligned}
          & \braket{-l_2|m_3} D_{m_1 st; m_2(-l_1)}
          + \braket{ l_1|m_3} D_{m_1 st; m_2  l_2 } \\ &
          = \begin{cases}
            \quad \braket{l_1 l_2} \braket{m_1 m_3}
                                   \braket{m_1|P_{m_1+1,s-1}|P_{s,t-1}|m_2} &
                  \text{if } \{s,t\} \in \mathcal{A} \\
            \quad P_{s,t-1}^2 \braket{l_1 l_2} \braket{m_1 m_2} \braket{m_3 m_1} &
                  \text{if } \{s,t\} \in \mathcal{B} \\
                - \big( \braket{l_1 l_2} \braket{m_2 m_3}
                        \braket{m_1|P_{1,s-1}|P_{s,m_1-1}|m_1} \\ \,\,\,
                      + \braket{m_1 m_2} \braket{m_3 l_2}
                        \braket{l_1|P_{1,s-1}|P_{s,m_1-1}|m_1} &
                  \text{if } \{s,t\} \in \mathcal{C} \\ \,\,\,
                      - \braket{m_2 m_3} \braket{l_1 m_1}
                        \braket{m_1|P_{m_1+1,s-1}|P_{1,k}|l_2}
                  \big) \\
                - \braket{m_1 m_2} \big( \braket{l_1 m_1}
                                    \braket{m_3|P_{m_1,s-1}|P_{s,k}|l_2} &
                  \text{if } \{s,t\} \in \mathcal{D} \\
                             \quad \quad \quad \quad \,
                      + \braket{m_3 m_1}
                        \braket{l_1|P_{1,m_1-1}|P_{s,k}|l_2} \big) \\
                - \braket{l_1 l_2} \braket{m_2 m_3}
                                   \braket{m_1|P_{t,m_1-1}|P_{m_1+1,s-1}|m_1} &
                  \text{if } \{s,t\} \in \mathcal{E} \\
            \quad \braket{l_1 l_2} \braket{m_1 m_2}
                                   \braket{m_3|P_{t,s-1}|P_{t,m_1-1}|m_1} &
                  \text{if } \{s,t\} \in \mathcal{F} .
            \end{cases}
      \label{chsum}
      \end{aligned} \end{equation}

      Finally, we write three distinct cases for the $R_{m_1 st}$ factor:
      \begin{equation} \begin{aligned}
          & R_{m_1 st} \\ & = \!
            \begin{cases}
                - \braket{s\!-\!1|s} \braket{t\!-\!1|t} \!/
                  \big( P_{s,t-1}^2 \!\!\!\!\!\!
                      & \braket{m_1|P_{m_1+1,t-1}|P_{s,t-1}|s\!-\!1}
                        \braket{m_1|P_{m_1+1,t-1}|P_{s+1,t-1}|s} \\ \!\!\!\!\!
                      & \braket{m_1|P_{m_1+1,s-1}|P_{s,t-2}|t\!-\!1}
                        \braket{m_1|P_{m_1+1,s-1}|P_{s,t-1}|t} \!
                  \big) \hspace{13pt}
                  \text{if } \{s,t\} \in \mathcal{A} \cup \mathcal{B} \\ \quad
                  \braket{s\!-\!1|s} \braket{l_1 l_2} \!/
                  \big( P_{-l_1,s-1}^2 \!\!\!\!\!\!
                      & \braket{m_1|P_{-l_1,m_1-1}|P_{-l_1,s-2}|s\!-\!1}
                        \braket{m_1|P_{-l_1,m_1-1}|P_{-l_1,s-1}|s} \\ \!\!\!\!\!
                      & \braket{l_1|P_{1,s-1}|P_{m_1+1,s-1}|m_1}
                        \braket{m_1|P_{m_1+1,s-1}|P_{s,k}|l_2} \!
                  \big) \hspace{28pt}
                  \text{if } \{s,t\} \in \mathcal{C} \cup \mathcal{D} \\ \quad
                  \braket{s\!-\!1|s} \braket{t|t\!-\!1} \!/
                  \big( P_{t,s-1}^2 \!\!\!\!\!\!
                      & \braket{m_1|P_{t,m_1-1}|P_{t,s-2}|s\!-\!1}
                        \braket{m_1|P_{t,m_1-1}|P_{t,s-1}|s} \\ \!\!\!\!\!
                      & \braket{m_1|P_{m_1+1,s-1}|P_{t+1,s-1}|t}
                        \braket{m_1|P_{m_1+1,s-1}|P_{t,s-1}|t\!-\!1} \!
                  \big) \hspace{2pt}
                  \text{if } \{s,t\} \in \mathcal{E} \cup \mathcal{F} .
            \end{cases}
      \label{Rcases}
      \end{aligned} \end{equation}
Here, the first and the third cases can develop simple loop dependence
on the borders of their respective domains:
in the first case $\ket{t}$ can be become includes $\ket{l_2}$, whereas
the third case includes $s=l_2$ and $t=1 \Rightarrow t-1=-l_1$
which have even a non-trivial overlap.
These subcases can only lead to loop spinors appearing
on the edges of spinor products and we will deal with these cases along the way.

      Of course, in some particular lower-point cases these formulae
can be simplified further using momentum conservation and Schouten identities,
but they are simple enough for us to proceed in full generality.

\subsection{NMHV pole structure}
\label{poles}

      In principle, to obtain explicit bubble coefficients formulas,
all that remains to do is
to make loop-variable change in the cut integrand (\ref{genericcut})
and plug it into our simplified master formula (\ref{CbubN2})
in which the only non-trivial operation is taking spinor residues
with respect to $\la$.
We need to do it separately for different cases $\mathcal{A}$ through $\mathcal{F}$
and their subcases with slightly modified loop dependence and then sum over the cases.
Thus, we write \emph{a frame formula for a generic NMHV bubble coefficient}:
\begin{equation} \begin{aligned}
      C^{\text{bub},P_{1k}}_{\mathcal{N}=1 \text{ chiral}}
        (m_1^-,m_2^-\!\in\!\{1,\dots,k\}, m_3^-\!\in&\{k\!+\!1,\dots,n\}) \\
          = \sum_{ \{s,t\} \in \mathcal{A} } R_{\mathcal{A}}^{s,t}
          + \sum_{ \{s,t\} \in \mathcal{B} } R_{\mathcal{B}}^{s,t}
 +\!\!\!\!\!\!\sum_{ \{s,t=-l_1\} \in \mathcal{C} } \!\!\!\!\! & R_{\mathcal{C}}^s
 +\!\!\!\!\!\!\sum_{ \{s,t=-l_1\} \in \mathcal{D} } \!\!\!\!\! R_{\mathcal{D}}^s
          + \sum_{ \{s,t\} \in \mathcal{E} } R_{\mathcal{E}}^{s,t}
          + \sum_{ \{s,t\} \in \mathcal{F} } R_{\mathcal{F}}^{s,t} ,
\label{CbubN1frame}
\end{aligned} \end{equation}
where we introduced a shorthand notation for residue sums
of each individual contribution to the cut (\ref{genericcut}).

      However, it is well known
\cite{Feng:2008ju,Hodges:2009hk,ArkaniHamed:2010gg,ArkaniHamed:2012nw}
that, in contrast to the Parke-Taylor MHV amplitudes \cite{Parke:1986gb},
the tree-level NMHV amplitudes derived from BCFW recursion contain spurious poles,
i.~e. poles that do not correspond to any physical propagator.
They can be viewed as an artifact of the on-shell derivation,
or as a price to pay to have more compact expressions
than what one would obtain from Feynman diagram calculations.
These poles obtain a geometrical meaning in (momentum) twistor variables
\cite{Hodges:2009hk,ArkaniHamed:2010gg,ArkaniHamed:2012nw}.

      Fortunately, by definition \emph{spurious poles have zero residues},
so we can just omit them in our calculation of bubble coefficients.
To do this, we need to tell them apart from physical poles.
As already mentioned, the common MHV prefactor of (\ref{NMHVtree})
contains only physical poles.
Evidently, spurious poles come from denominators
of different $\mathcal{R}$-invariants.
Each term can have a non-zero spurious residue,
but they are bound to cancel in a sum over $s$ and $t$.

\begin{figure}[h]
      \centering
      \parbox{127pt}{ \begin{fmffile}{graph5}
      \fmfframe(10,10)(10,10){ \begin{fmfgraph*}(100,60)
            \fmflabel{$l_1$}{bottom}
            \fmflabel{$\cdots 1^+$}{g1}
            \fmflabel{$m_1^-$}{gm1}
            \fmflabel{$\dots$}{dots1}
            \fmflabel{$j^+$}{gj}
            \fmflabel{$(j\!+\!1)^+$}{gj1}
            \fmflabel{$\dots$}{dots2}
            \fmflabel{$m_2^-$}{gm2}
            \fmflabel{$\cdots k^+$}{gk}
            \fmflabel{$l_2$}{top}
            \fmflabel{$(k\!+\!1)^+$}{gk1}
            \fmflabel{$\dots$}{dots3}
            \fmflabel{$m_3^-$}{gm3}
            \fmflabel{$\dots$}{dots4}
            \fmflabel{$n^+$}{gn}
            \fmfleft{,gm1,dots1,,gj,,left,,gj1,,dots2,gm2,}
            \fmfright{gn,dots4,gm3,dots3,gk1}
            \fmftop{,,gk,top,,,}
            \fmfbottom{,,g1,bottom,,,}
            \fmf{wiggly,tension=5.0}{g1,vleft}
            \fmf{wiggly}{gm1,vleft}
            \fmf{wiggly,tension=0.5}{gj,vleft}
            \fmf{dashes,tension=0}{left,vleft,center}
            \fmf{wiggly,tension=0.5}{gj1,vleft}
            \fmf{wiggly}{gm2,vleft}
            \fmf{wiggly,tension=5.0}{gk,vleft}
            \fmf{wiggly}{gk1,vright}
            \fmf{wiggly}{gm3,vright}
            \fmf{wiggly}{gn,vright}
            \fmf{dashes,tension=5.0}{top,center,bottom}
            \fmf{plain,left,tension=1.0}{vleft,vright}
            \fmf{plain,right,tension=1.0}{vleft,vright}
            \fmfblob{0.17w}{vleft}
            \fmfblob{0.17w}{vright}
      \end{fmfgraph*} }
      \end{fmffile} }
      \caption{$P_{1k}$-channel cut for
               $ A_{\mathcal{N}=1 \text{ chiral}}^{\text{1-loop,NMHV}}
                   (m_1^-,m_2^-\!\in\!\{1,\dots,k\},
                          m_3^-\!\in\!\{k\!+\!1,\dots,n\}) $
               has non-vanishing three-mass triple cuts
               only between $m_1$ and $m_2$.
               \label{double2triple}}
      \end{figure}
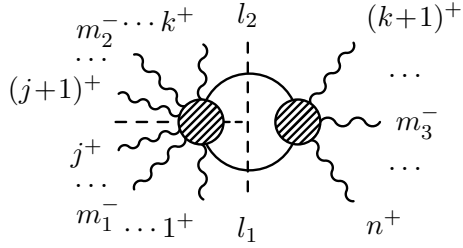

      Of course, for our one-loop calculation
we are only concerned by telling apart poles that depend on the loop momentum.
The common MHV denominator in (\ref{genericcut}) already captures
four massless physical poles:
$\braket{l_1|1} \Rightarrow (l_1-p_1)^2$,
$\braket{k|l_2} \Rightarrow (l_2+p_k)^2$,
$\braket{k\!+\!1|l_2} \Rightarrow (l_2-p_{k+1})^2$,
$\braket{l_1|n} \Rightarrow (l_1+p_n)^2$.
So what we seek is a physical massive pole
that has to look like
\begin{equation}
      P_{-l_1,j}^2 = (l_1 - P_{1,j})^2 = (l_2 + P_{j+1,k})^2 = P_{j+1,l_2}^2 .
\label{massivepole}
\end{equation}
Moreover, the presence of such a pole means that one can cut it and obtain
a non-zero three-mass triple cut, which can only occur
if we cut between the two minus-helicity gluons
on the left-hand side of the double cut, see fig. \ref{double2triple}.
Therefore, $j \in \{m_1,\dots,m_2\!-\!1\} \cap \{2,\dots,k\!-\!2\} $.

Let us then examine one-by-one each of the five denominators
of $R_{m_1 st}$:

\begin{enumerate}
\item $ P_{\mathcal{S}}^2 = P_{s,t-1}^2 $
      produces massive physical poles, unless it is canceled by the numerator;
      it develops the desired loop-momentum dependence (\ref{massivepole})
      for either $\{s=-l_1, t=j\!+\!1\}$ or $\{s=j\!+\!1, t=-l_1\}$.
      However, the position of $s$ in the former also constrains
      $t$ to be in $\{2,\dots,m_1\!-\!1\}$,
      which is inconsistent with $j = t\!-\!1 \in \{m_1,\dots,m_2\!-\!1\}$,
      so only the latter case is meaningful.
\item $ \braket{s\!-\!1|P_{\mathcal{S}}|P_{\mathcal{T}}|r}
      = \braket{m_1|P_{t,m_1}|P_{t,s-1}|s\!-\!1} $
      can obviously produce a non-zero momentum square only
      if the two spinor arguments become adjacent.
      With $s \in \{m_1\!+\!2,\dots,m_1\!-\!3\}$
      it is only possible in case $s\!-\!1=m_1\!+\!1$.
      Moreover, to obtain the right loop dependence (\ref{massivepole}),
      we need to have $t=-l_1$, for which this denominator becomes
      $ \braket{m_1|m_1\!+\!1} P_{-l_1,m_1}^2 $.
\item $ \braket{s|P_{\mathcal{S}}|P_{\mathcal{T}}|r}
      = \braket{m_1|P_{t,m_1}|P_{t,s}|s} $
      cannot produce a momentum square
      as $s$ is never adjacent to $m_1$.
\item $ \braket{t\!-\!1|P_{\mathcal{S}}|P_{\mathcal{R}}|r}
      = \braket{m_1|P_{s,m_1}|P_{s,t-1}|t\!-\!1} $
      cannot produce a momentum square
      because $t\!-\!1$ is never adjacent to $m_1$.
\item $ \braket{t|P_{\mathcal{S}}|P_{\mathcal{R}}|r}
      = \braket{m_1|P_{s,m_1}|P_{s,t}|t} $
      can be factorized with a momentum square as
      $ \braket{m_1|m_1\!-\!1} P_{s,m_1-1}^2 $ for $t=m_1\!-\!1$,
      but it cannot result in
      the desired loop-momentum dependence (\ref{massivepole}) for any $s$.
\end{enumerate}

      Thus we have only two potential sources of physical massive poles:
the first one, $P_{-l_1,m_1}^2$, comes from factorizing
$ \braket{s\!-\!1|P_{\mathcal{S}}|P_{\mathcal{T}}|r} $ for $s=m_1\!+\!2$,
while all subsequent poles come simply from $P_{\mathcal{S}}^2$
for $s \in \{m_1\!+\!2,\dots,\min(m_2,k\!-\!1)\}$.
In both cases $t$ remains equal to $-l_1$,
which corresponds to cases $\mathcal{C}$ and $\mathcal{D}$.
Moreover, the only way a massive pole can occur in case $\mathcal{D}$
is having the minus-helicity gluons adjacent to each other: $m_2=m_1\!+\!1$,
so that $s=m_2\!+\!1=m_1\!+\!2$.

      To sum up, for a generic helicity configuration,
case $\mathcal{C}$ contains all physical massive poles:
\begin{itemize}
\item $R_{m_1(m_1\!+\!2)(-l_1)}$ generate two poles
      $P_{-l_1,m_1}^2$ and $P_{-l_1,m_1\!+\!1}^2$;
\item subsequent $R_{m_1 s (-l_1)}$ with $s \in \{m_1\!+\!3,\dots,m_2\}$
      each have only one pole $P_{-l_1,s-1}^2$.
\end{itemize}
The configuration with two adjacent minus-helicity gluons
generates a single physical massive pole $P_{-l_1,m_1}^2$
through $R_{m_1(m_2\!+\!1)(-l_1)}$ which belongs to case $\mathcal{D}$.
All other non-MHV-like loop-dependent poles are spurious
and thus can be omitted in the sum over residues.

\subsection{Massive pole residues}
\label{massivepoles}

      In this section, we specify how we take residues of massive poles.
If we have such a pole
\begin{equation}
      (l_1-P_{1,i})^2 =  P_{1k}^2 \frac{ \bra{\la}\!Q_i|\lb] }
                                       { \bra{\la}\!P_{1k}|\lb] } ,
\label{pole1i}
\end{equation}
after using \eqref{CbubN2}, it becomes proportional to $\braket{\la|Q_i|K|\la}$ .
Then from the definitions of $Q_i$ (\ref{Qi4dim}),
$P^i_\pm$ (\ref{Ptri}) and $x^i_\pm$ (\ref{Xtri}),
one can deduce that
      \begin{equation}
            \braket{\la|Q_i|K|\la}
              = - \frac{ \braket{\la|\la^i_+}[\lb^i_+|\lb^i_-]
                         \braket{\la^i_-|\la} }
                       { x^i_+ - x^i_- } .
      \label{trianglepoles}
	\end{equation}
This lets us split a massive pole into two massless ones,
which is why we introduce momenta $P^i_\pm$ in the first place.
So after taking the residues in the standard way (\ref{simplespinorpole})
and doing some simplifications we obtain the following simple prescription:
      \begin{equation}
            \Res_{\lambda=\la^i_\pm} \frac{ F(\la,\lb) }
                                          { \bra{\la}\!P_{1k}|\lb]
                                            \braket{\la|Q_i|P_{1k}|\la} }
                  = - \frac{  F(\la^i_\pm,\lb^i_\pm) }
                           { 4 \left( ( P_{1,i-1} \cdot P_{i,k} )^2
                                      - P_{1,i-1}^2 P_{i,k}^2
                               \right) } .
      \label{massivepoleres}
      \end{equation}

      The drawback of our method is that it introduces a superficial non-rationality
in otherwise rational coefficient formulas.
Indeed, massless momenta $P^i_\pm$ are defined in (\ref{Ptri}) through $x^i_\pm$
which contain a non-trivial square root
$\sqrt{ (K \cdot Q_i)^2 - K^2 Q_i^2 }
=\sqrt{ ( P_{1,i-1} \cdot P_{i,k} )^2 - P_{1,i-1}^2 P_{i,k}^2 }$.
However, this square root dependence is guaranteed to effectively cancel
in the sum over $\pm$-solutions.

      Other methods may produce explicitly rational expressions, such as
the three-mass triangle formula from \cite{BjerrumBohr:2007vu,Dunbar:2009ax}
given in Appendix \ref{app:triangles},
where our approach (\ref{Ctri}) would generate superficially non-rational results.
We leave dealing with this minor issue for future work.

\section{n-point bubble coefficients}
\label{nmhvbubbles}

      In this section, we obtain our main results ---
analytic formulas for each term
in the NMHV bubble coefficient formula \eqref{CbubN1frame}.
We go through all the cases, starting with those which can contain massive poles.

\subsection{Case $\mathcal{D}$}
\label{case:d}

      First, we study case $\mathcal{D}$, because it only contains a massive pole
in a particular subcase of two adjacent minus-helicity gluons,
which can perfectly illustrate our method
in presence of spurious and massive poles.

\subsubsection{Special $\mathcal{D}$-case contribution
               for adjacent negative helicities}
\label{subcase:adjacentd}

      In this section, we consider in detail case $\mathcal{D}$
for the configuration with adjacent negative-helicity gluons: $m_2=m_1+1$.
As was shown in Section \ref{poles},
the term containing the only physical massive pole in that bubble coefficient
is generated by
      \begin{equation} \begin{aligned}
          & R_{m_1(m_1+2)(-l_1)} \\ & =
            \frac{ \braket{m_1\!+\!1|m_1\!+\!2} \braket{l_1|l_2} }
                 { P_{-l_1,m_1}^2 P_{-l_1,m_1+1}^2 \!
                   \braket{m_1|m_1\!+\!1}^3 \!
                   \braket{m_1|P_{-l_1,m_1-1}|P_{-l_1,m_1+1}|m_1\!+\!2} 
                   \bra{l_1}\!P_{1,m_1}|m_1\!+\!1]
                   [m_1\!+\!1|P_{m_1+2,k}\!\ket{l_2} } .
      \label{R:adjacent}
      \end{aligned} \end{equation}
We can simplify other factors further with respect to the corresponding expressions
in (\ref{D2cases}) and (\ref{D1cases}):
      \begin{subequations} \begin{align}
            D_{m_1(m_1+2)(-l_1); m_2 l_2} & =
                  P_{-l_1,m_1+1}^2 \braket{m_1|m_1\!+\!1} \braket{-l_1|m_1} , \\
            D_{m_1(m_1+2)(-l_1); m_2(-l_1)} & =
                  \braket{m_1|m_1\!+\!1}
                  \braket{m_1|P_{-l_1,m_1-1}|P_{m_1+2,k}|l_2} ,
      \end{align} \label{Dadjacent} \end{subequations}
as well as the chiral sum (\ref{chsum}):
      \begin{equation} \begin{aligned}
                  \braket{-l_2|m_3} D_{m_1(m_1+2)(-l_1); m_2 l_2}
              + & \braket{ l_1|m_3} D_{m_1(m_1+2)(-l_1); m_2(-l_1)} \\
              = - \braket{m_1|m_1\!+\!1}
                  \big(\!\braket{m_1|m_3} &
                         \braket{l_1|P_{1,m_1+1}|P_{m_1+2,k}|l_2}
                       - \braket{l_1|m_3}
                         \braket{m_1|m_1\!+\!1|P_{m_1+2,k}|l_2}\!
                  \big) .
      \label{chsum:adjacent}
      \end{aligned} \end{equation}
When we plug these expressions into the master formula (\ref{genericcut}),
the second massive pole $P_{-l_1,m_1+1}^2$ in (\ref{R:adjacent}) cancels out,
and, after introducing homogeneous variables (\ref{lvariables}),
we get the following cut expression:
\begin{equation} \begin{aligned}
\sum_{h_1,h_2} \! A_1 A_2 = \dots
    - \frac{ \braket{m_1|m_1\!+\!1} \! \braket{m_1\!+\!1|m_1\!+\!2} }
           { \braket{12} \dots \braket{k\!-\!1|k} \!
             \braket{k\!+\!1|k\!+\!2} \dots \braket{n\!-\!1|n} }
      \frac{ \braket{\la|m_1} \! \braket{\la|m_3} }
           { \braket{\la|1} \! \braket{\la|n} \!
             \bra{k}\!P_{1k}|\lb] \! \bra{k\!+\!1}\!P_{1k}|\lb] }
      \hspace{60pt} & \\ \times
      \frac{ \bra{m_3}\!P_{1k}|\lb]
             \big(\!\bra{m_1}\!P_{1,m_1-1}|P_{m_1+2,k}|P_{1k}|\lb]
                    \bra{\la}\!P_{1k}|\lb]
                  - P_{1k}^2 \! \braket{m_1|\la} [\lb|P_{m_1+2,k}|P_{1k}|\lb]
             \big) }
           { P_{1k}^2 \! \bra{\la}\!Q_{m_1+1}|\lb] \!
             \bra{\la}\!P_{1,m_1}|m_1\!+\!1]
             [m_1\!+\!1|P_{m_1+2,k}|P_{1k}|\lb] } & \\ \times
      \frac{ \big(\!\braket{m_1|m_3}
                    \bra{\la}\!P_{1,m_1+1}|P_{m_1+2,k}|P_{1k}|\lb]
                  - \braket{\la|m_3}
                    \bra{m_1}\!m_1\!+\!1|P_{m_1+2,k}|P_{1k}|\lb]
             \big)^2 }
           { \braket{m_1|P_{1,m_1-1}|P_{1,m_1+1}|m_1\!+\!2} \! \bra{\la}\!P_{1,k}|\lb]
           -\! P_{1k}^2 \! \braket{m_1|\la} \! [\lb|P_{1,m_1+1}\!\ket{m_1\!+\!2}
           -\! P_{1k}^2 \! \bra{m_1}\!P_{1,m_1-1}|\lb] \! \braket{\la|m_1\!+\!2} } & ,
\label{cut:adjacent}
\end{aligned} \end{equation}
where by the dots in the beginning we indicated that
this is just one of the contributions to the cut.

      Here it becomes clear that all overall factors $\bra{\la}\!P_{1k}|\lb]$ cancel,
as promised in Section \ref{simplifiedbubble}.
Therefore, we can use our simplified bubble formula (\ref{CbubN2}).
As for $\bra{\la}\!P_{1k}|\lb]$ inside complicated factors,
they vanish when we take $|\lb]=|P_{1k}\!\ket{\la}$.
After slight modification of the denominator, we obtain:
\begin{equation} \begin{aligned}
C^{\text{bub},P_{1k}}_{\mathcal{N}=1 \text{ chiral}} = \dots +
\frac{P_{1k}^2}{(4\pi)^{\frac{d}{2}}i}
\frac{ \braket{m_1|m_1\!+\!1} \braket{m_1\!+\!1|m_1\!+\!2} }
     { \braket{12} \dots \braket{k\!-\!1|k}
       \braket{k\!+\!1|k\!+\!2} \dots \braket{n\!-\!1|n} } \!
\sum_{\text{residues}}
\frac{ [\lb|q] }{ \bra{\la}\!K|\lb] \bra{\la}\!K|q] } & \\ \times
\frac{ \braket{\la|m_1}^2 \braket{\la|m_3}^2 }
     { \braket{\la|1} \braket{\la|k} \braket{\la|k\!+\!1} \braket{\la|n}
       \braket{\la|Q_{m_1+1}|P_{1k}|\la} }
\frac{ \braket{\la|P_{1,m_1+1}|P_{m_1+2,k}|\la} }
     { \bra{\la}\!P_{1,m_1}|m_1\!+\!1]
       \bra{\la}\!P_{m_1+2,k}|m_1\!+\!1] } & \\ \times
\frac{ \big(\!\braket{m_1|m_3} \braket{\la|P_{1,m_1+1}|P_{m_1+2,k}|\la}
            - \braket{\la|m_3} \braket{m_1|m_1\!+\!1|P_{m_1+2,k}|\la} \!
       \big)^2 }
     { \braket{\la|m_1\!+\!2} \braket{m_1|P_{m_1+1,m_1+2}|P_{1k}|\la}
     + \braket{m_1|m_1\!+\!2} \braket{\la|P_{m_1+3,k}|P_{1k}|\la} } & .
\label{cbub:adjacent}
\end{aligned} \end{equation}
Here, the first fraction in the second line contains all physical poles:
four MHV-like ones and one massive pole that we split into two simple poles
using formula (\ref{trianglepoles}).
In addition to that, $\bra{\la}\!K|q]$ gives another simple pole.
All subsequent poles are spurious.
So we write the sum over only non-spurious poles:
\begin{equation} \begin{aligned}
R_{\mathcal{D}}^{s=m_1+2=m_2+1} \! =
\frac{P_{1k}^2}{(4\pi)^{\frac{d}{2}}i}
\frac{ \braket{m_1|m_1\!+\!1} \braket{m_1\!+\!1|m_1\!+\!2} }
     { \braket{12} \dots \braket{k\!-\!1|k}
       \braket{k\!+\!1|k\!+\!2} \dots \braket{n\!-\!1|n} } \hspace{180pt} \\ \times
\bigg\{
\frac{ F_{\mathcal{D}}^{s=m_1+2}(\la_1,\lb_1) }
     { \braket{1|k} \braket{1|k\!+\!1} \braket{1|n} }
+
\frac{ F_{\mathcal{D}}^{s=m_1+2}(\la_k,\lb_k) }
     { \braket{k|1} \braket{k|k\!+\!1} \braket{k|n} }
+
\frac{ F_{\mathcal{D}}^{s=m_1+2}(\la_{k+1},\lb_{k+1}) }
     { \braket{k\!+\!1|1} \braket{k\!+\!1|k} \braket{k\!+\!1|n} }
+
\frac{ F_{\mathcal{D}}^{s=m_1+2}(\la_n,\lb_n) }
     { \braket{n|1} \braket{n|k} \braket{n|k\!+\!1} } \\
-
\frac{ \bra{m_1}\!P_{1k}|q]^2 \bra{m_3}\!P_{1k}|q]^2
       [q|P_{1,m_1+1}|P_{m_1+2,k}|q] }
     { P_{1k}^4 \! \bra{1}\!P_{1k}|q] \! \bra{k}\!P_{1k}|q] \!
       \bra{k\!+\!1}\!P_{1k}|q] \! \bra{n}\!P_{1k}|q]
       [q|P_{1,m_1}|P_{m_1+1,k}|q]
       [m_1\!+\!1|P_{1,m_1}|P_{1k}|q]
       [m_1\!+\!1|P_{m_1+2,k}|P_{1k}|q] } \\ \times
\frac{ \big(\!\braket{m_1|m_3} [q|P_{1k}|P_{1,m_1+1}|P_{m_1+2,k}|P_{1k}|q]
            - \bra{m_3}\!P_{1k}|q] \bra{m_1}\!m_1\!+\!1|P_{m_1+2,k}|P_{1k}|q]
       \big)^2 }
     { \bra{m_1}\!P_{m_1+1,m_1+2}|q] \bra{m_1\!+\!2}\!P_{1k}|q]
     - \braket{m_1|m_1\!+\!2} [q|P_{m_1+3,k}|P_{1k}|q] } \\
+ \, M_{\mathcal{D}}^{s=m_1+2=m_2+1}(\la^{m_1+1}_+,\lb^{m_1+1}_+)
   + M_{\mathcal{D}}^{s=m_1+2=m_2+1}(\la^{m_1+1}_-,\lb^{m_1+1}_-) &
\bigg\} ,
\label{PolesAdjacentD}
\end{aligned} \end{equation}
where we introduce a shorthand notation for an expression
that occurs in all MHV-like poles:
\begin{equation} \begin{aligned}
F_{\mathcal{D}}^{s=m_1+2}(\la,\lb) =
\frac{ \braket{\la|m_1}^2 \braket{\la|m_3}^2
       \braket{\la|P_{1,m_1+1}|P_{m_1+2,k}|\la} [\lb|q] }
     { \braket{\la|P_{1,m_1}|P_{m_1+1,k}|\la}
       \bra{\la}\!P_{1,m_1}|m_1\!+\!1] \bra{\la}\!P_{m_1+2,k}|m_1\!+\!1]
       \bra{\la}\!P_{1k}|\lb] \bra{\la}\!P_{1k}|q] } & \\ \times
\frac{ \big(\!\braket{m_1|m_3} \braket{\la|P_{1,m_1+1}|P_{m_1+2,k}|\la}
            - \braket{\la|m_3} \braket{m_1|m_1\!+\!1|P_{m_1+2,k}|\la} \!
       \big)^2 }
     { \braket{\la|m_1\!+\!2} \braket{m_1|P_{m_1+1,m_1+2}|P_{1k}|\la}
     + \braket{m_1|m_1\!+\!2} \braket{\la|P_{m_1+3,k}|P_{1k}|\la} } &.
\label{FadjacentD}
\end{aligned} \end{equation}
The last line in (\ref{PolesAdjacentD}) contains contributions
from massive poles $\la^{m_1+1}_\pm$, denoted by $M_{\mathcal{D}}$,
which, according to definitions (\ref{Qi4dim}) and (\ref{Ptri}),
correspond to massless linear combinations of $P_{1,m_1}$ and $P_{1,k}$.
We denote them as
\begin{equation} \begin{aligned}
M_{\mathcal{D}}^{s=m_1+2=m_2+1}(\la,\lb) =
- & \frac{1}{4( (P_{1,m_1} \cdot P_{m_1+1,k})^2
              - P_{1,m_1}^2 P_{m_1+1,k}^2) )} \\ \times &
\frac{ \braket{\la|m_1}^2 \braket{\la|m_3}^2
       \braket{\la|P_{1,m_1+1}|P_{m_1+2,k}|\la} [\lb|q] }
     { \braket{\la|1} \braket{\la|k} \braket{\la|k\!+\!1} \braket{\la|n}
       \bra{\la}\!P_{1,m_1}|m_1\!+\!1]
       \bra{\la}\!P_{m_1+2,k}|m_1\!+\!1] \bra{\la}\!P_{1k}|q] } \\ \times &
\frac{ \big(\!\braket{m_1|m_3} \braket{\la|P_{1,m_1+1}|P_{m_1+2,k}|\la}
            - \braket{\la|m_3} \braket{m_1|m_1\!+\!1|P_{m_1+2,k}|\la}\!
       \big)^2 }
     { \braket{\la|m_1\!+\!2} \braket{m_1|P_{m_1+1,m_1+2}|P_{1k}|\la}
     + \braket{m_1|m_1\!+\!2} \braket{\la|P_{m_1+3,k}|P_{1k}|\la} } .
\label{MassivePoleAdjacentD}
\end{aligned} \end{equation}

\subsubsection{Generic $\mathcal{D}$-case contribution}
\label{subcase:standardd}

      In this and subsequent sections, we will only present final formulas
using analogous notation for residue sums.
The applicability of the simplified bubble formula (\ref{CbubN2}),
i.~e. the cancellation of all overall factors of $\bra{\la}\!P_{1k}|\lb]$,
was verified during the derivation.
To get the full bubble coefficient, contributions from all non-vanishing cases
are to be summed over using the frame formula (\ref{CbubN1frame}).

      For example, a standard $\mathcal{D}$-case $R_{m_1 s (-l_1)}$
generates the following contribution:
\begin{equation} \begin{aligned}
R_{\mathcal{D}}^{s} =
-\frac{P_{1k}^2}{(4\pi)^{\frac{d}{2}}i}
\frac{ \braket{m_1|m_2}^4 \braket{s\!-\!1|s} }
     { \braket{12} \dots \braket{k\!-\!1|k}
       \braket{k\!+\!1|k\!+\!2} \dots \braket{n\!-\!1|n} } \hspace{194pt} & \\ \times
\bigg\{
\frac{ F_{\mathcal{D}}^{s}(\la_1,\lb_1) }
     { \braket{1|k} \braket{1|k\!+\!1} \braket{1|n} }
+
\frac{ F_{\mathcal{D}}^{s}(\la_k,\lb_k) }
     { \braket{k|1} \braket{k|k\!+\!1} \braket{k|n} }
+
\frac{ F_{\mathcal{D}}^{s}(\la_{k+1},\lb_{k+1}) }
     { \braket{k\!+\!1|1} \braket{k\!+\!1|k} \braket{k\!+\!1|n} }
+
\frac{ F_{\mathcal{D}}^{s}(\la_n,\lb_n) }
     { \braket{n|1} \braket{n|k} \braket{n|k\!+\!1} } & \\
-
\frac{ \bra{m_1}\!P_{1k}|q]^2 \bra{m_3}\!P_{1k}|q]^2
       [q|P_{1k}|P_{1,s-1}|P_{s,k}|P_{1k}|q] }
     { P_{1k}^6 \! \bra{1}\!P_{1k}|q] \! \bra{k}\!P_{1k}|q] \!
       \bra{k\!+\!1}\!P_{1k}|q] \! \bra{n}\!P_{1k}|q] \!
       \bra{m_1}\!P_{m_1+1,s-1}|P_{s,k}|P_{1k}|q] \!
       \bra{m_1}\!P_{m_1+1,s-1}|P_{1,s-1}|P_{1k}|q] } & \\ \times
( \braket{m_1|m_3} [q|P_{1k}|P_{1,m_1-1}|P_{s,k}|P_{1k}|q]
- \bra{m_1}\!P_{1k}|q] \bra{m_3}\!P_{m_1,s-1}|P_{s,k}|P_{1k}|q])^2
\hspace{14pt} & \\ /
\Big(
\big(\!\bra{m_1}\!P_{1k}|q] \bra{s\!-\!1}\!P_{1,s-2}|q]
     - \bra{s\!-\!1}\!P_{1k}|q] \bra{m_1}\!P_{1,m_1-1}|q]
\big) \quad \hspace{7pt} & \\ \times
\big(\!\bra{m_1}\!P_{1k}|q] \bra{s}\!P_{1,s-1}|q]
     - \bra{s}\!P_{1k}|q] \bra{m_1}\!P_{1,m_1-1}|q]
\big)
\Big)
\bigg\} , &
\label{PolesStandardD}
\end{aligned} \end{equation}
where
\begin{equation} \begin{aligned}
F_{\mathcal{D}}^{s}(\la,\lb) = 
\frac{ \braket{\la|m_1}^2 \braket{\la|m_3}^2
       \braket{\la|P_{1,s-1}|P_{s,k}|\la} [\lb|q] }
     { \braket{\la|P_{1,s-1}|P_{m_1+1,s-1}|m_1}
       \braket{\la|P_{s,k}|P_{m_1+1,s-1}|m_1}
       \bra{\la}\!P_{1k}|\lb] \bra{\la}\!P_{1k}|q] } & \\ \times
\big(\!\braket{m_1|m_3} \braket{\la|P_{1,m_1-1}|P_{s,k}|\la}
     - \braket{\la|m_1} \braket{m_3|P_{m_1,s-1}|P_{s,k}|\la}\!\big)^2\; & \\ /
\Big(
\big(\!\braket{\la|m_1} \braket{\la|P_{1k}|P_{1,s-2}|s\!-\!1}
     - \braket{\la|s\!-\!1} \braket{\la|P_{1k}|P_{1,m_1-1}|m_1}\!\big)
\hspace{7pt} & \\ \times
\big(\!\braket{\la|m_1} \braket{\la|P_{1k}|P_{1,s-1}|s}
     - \braket{\la|s} \braket{\la|P_{1k}|P_{1,m_1-1}|m_1}\!\big)
\Big) &.
\label{FstandardD}
\end{aligned} \end{equation}

      These formulas apply either for $s \in \{m_2\!+\!1,\dots,k\}$
or starting from $m_2\!+\!3$ for for the configuration
with two adjacent negative-helicity gluons $m_1$ and $m_2=m_1\!+\!1$.
As expected from our analysis in Section \ref{poles},
normally, case $\mathcal{D}$ generates no massive poles.

\subsection{Case $\mathcal{C}$}
\label{case:c}

      In this section, we consider case $\mathcal{C}$ which normally contains
all massive poles (unless $m_2=m_1\!+\!1$).

\subsubsection{First $\mathcal{C}$-case contribution}
\label{subcase:firstc}

      We start by presenting residue contributions
generated by $R_{m_1(m_1+2)(-l_1)}$ which, as we concluded in Section \ref{poles},
contains two massive poles at the same time:
\begin{equation} \begin{aligned}
R_{\mathcal{C}}^{s=m_1+2} \! = \!
\frac{P_{1k}^2}{(4\pi)^{\frac{d}{2}}i}
                  \frac{ \braket{m_1|m_1\!+\!1} \!
                         \braket{m_1\!+\!1|m_1\!+\!2} }
                       { \braket{12} \dots \braket{k\!-\!1|k} \!
                         \braket{k\!+\!1|k\!+\!2} \dots \braket{n\!-\!1|n} }
\bigg\{
\frac{ F_{\mathcal{C}}^{s=m_1+2}(\la_1,\lb_1) }
     { \braket{1|k} \! \braket{1|k\!+\!1} \! \braket{1|n} }
\! + \!
\frac{ F_{\mathcal{C}}^{s=m_1+2}(\la_k,\lb_k) }
     { \braket{k|1} \! \braket{k|k\!+\!1} \! \braket{k|n} } & \\
+
\frac{ F_{\mathcal{C}}^{s=m_1+2}(\la_{k+1},\lb_{k+1}) }
     { \braket{k\!+\!1|1} \! \braket{k\!+\!1|k} \! \braket{k\!+\!1|n} }
\! + \!
\frac{ F_{\mathcal{C}}^{s=m_1+2}(\la_n,\lb_n) }
     { \braket{n|1} \! \braket{n|k} \! \braket{n|k\!+\!1} }
-
\frac{ \bra{m_1}\!P_{1k}|q]^2 \bra{m_2}\!P_{1k}|q]^2 \bra{m_3}\!P_{1k}|q]^2 }
     { P_{1k}^4 \! \bra{1}\!P_{1k}|q] \! \bra{k}\!P_{1k}|q] \!
       \bra{k\!+\!1}\!P_{1k}|q] \! \bra{n}\!P_{1k}|q] } & \\ \times
\frac{ [m_1\!+\!1|q]^2 }
     { [q|P_{1,m_1}|P_{m_1+1,k}|q]
       [q|P_{1,m_1+1}|P_{m_1+2,k}|q]
       [m_1\!+\!1|P_{1,m_1}|P_{1k}|q]
       [m_1\!+\!1|P_{m_1+2,k}|P_{1k}|q] } & \\ \times
\frac{ \big( P_{1k}^2 \braket{m_3|m_1} \bra{m_2}\!P_{1k}|q] [m_1\!+\!1|q]
     + \braket{m_1|m_2} \bra{m_3}\!P_{1k}|q] [m_1\!+\!1|P_{m_1+2,k}|P_{1k}|q]
       \big)^2 }
     { \bra{m_1}\!P_{m_1+1,m_1+2}|q] \bra{m_1\!+\!2}\!P_{1k}|q]
     - \braket{m_1|m_1\!+\!2} [q|P_{m_1+3,k}|P_{1k}|q] } & \\
+ \, \tilde{M}_{\mathcal{C}}^{s=m_1+2}(\la^{m_1+1}_+,\lb^{m_1+1}_+)
   + \tilde{M}_{\mathcal{C}}^{s=m_1+2}(\la^{m_1+1}_-,\lb^{m_1+1}_-) \\
+ \, M_{\mathcal{C}}^{s=m_1+2}(\la^{m_1+2}_+,\lb^{m_1+2}_+)
   + M_{\mathcal{C}}^{s=m_1+2}(\la^{m_1+2}_-,\lb^{m_1+2}_-) &
\bigg\} ,
\label{PolesFirstC}
\end{aligned} \end{equation}
where the MHV-like residues are written with the help of
\begin{equation} \begin{aligned}
F_{\mathcal{C}}^{s=m_1+2}(\la,\lb) \hspace{387pt} \\ = \!
\frac{ \braket{\la|m_1}^2 \braket{\la|m_2}^2 \braket{\la|m_3}^2
       \bra{\la}\!P_{1k}|m_1\!+\!1]^2 [\lb|q] }
     { \braket{\la|P_{1,m_1}|P_{m_1+1,k}|\la} \!
       \braket{\la|P_{1,m_1+1}|P_{m_1+2,k}|\la} \!
       \bra{\la}\!P_{1,m_1}|m_1\!+\!1] \!
       \bra{\la}\!P_{m_1+2,k}|m_1\!+\!1] \!
       \bra{\la}\!P_{1k}|\lb] \! \bra{\la}\!P_{1k}|q] } \\ \times
\frac{ \big(\!\braket{m_3|m_1} \braket{\la|m_2} \bra{\la}\!P_{1k}|m_1\!+\!1]
     + \braket{m_1|m_2} \braket{\la|m_3} \bra{\la}\!P_{m_1+2,k}|m_1\!+\!1]
       \big)^2 }
     { \braket{\la|m_1\!+\!2} \braket{m_1|P_{m_1+1,m_1+2}|P_{1k}|\la}
     + \braket{m_1|m_1\!+\!2} \braket{\la|P_{m_1+3,k}|P_{1k}|\la} } , \!
\label{FfirstC}
\end{aligned} \end{equation}
and the two massive poles generate contributions of the form:
\begin{equation} \begin{aligned}
\tilde{M}_{\mathcal{D}}^{s=m_1+2}(\la,\lb) =
-\frac{1}{4( (P_{1,m_1} \cdot P_{m_1+1,k})^2
            - P_{1,m_1}^2 P_{m_1+1,k}^2) )} \hspace{173pt} \\ \times
\frac{ \braket{\la|m_1}^2 \braket{\la|m_2}^2 \braket{\la|m_3}^2
       \bra{\la}\!P_{1k}|m_1\!+\!1]^2 [\lb|q] }
     { \braket{\la|1} \braket{\la|k} \braket{\la|k\!+\!1} \braket{\la|n}
       \braket{\la|P_{1,m_1+1}|P_{m_1+2,k}|\la}
       \bra{\la}\!P_{1,m_1}|m_1\!+\!1]
       \bra{\la}\!P_{m_1+2,k}|m_1\!+\!1] \bra{\la}\!P_{1k}|q] } \\ \times
\frac{ \big(\!\braket{m_3|m_1} \braket{\la|m_2} \bra{\la}\!P_{1k}|m_1\!+\!1]
            + \braket{m_1|m_2} \braket{\la|m_3} \bra{\la}\!P_{m_1+2,k}|m_1\!+\!1]
       \big)^2 }
     { \braket{\la|m_1\!+\!2} \braket{m_1|P_{m_1+1,m_1+2}|P_{1k}|\la}
     + \braket{m_1|m_1\!+\!2} \braket{\la|P_{m_1+3,k}|P_{1k}|\la} } , \!
\label{MassivePoleFirst}
\end{aligned} \end{equation}
and
\begin{equation} \begin{aligned}
M_{\mathcal{D}}^{s=m_1+2}(\la,\lb) =
-\frac{1}{4( (P_{1,m_1+1} \cdot P_{m_1+2,k})^2
             -P_{1,m_1+1}^2 P_{m_1+2,k}^2) )} \hspace{141pt} \\ \times
\frac{ \braket{\la|m_1}^2 \braket{\la|m_2}^2 \braket{\la|m_3}^2
       \bra{\la}\!P_{1k}|m_1\!+\!1]^2 [\lb|q] }
     { \braket{\la|1} \braket{\la|k} \braket{\la|k\!+\!1} \braket{\la|n}
       \braket{\la|P_{1,m_1}|P_{m_1+1,k}|\la}
       \bra{\la}\!P_{1,m_1}|m_1\!+\!1]
       \bra{\la}\!P_{m_1+2,k}|m_1\!+\!1] \bra{\la}\!P_{1k}|q] } \\ \times
\frac{ \big(\!\braket{m_3|m_1} \braket{\la|m_2} \bra{\la}\!P_{1k}|m_1\!+\!1]
            + \braket{m_1|m_2} \braket{\la|m_3} \bra{\la}\!P_{m_1+2,k}|m_1\!+\!1]
       \big)^2 }
     { \braket{\la|m_1\!+\!2} \braket{m_1|P_{m_1+1,m_1+2}|P_{1k}|\la}
     + \braket{m_1|m_1\!+\!2} \braket{\la|P_{m_1+3,k}|P_{1k}|\la} } . \!
\label{MassivePoleSecond}
\end{aligned} \end{equation}

\subsubsection{Generic $\mathcal{C}$-case contribution}
\label{subcase:standardc}

      Each subsequent $R_{m_1 s(-l_1)}$ with $s \in \{m_1\!+\!3, \dots,m_2\}$
generates a single massive pole.
\begin{equation} \begin{aligned}
R_{\mathcal{C}}^{s} = \! -
\frac{P_{1k}^2}{(4\pi)^{\frac{d}{2}}i}
\frac{ \braket{s\!-\!1|s} }
     { \braket{12} \dots \braket{k\!-\!1|k} \!
       \braket{k\!+\!1|k\!+\!2} \dots \braket{n\!-\!1|n} }
\bigg\{
\frac{ F_{\mathcal{C}}^{s}(\la_1,\lb_1) }
     { \braket{1|k} \! \braket{1|k\!+\!1} \! \braket{1|n} }
+
\frac{ F_{\mathcal{C}}^{s}(\la_k,\lb_k) }
     { \braket{k|1} \! \braket{k|k\!+\!1} \! \braket{k|n} } & \\
+
\frac{ F_{\mathcal{C}}^{s}(\la_{k+1},\lb_{k+1}) }
     { \braket{k\!+\!1|1} \! \braket{k\!+\!1|k} \! \braket{k\!+\!1|n} }
+
\frac{ F_{\mathcal{C}}^{s}(\la_n,\lb_n) }
     { \braket{n|1} \! \braket{n|k} \! \braket{n|k\!+\!1} }
+
\frac{ \bra{m_1}\!P_{1k}|q]^2 \bra{m_2}\!P_{1k}|q]^2 \bra{m_3}\!P_{1k}|q]^2 }
     { P_{1k}^4 \bra{1}\!P_{1k}|q] \bra{k}\!P_{1k}|q]
       \bra{k\!+\!1}\!P_{1k}|q] \bra{n}\!P_{1k}|q] } & \\ \times
\frac{ \bra{m_1}\!P_{m_1+1,s-1}|q]^2 }
     { [q|P_{1,s-1}|P_{s,k}|q]
       \bra{m_1}\!P_{m_1+1,s-1}|P_{1,s-1}|P_{1k}|q]
       \bra{m_1}\!P_{m_1+1,s-1}|P_{s,k}|P_{1k}|q] } & \\ \times
 \big( P_{1k}^2 \braket{m_3|m_1} \bra{m_2}\!P_{1k}|q]
       \bra{m_1}\!P_{m_1,s-1}|q]
     + \braket{m_1|m_2} \bra{m_3}\!P_{1k}|q]
       \bra{m_1}\!P_{m_1,s-1}|P_{s,k}|P_{1k}|q]
 \big)^2 & \\ /
 \Big(
     \big(\!\braket{m_1|s\!-\!1} [q|P_{1,m_1}|P_{m_1+1,k}|q]
          + \bra{m_1}\!P_{1k}|q] \bra{s\!-\!1}\!P_{m_1+1,s-2}|q] \big)
     \hspace{6pt} \\ \times
     \big(\!\braket{m_1|s} [q|P_{1,m_1}|P_{m_1+1,k}|q]
          + \bra{m_1}\!P_{1k}|q] \bra{s}\!P_{m_1+1,s-1}|q] \big)
 \Big) \\
+ \, M_{\mathcal{C}}^{s}(\la_s^+,\lb_s^+)
   + M_{\mathcal{C}}^{s}(\la_s^-,\lb_s^-) &
\bigg\} ,
\label{PolesStandardC}
\end{aligned} \end{equation}
where 
\begin{equation} \begin{aligned}
F_{\mathcal{C}}^{s}(\la,\lb) =
\frac{ \braket{\la|m_1}^2 \braket{\la|m_2}^2 \braket{\la|m_3}^2
       \braket{\la|P_{1k}|P_{m_1+1,s-1}|m_1}^2 [\lb|q] }
     { \braket{\la|P_{1,s-1}|P_{s,k}|\la}
       \braket{\la|P_{1,s-1}|P_{m_1+1,s-1}|m_1}
       \braket{\la|P_{s,k}|P_{m_1+1,s-1}|m_1} 
       \bra{\la}\!P_{1k}|\lb] \bra{\la}\!P_{1k}|q] } & \\ \times
 \big(\!\braket{\la|m_2} \braket{m_3|m_1} \braket{m_1|P_{m_1+1,s-1}|P_{1k}|\la}
      + \braket{\la|m_3} \braket{m_1|m_2} \braket{m_1|P_{m_1+1,s-1}|P_{s,k}|\la}\!
 \big)^2 & \\ /
 \Big(
     \big(\!\braket{m_1|s\!-\!1} \braket{\la|P_{1,m_1}|P_{m_1+1,k}|\la}
          + \braket{\la|m_1} \braket{\la|P_{1k}|P_{m_1+1,s-2}|s\!-\!1}\!\big)
     \hspace{6pt} & \\
     \big(\!\braket{m_1|s} \braket{\la|P_{1,m_1}|P_{m_1+1,k}|\la}
          + \braket{\la|m_1} \braket{\la|P_{1k}|P_{m_1+1,s-1}|s}\!\big)
 \Big) & ,
\label{FstandardC}
\end{aligned} \end{equation}
and
\begin{equation} \begin{aligned}
M_{\mathcal{C}}^{s}(\la,\lb) =
-\frac{1}{4( (P_{1,s-1} \cdot P_{s,k})^2
            - P_{1,s-1}^2 P_{s,k}^2) )} \hspace{227pt} \\ \times
\frac{ \braket{\la|m_1}^2 \braket{\la|m_2}^2 \braket{\la|m_3}^2
       \braket{\la|P_{1k}|P_{m_1+1,s-1},m_1}^2 [\lb|q] }
     { \braket{\la|1} \braket{\la|k} \braket{\la|k\!+\!1} \braket{\la|n}
       \braket{\la|P_{1,s-1}|P_{m_1+1,s-1}|m_1}
       \braket{\la|P_{s,k}|P_{m_1+1,s-1}|m_1} \bra{\la}\!P_{1k}|q] } & \\ \times
 \big(\!\braket{\la|m_2} \braket{m_3|m_1} \braket{m_1|P_{m_1+1,s-1}|P_{1k}|\la}
      + \braket{\la|m_3} \braket{m_1|m_2} \braket{m_1|P_{m_1+1,s-1}|P_{s,k}|\la}\!
 \big)^2 & \\ /
 \Big(
     \big(\!\braket{m_1|s\!-\!1} \braket{\la|P_{1,m_1}|P_{m_1+1,k}|\la}
          + \braket{\la|m_1} \braket{\la|P_{1k}|P_{m_1+1,s-2}|s\!-\!1}\!\big)
     \hspace{6pt} & \\
     \big(\!\braket{m_1|s} \braket{\la|P_{1,m_1}|P_{m_1+1,k}|\la}
          + \braket{\la|m_1} \braket{\la|P_{1k}|P_{m_1+1,s-1}|s}\!\big)
 \Big) & .
\label{MassivePoleNonAdjacent}
\end{aligned} \end{equation}

\subsection{Cases $\mathcal{A}$ and $\mathcal{B}$}
\label{case:ab}

      Cases $\mathcal{A}$ and $\mathcal{B}$ can be considered together
by introducing a vector which encodes the only difference between them:
      \begin{equation}
            B = \begin{cases}
                  P_{m_1+1,s-1} &
                  \text{if } \{s,t\} \in \mathcal{A} \\
                - P_{s,t-1} &
                  \text{if } \{s,t\} \in \mathcal{B}
            \end{cases}
      \label{Bmomentum}
      \end{equation}
so that one can define residue contributions for both cases using one function
\begin{subequations} \begin{align}
      R_{\mathcal{A}}^{s,t} & = R_{\mathcal{AB}}^{s,t}(P_{m_1+1,s-1}) , \\
      R_{\mathcal{B}}^{s,t} & = R_{\mathcal{AB}}^{s,t}(-P_{s,t-1}) .
\label{RAB}
\end{align} \end{subequations}

      Both cases include a subcase in which $t=l_2$ and so the loop-momentum dependence of $R_{m_1 s t}$ is slightly modified,
as compared to the generic situation $t \in \{m_2+1,\dots,k\}$.
We study these subcases separately.

\subsubsection{Generic $\mathcal{A}$- and $\mathcal{B}$-case contributions}
\label{subcase:standardab}

      For $s \in \{m_1+2,\dots,k-2\}$ and $t \in \{m_2+1,\dots,k\}$,
$R_{m_1 s t}$ is independent of loop momenta,
so all the corresponding residues are MHV-like:
\begin{equation} \begin{aligned}
R_{\mathcal{AB}}^{s,t}(B) =
\frac{P_{1k}^2}{(4\pi)^{\frac{d}{2}}i}
\frac{ \braket{m_1|m_3}^2
       \braket{m_1|B|P_{s,t-1}|m_2}^4 }
     { \braket{12} \dots \braket{k\!-\!1|k}
       \braket{k\!+\!1|k\!+\!2} \dots \braket{n\!-\!1|n} }
\hspace{140pt} & \\ \times
\braket{s\!-\!1|s} \braket{t\!-\!1|t} \! /
\big(P_{s,t-1}^2  \braket{m_1|P_{m_1+1,t-1}|P_{s,t-1}|s\!-\!1}
       \braket{m_1|P_{m_1+1,t-1}|P_{s+1,t-1}|s} & \\ \times
       \braket{m_1|P_{m_1+1,s-1}|P_{s,t-2}|t\!-\!1}
       \braket{m_1|P_{m_1+1,s-1}|P_{s,t-1}|t} & \big) \\ \times
\bigg\{
\frac{ F_{\mathcal{AB}}(\la_1,\lb_1) }
     { \braket{1|k} \braket{1|k\!+\!1} \braket{1|n} }
+
\frac{ F_{\mathcal{AB}}(\la_k,\lb_k) }
     { \braket{k|1} \braket{k|k\!+\!1} \braket{k|n} }
+
\frac{ F_{\mathcal{AB}}(\la_{k+1},\lb_{k+1}) }
     { \braket{k\!+\!1|1} \braket{k\!+\!1|k} \braket{k\!+\!1|n} } & \\
+
\frac{ F_{\mathcal{AB}}(\la_{n},\lb_n) }
     { \braket{n|1} \braket{n|k} \braket{n|k\!+\!1} }
+
\frac{ \bra{m_1}\!P_{1k}|q]^2 \bra{m_3}\!P_{1k}|q]^2 }
     { P_{1k}^2 \bra{1}\!P_{1k}|q] \bra{k}\!P_{1k}|q]
       \bra{k\!+\!1}\!P_{1k}|q] \bra{n}\!P_{1k}|q] } &
\bigg\} ,
\label{StandardPolesAB}
\end{aligned} \end{equation}
where
\begin{equation}
F_{\mathcal{AB}}(\la,\lb) = 
\frac{ \braket{\la|m_1}^2 \braket{\la|m_3}^2 [\lb|q] }
     { \bra{\la}\!P_{1k}|\lb] \bra{\la}\!P_{1k}|q] } .
\label{FstandardAB}
\end{equation}

\subsubsection{Special $\mathcal{A}$- and $\mathcal{B}$-case contributions}
\label{subcase:lastab}

      When $t$ becomes equal to $l_2$,
both cases $\mathcal{A}$ and $\mathcal{B}$ get modified
as the numerator and the denominator of $R_{m_1 s l_2}$ begin to depend
on the loop momentum.
However, the new pole in the denominator remains spurious,
so for these subcases we have formulas
only slightly different from those in the previous section:
\begin{equation} \begin{aligned}
R_{\mathcal{AB}}^{s,t=l_2}(B) =
\frac{P_{1k}^2}{(4\pi)^{\frac{d}{2}}i}  
\frac{ \braket{m_1|m_3}^2
       \braket{m_1|B|P_{s,k}|m_2}^4 }
     { \braket{12} \dots \braket{k\!-\!1|k}
       \braket{k\!+\!1|k\!+\!2} \dots \braket{n\!-\!1|n} }
\hspace{127pt} & \\ \times
\frac{ \braket{s\!-\!1|s} }
     { P_{s,k}^2 \braket{m_1|P_{m_1+1,k}|P_{s,k}|s\!-\!1}
      \braket{m_1|P_{m_1+1,k}|P_{s+1,k}|s}
      \braket{m_1|P_{m_1+1,s-1}|P_{s,k-1}|k} } & \\ \times
\bigg\{
\frac{ F_{\mathcal{AB}}^{s,t=l_2}(\la_1,\lb_1) }
     { \braket{1|k\!+\!1} \braket{1|n} }
+
\frac{ F_{\mathcal{AB}}^{s,t=l_2}(\la_{k+1},\lb_{k+1}) }
     { \braket{k\!+\!1|1} \braket{k\!+\!1|n} }
+
\frac{ F_{\mathcal{AB}}^{s,t=l_2}(\la_n,\lb_n) }
     { \braket{n|1} \braket{n|k\!+\!1} } \hspace{54pt} \\
+
\frac{ \bra{m_1}\!P_{1k}|q]^2 \bra{m_3}\!P_{1k}|q]^2 }
     { P_{1k}^2 \bra{1}\!P_{1k}|q] \bra{k\!+\!1}\!P_{1k}|q]
       \bra{n}\!P_{1k}|q] \bra{m_1}\!P_{m_1+1,s-1}|P_{s,k}|P_{1k}|q] }
\bigg\} & ,
\label{PolesLastAB}
\end{aligned} \end{equation}
where
\begin{equation}
F_{\mathcal{AB}}^{s,t=l_2}(\la,\lb) = 
\frac{ \braket{\la|m_1}^2 \braket{\la|m_3}^2 [\lb|q] }
     { \braket{\la|P_{s,k}|P_{m_1+1,s-1}|m_1}
       \bra{\la}\!P_{1k}|\lb] \bra{\la}\!P_{1k}|q] } ,
\label{FlastAB}
\end{equation}
and $s \in \{m_1+2,\dots,k-1\}$.
We note that, along with introducing a harmless unphysical pole,
one of the physical MHV-like poles got canceled by
$\braket{t\!-\!1|t} \Rightarrow \braket{k|\la}$ in $R_{m_1 s \, l_2}$.

\subsection{Case $\mathcal{E}$}
\label{case:e}

      Case $\mathcal{E}$ develops a subcase for $t=1$,
while all further values of $t \in \{2,\dots,m_1-1\}$ form a generic subcase.

\subsubsection{Generic $\mathcal{E}$-case contribution}
\label{subcase:standarde}

      For a standard $\mathcal{E}$-case contribution,
i.~e. for $s \in \{m_1+2,\dots,m_2\}$ and $t \in \{2,\dots,m_1-1\}$,
$R_{m_1 s t}$ has no loop-momentum dependence,
so we encounter only usual MHV-like poles:
$\ket{\la}=\ket{1}$, $\ket{\la}=\ket{k}$, $\ket{\la}=\ket{k\!+\!1}$,
$\ket{\la}=\ket{n}$ and $\ket{\la}=|P_{1k}|q]$. Their residues are
\begin{equation} \begin{aligned}
R_{\mathcal{E}}^{s,t} =
\frac{P_{1k}^2}{(4\pi)^{\frac{d}{2}}i}
\frac{ \braket{m_2|m_3}^2
       \braket{m_1|P_{t,m_1-1}|P_{m_1+1,s-1}|m_1}^4 }
     { \braket{12} \dots \braket{k\!-\!1|k}
       \braket{k\!+\!1|k\!+\!2} \dots \braket{n\!-\!1|n} }
\hspace{107pt} \\ \times
\braket{s\!-\!1|s} \braket{t\!-\!1|t} \! /
\big( P_{t,s-1}^2 \braket{m_1|P_{t,m_1-1}|P_{t,s-2}|s\!-\!1}
       \braket{m_1|P_{t,m_1-1}|P_{t,s-1}|s} & \\
       \braket{m_1|P_{m_1+1,s-1}|P_{t,s-1}|t\!-\!1}
       \braket{m_1|P_{m_1+1,s-1}|P_{t+1,s-1}|t} & \big) \\ \times
\bigg\{
\frac{ F_{\mathcal{E}}(\la_1,\lb_1) }
     { \braket{1|k} \braket{1|k\!+\!1} \braket{1|n} }
+
\frac{ F_{\mathcal{E}}(\la_k,\lb_k) }
     { \braket{k|1} \braket{k|k\!+\!1} \braket{k|n} }
+
\frac{ F_{\mathcal{E}}(\la_{k+1},\lb_{k+1}) }
     { \braket{k\!+\!1|1} \braket{k\!+\!1|k} \braket{k\!+\!1|n} } & \\
+
\frac{ F_{\mathcal{E}}(\la_n,\lb_n) }
     { \braket{n|1} \braket{n|k} \braket{n|k\!+\!1} }
+
\frac{ \bra{m_2}\!P_{1k}|q]^2 \bra{m_3}\!P_{1k}|q]^2 }
     { P_{1k}^2 \bra{1}\!P_{1k}|q] \bra{k}\!P_{1k}|q]
       \bra{k\!+\!1}\!P_{1k}|q] \bra{n}\!P_{1k}|q] } &
\bigg\} ,
\label{PolesStandardE}
\end{aligned} \end{equation}
where
\begin{equation}
F_{\mathcal{E}}(\la,\lb) = 
\frac{ \braket{\la|m_2}^2 \braket{\la|m_3}^2 [\lb|q] }
     { \bra{\la}\!P_{1k}|\lb] \bra{\la}\!P_{1k}|q] } .
\label{FstandardE}
\end{equation}

\subsubsection{Special $\mathcal{E}$-case contribution}
\label{subcase:firste}

      At $t=1$, $R_{m_1 s\, 1}$ develops a simple loop-momentum dependence
through spinor $\ket{t\!-\!1} = \ket{-l_1}$.
The contribution from this subcase becomes
\begin{equation} \begin{aligned}
R_{\mathcal{E}}^{s,t=1} =
-\frac{P_{1k}^2}{(4\pi)^{\frac{d}{2}}i} 
\frac{ \braket{m_2|m_3}^2
       \braket{m_1|P_{1,m_1-1}|P_{m_1+1,s-1}|m_1}^4 }
     { \braket{12} \dots \braket{k\!-\!1|k}
       \braket{k\!+\!1|k\!+\!2} \dots \braket{n\!-\!1|n} }
\hspace{137pt} \\ \times
\frac{ \braket{s\!-\!1|s} }
     { P_{1,s-1}^2 \braket{m_1|P_{1,m_1-1}|P_{1,s-2}|s\!-\!1}
       \braket{m_1|P_{1,m_1-1}|P_{1,s-1}|s}
       \braket{m_1|P_{m_1+1,s-1}|P_{2,s-1}|1} } \\ \times
\bigg\{
\frac{ F_{\mathcal{E}}^{s,t=1}(\la_1,\lb_1) }
     { \braket{k|k\!+\!1} \braket{k|n} }
+
\frac{ F_{\mathcal{E}}^{s,t=1}(\la_{k+1},\lb_{k+1}) }
     { \braket{k\!+\!1|k} \braket{k\!+\!1|n} }
+
\frac{ F_{\mathcal{E}}^{s,t=1}(\la_n,\lb_n) }
     { \braket{n|k} \braket{n|k\!+\!1} } \hspace{51pt} & \\
+
\frac{ \bra{m_2}\!P_{1k}|q]^2 \bra{m_3}\!P_{1k}|q]^2 }
     { P_{1k}^2 \bra{k}\!P_{1k}|q] \bra{k\!+\!1}\!P_{1k}|q] \bra{n}\!P_{1k}|q]
       \bra{m_1}\!P_{m_1+1,s-1}|P_{1,s-1}|P_{1k}|q] }
\bigg\} & ,
\label{PolesFirstE}
\end{aligned} \end{equation}
where
\begin{equation}
F_{\mathcal{E}}^{s,t=1}(\la,\lb) = 
\frac{ \braket{\la|m_2}^2 \braket{\la|m_3}^2 [\lb|q] }
     { \braket{\la|P_{1,s-1}|P_{m_1+1,s-1}|m_1}
       \bra{\la}\!P_{1k}|\lb] \bra{\la}\!P_{1k}|q] } .
\label{FfirstE}
\end{equation}

\subsection{Case $\mathcal{F}$}
\label{case:f}

      For $s \in \{m_2+1,\dots,l_2\}$ and $t \in \{1,\dots,m_1-1\}$, $R_{m_1 s t}$ is independent of loop momenta only if $s \neq l_2$ and $t \neq 1$.
Otherwise, we have three special subcases: $\{s = l_2, t \neq 1\}$,
$\{s \neq l_2, t = 1\}$ and $\{s = l_2, t = 1\}$.

\subsubsection{Generic $\mathcal{F}$-case contribution}
\label{subcase:standardf}

      For $s \in \{m_2+1,\dots,k\}$ and $t \in \{2,\dots,m_1-1\}$,
case $\mathcal{F}$ contains only standard MHV-like poles:
\begin{equation} \begin{aligned}
R_{\mathcal{F}}^{s,t} =
\frac{P_{1k}^2}{(4\pi)^{\frac{d}{2}}i}
\frac{ \braket{m_1|m_2}^4
       \braket{m_1|P_{t,m_1-1}|P_{t,s-1}|m_3}^2 }
     { \braket{12} \dots \braket{k\!-\!1|k}
       \braket{k\!+\!1|k\!+\!2} \dots \braket{n\!-\!1|n} }
\hspace{100pt} \\ \times
\braket{s\!-\!1|s} \braket{t\!-\!1|t} \! /
\big( P_{t,s-1}^2 \braket{m_1|P_{t,m_1-1}|P_{t,s-2}|s\!-\!1}
       \braket{m_1|P_{t,m_1-1}|P_{t,s-1}|s} & \\
       \braket{m_1|P_{m_1+1,s-1}|P_{t,s-1}|t\!-\!1}
       \braket{m_1|P_{m_1+1,s-1}|P_{t+1,s-1}|t} & \big) \\ \times
\bigg\{
\frac{ F_{\mathcal{F}}^{s,t}(\la_1,\lb_1) }
     { \braket{1|k} \braket{1|k\!+\!1} \braket{1|n} }
+
\frac{ F_{\mathcal{F}}^{s,t}(\la_k,\lb_k) }
     { \braket{k|1} \braket{k|k\!+\!1} \braket{k|n} }
+
\frac{ F_{\mathcal{F}}^{s,t}(\la_{k+1},\lb_{k+1}) }
     { \braket{k\!+\!1|1} \braket{k\!+\!1|k} \braket{k\!+\!1|n} } & \\
+
\frac{ F_{\mathcal{F}}^{s,t}(\la_n,\lb_n) }
     { \braket{n|1} \braket{n|k} \braket{n|k\!+\!1} }
+
\frac{ \bra{m_3}\!P_{1k}|q]^2 \bra{m_1}\!P_{t,m_1-1}|P_{t,s-1}|P_{1k}|q]^2 }
     { P_{1k}^2 \bra{1}\!P_{1k}|q] \bra{k}\!P_{1k}|q]
       \bra{k\!+\!1}\!P_{1k}|q] \bra{n}\!P_{1k}|q] } &
\bigg\} ,
\label{PolesStandardF}
\end{aligned} \end{equation}
where
\begin{equation}
F_{\mathcal{F}}^{s,t}(\la,\lb) = 
\frac{ \braket{\la|m_3}^2 \braket{\la|P_{t,s-1}|P_{t,m_1-1}|m_1}^2 [\lb|q] }
     { \bra{\la}\!P_{1k}|\lb] \bra{\la}\!P_{1k}|q] } .
\label{FstandardF}
\end{equation}

\subsubsection{Special $\mathcal{F}$-case contributions}
\label{subcase:speciale}

However, for this case $R_{m_1 s t}$ can get loop-momentum dependence
from both $s=l_2$ and $t=1$.
As a result, for the first subcase we obtain
\begin{equation} \begin{aligned}
& R_{\mathcal{F}}^{s=l_2,t} =
\frac{P_{1k}^2}{(4\pi)^{\frac{d}{2}}i}
\frac{ \braket{m_1|m_2}^4
       \braket{m_1|P_{t,m_1-1}|P_{t,k}|m_3}^2 }
     { \braket{12} \dots \braket{k\!-\!1|k}
       \braket{k\!+\!1|k\!+\!2} \dots \braket{n\!-\!1|n} } \\ & \times
\frac{ \braket{t\!-\!1|t} }
     { P_{t,k}^2 \braket{m_1|P_{t,m_1-1}|P_{t,k-1}|k}
       \braket{m_1|P_{m_1+1,k}|P_{t,k}|t\!-\!1}
       \braket{m_1|P_{m_1+1,k}|P_{t+1,k}|t} } \\ & \times \!
\bigg\{
\frac{ F_{\mathcal{F}}^{s=l_2,t}(\la_1,\lb_1) }
     { \braket{1|k\!+\!1} \braket{1|n} }
\! + \!
\frac{ F_{\mathcal{F}}^{s=l_2,t}(\la_{k+1},\lb_{k+1}) }
     { \braket{k\!+\!1|1} \braket{k\!+\!1|n} }
\! + \!
\frac{ F_{\mathcal{F}}^{s=l_2,t}(\la_n,\lb_n) }
     { \braket{n|1} \braket{n|k\!+\!1} }
\! + \!
\frac{ \bra{m_3}\!P_{1k}|q]^2 \!
       \bra{m_1}\!P_{t,m_1-1}|P_{t,k}|P_{1k}|q] }
     { P_{1k}^2 \! \bra{1}\!P_{1k}|q] \!
       \bra{k\!+\!1}\!P_{1k}|q] \! \bra{n}\!P_{1k}|q] }
\bigg\} ,
\label{PolesLastF}
\end{aligned} \end{equation}
where
\begin{equation}
F_{\mathcal{F}}^{s=l_2,t}(\la,\lb) = 
\frac{ \braket{\la|m_3}^2 \braket{\la|P_{t,k}|P_{t,m_1-1}|m_1} [\lb|q] }
     { \bra{\la}\!P_{1k}|\lb] \bra{\la}\!P_{1k}|q] } ,
\label{FlastF}
\end{equation}
and $t \in \{2,\dots,m_1-1\}$.

      For the second subcase we have
\begin{equation} \begin{aligned}
R_{\mathcal{F}}^{s,t=1} =
-\frac{P_{1k}^2}{(4\pi)^{\frac{d}{2}}i}
\frac{ \braket{m_1|m_2}^4
       \braket{m_1|P_{1,m_1-1}|P_{1,s-1}|m_3}^2 }
     { \braket{12} \dots \braket{k\!-\!1|k}
       \braket{k\!+\!1|k\!+\!2} \dots \braket{n\!-\!1|n} }
\hspace{137pt} \\ \times
\frac{ \braket{s\!-\!1|s} }
     { P_{1,s-1}^2 \braket{m_1|P_{1,m_1-1}|P_{1,s-2}|s\!-\!1}
       \braket{m_1|P_{1,m_1-1}|P_{1,s-1}|s}
       \braket{m_1|P_{m_1+1,s-1}|P_{2,s-1}|1} } & \\ \times
\bigg\{
\frac{ F_{\mathcal{F}}^{s,t=1}(\la_k,\lb_k) }
     { \braket{k|k\!+\!1} \braket{k|n} }
+
\frac{ F_{\mathcal{F}}^{s,t=1}(\la_{k+1},\lb_{k+1}) }
     { \braket{k\!+\!1|k} \braket{k\!+\!1|n} }
+
\frac{ F_{\mathcal{F}}^{s,t=1}(\la_n,\lb_n) }
     { \braket{n|k} \braket{n|k\!+\!1} } \hspace{52pt} \\
+
\frac{ \bra{m_3}\!P_{1k}|q]^2
       \bra{m_1}\!P_{1,m_1-1}|P_{1,s-1}|P_{1k}|q]^2 }
     { P_{1k}^2 \bra{k}\!P_{1k}|q] \bra{k\!+\!1}\!P_{1k}|q]
       \bra{n}\!P_{1k}|q] \bra{m_1}\!P_{m_1+1,s-1}|P_{1,s-1}|P_{1k}|q] } &
\bigg\} ,
\label{PolesFirstF}
\end{aligned} \end{equation}
where
\begin{equation}
F_{\mathcal{F}}^{s,t=1}(\la,\lb) = 
\frac{ \braket{\la|m_3}^2 \braket{\la|P_{1,s-1}|P_{1,m_1-1}|m_1}^2 [\lb|q] }
     { \braket{\la|P_{1,s-1}|P_{m_1+1,s-1}|m_1}
       \bra{\la}\!P_{1k}|\lb] \bra{\la}\!P_{1k}|q] } ,
\label{FfirstF}
\end{equation}
which is valid for $s \in \{m_2+1,\dots,k\}$.

      Finally, when both subcases coincide,
$R_{m_1 l_2 \, 1}$ cancels not just one, but two physical poles
$\ket{\la}=\ket{1}$ and $\ket{\la}=\ket{k}$,
and we are left with only three residues:
\begin{equation} \begin{aligned}
R_{\mathcal{F}}^{s=l_2,t=1} \! = \! - 
\frac{1}{(4\pi)^{\frac{d}{2}}i}
\frac{ \braket{m_1|m_2}^4
       \braket{m_1|P_{1,m_1-1}|P_{1k}|m_3}^2 }
     { \braket{12} \dots \braket{k\!-\!1|k} \!
       \braket{k\!+\!1|k\!+\!2} \dots \braket{n\!-\!1|n} }
\frac{1}{ \braket{m_1|P_{1,m_1-1}|P_{1k}|k} \!
          \braket{m_1|P_{m_1+1,k}|P_{1k}|1} } & \\ \times
\bigg\{
\frac{ F_{\mathcal{F}}^{s=l_2,t=1}(\la_{k+1},\lb_{k+1}) }
     { \braket{k\!+\!1|n} }
\! + \!
\frac{ F_{\mathcal{F}}^{s=l_2,t=1}(\la_n,\lb_n) }
     { \braket{n|k\!+\!1} }
\! + \!
\frac{ \bra{m_3}\!P_{1k}|q]^2 \bra{m_1}\!P_{1,m_1-1}|q] }
     { P_{1k}^2 \! \bra{k\!+\!1}\!P_{1k}|q] \!
      \bra{n}\!P_{1k}|q] \! \bra{m_1}\!P_{m_1+1,k}|q] }
\bigg\} & ,
\label{PolesFinalF}
\end{aligned} \end{equation}
where
\begin{equation}
F_{\mathcal{F}}^{s=l_2,t=1}(\la,\lb) = 
\frac{ \braket{\la|m_3}^2 \braket{\la|P_{1k}|P_{1,m_1-1}|m_1} [\lb|q] }
     { \braket{\la|P_{1k}|P_{m_1+1,k}|m_1}
       \bra{\la}\!P_{1k}|\lb] \bra{\la}\!P_{1k}|q] } .
\label{FfinalF}
\end{equation}

\subsubsection{Remark}
\label{subcase:remarke}

      If we go back to the adjacent-helicity bubble coefficient
computed as an example earlier in section (\ref{simplebubbles}),
we see that in that case all non-zero contributions
come from case $\mathcal{F}$ with $s=l_2$ and $t \in \{1,\dots,m_2\!-\!1=k\!-\!2\}$,
and formula (\ref{simplebubbleq}) might just effictively sum over
contributions (\ref{PolesLastF}) and (\ref{PolesFinalF}).
Yet it turns out that for $m_1=k\!-\!1$, spurious pole
$\braket{\la|P_{1k}|P_{m_1+1,k}|m_1}=\bra{\la}\!P_{1k}|k]\braket{k|m_1}$
becomes physical. Formula (\ref{simplebubbleq}) takes account of that,
so it is complementary and can be considered as another subcase of $\mathcal{F}$.

\section{Box coefficients}
\label{nmhvboxes}

      In this section, we provide new all-$n$ formulas
for a family of 2-mass-easy and 1-mass NMHV box coefficients.

      In Section \ref{box}, we gave a general spinor integration formula
for box coefficients and explained that it is equivalent
to the quadruple cut method \cite{Britto:2004nc}.
Once again, the general implementation (\ref{Cbox}) contains square roots
within $P^{ij}_\pm$,
whereas coefficients are known to be rational functions of spinor products.
In principle, the sole purpose of $P^{ij}_\pm$ is
to define a kinematic configuration for on-shell momenta $l_1$ and $l_2$
in which two more loop momenta $l_3$ and $l_4$ will be on shell as well.
For all but four-mass quadruple cuts,
it is easy to arrange this without involving any square roots,
\cite{Risager:2008yz}
so we prefer to do this part of the calculation
without complicating things with homogeneous spinor variables.

      Now let us consider a quadruple cut of a NMHV gluon amplitude with
$\mathcal{N}=1$ chiral matter in the loop.
In each corner of the four corners of the cut,
we have a tree amplitude with all gluon legs except two matter legs.
If all gluons are have a positive helicity, it vanishes,
unless it is a three-point amplitude,
so each massive corner must contain a negative-helicity gluon.
As there are only three of them, four-mass cuts are zero.
Massless boxes are absent just because we consider $n>4$.
Thus, we need to consider three-mass, two-mass and one-mass boxes.

      It is now easy to see that many NMHV quadruple cuts are constructed
from only MHV and $\overline{\text{MHV}}$ tree amplitudes.
We refer to such box coefficients as MHV-constructible.
For example, all non-zero three-mass quadruple cuts contain
three MHV amplitudes and one three-point $\overline{\text{MHV}}$ amplitude.
For these and other MHV-constructible box coefficients
all-multiplicity formulas were already given in \cite{Dunbar:2009ax},
which we rewrite in Appendix \ref{app:boxes}.

      Thus, we need to consider only quadruple cuts that contain NMHV vertices,
i.~e. amplitudes with two negative-helicity gluons and two matter legs.
The last negative-helicity gluon provides a massive or massless MHV vertex,
while two remaining vertices have to be three-point $\overline{\text{MHV}}$.
Such three-point amplitudes are non-vanishing only for complex momenta
that satisfy spinor proportionality conditions
$\la_1 \propto \la_2 \propto \la_3$.
This prevents two three-point vertices of the same kind to be adjacent to each other
because otherwise that would constrain not only the loop momentum,
but also arbitrary external momenta.
Therefore, the $\overline{\text{MHV}}$ amplitudes
must be in the opposite corners of the box.
If the MHV vertex contains multiple gluons,
the cut corresponds to what is usually called two-mass-easy box.
In case it only has one (negative-helicity) gluon, it becomes one-mass box,
which can fortunately be considered just as a subcase of the former.

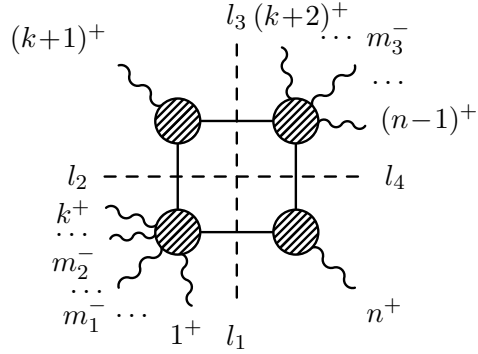
\begin{figure}[h]
      \centering
      \parbox{127pt}{ \begin{fmffile}{graph6}
      \fmfframe(10,10)(10,10){ \begin{fmfgraph*}(100,100)
            \fmflabel{$l_1$}{bottom}
            \fmflabel{$1^+$}{g1}
            \fmflabel{$\dots$}{dots1}
            \fmflabel{$m_1^-$}{gm1}
            \fmflabel{$\dots$}{dots2}
            \fmflabel{$m_2^-$}{gm2}
            \fmflabel{$\dots$}{dots3}
            \fmflabel{$k^+$}{gk}
            \fmflabel{$l_2$}{left}
            \fmflabel{$(k\!+\!1)^+$}{gk1}
            \fmflabel{$l_3$}{top}
            \fmflabel{$\quad(k\!+\!2)^+$}{gk2}
            \fmflabel{$\dots$}{dots4}
            \fmflabel{$m_3^-$}{gm3}
            \fmflabel{$\dots$}{dots5}
            \fmflabel{$(n\!-\!1)^+$}{gn1}
            \fmflabel{$l_4$}{right}
            \fmflabel{$n^+$}{gn}
            \fmfleft{,,,dots2,,,,gm2,dots3,,gk,,,left,,,,,,,,,,,,,}
            \fmfright{,,,,,,,,right,,,gn1,,dots5,,,}
            \fmftop{,gk1,,,,,,,,top,,,gk2,,dots4,,,gm3,}
            \fmfbottom{,gm1,,,dots1,,g1,,,bottom,,,,,,,,gn,}
            \fmf{wiggly,tension=0}{g1,v1}
            \fmf{wiggly,tension=1.0}{gm1,v1}
            \fmf{wiggly,tension=0}{gm2,v1}
            \fmf{wiggly,tension=0}{gk,v1}
            \fmf{wiggly,tension=1.0}{gk1,v2}
            \fmf{wiggly,tension=0}{gk2,v3}
            \fmf{wiggly,tension=1.0}{gm3,v3}
            \fmf{wiggly,tension=0}{gn1,v3}
            \fmf{wiggly,tension=1.0}{gn,v4}
            \fmf{dashes}{top,bottom}
            \fmf{dashes}{left,right}
            \fmf{plain,tension=0.50}{v1,v2}
            \fmf{plain,tension=0.50}{v2,v3}
            \fmf{plain,tension=0.50}{v3,v4}
            \fmf{plain,tension=0.50}{v4,v1}
            \fmfblob{0.17w}{v1}
            \fmfblob{0.17w}{v2}
            \fmfblob{0.17w}{v3}
            \fmfblob{0.17w}{v4}
      \end{fmfgraph*} }
      \end{fmffile} }
      \caption{$P_{1k}$-channel cut for
               $ A_{\mathcal{N}=1 \text{ chiral}}^{\text{1-loop,NMHV}}
                   (m_1^-,m_2^-\!\in\!\{1,\dots,k\},
                          m_3^-\!\in\!\{k\!+\!2,\dots,n\!-\!1\}) $
               promoted to a two-mass-easy quadruple cut
               by cutting $l_3=l_2-p_{k+1}$ and $l_4=l_1+p_n$
               \label{box2me}}
      \end{figure}

      Consequently, we only need to consider the cut shown in fig. \ref{box2me},
where we label the gluons in the NMHV corner as
$\{1^+,\dots,m_1^-,\dots,m_2^-,\dots,k^+\}$
in order to match our notation from the previous sections.
In fact, we can produce the two-mass-easy quadruple cut simply from the double cut
which we carefully constructed in Section \ref{cutconstruction},
by cutting two more propagators adjacent to the original cut, namely,
$l_3^2 \equiv (l_2-p_{k+1})^2 = -\braket{k\!+\!1|l_2} [l_2|k\!+\!1] $
and
$l_4^2 \equiv (l_1+p_n)^2 = \braket{l_1|n} [n|l_1] $.

      There are two complex-conjugate kinematic solutions
that put all four cut propagators on shell,
but only one of them corresponds to non-zero $\overline{\text{MHV}}$ vertices:
\begin{equation}
\begin{cases}
      l_1^{\mu} =  \dfrac{1}{2}
                   \dfrac{ \bra{n}\!\gamma^{\mu}|P_{1k}\!\ket{k\!+\!1} }
                         { \braket{n|k\!+\!1} } \\
      l_2^{\mu} = -\dfrac{1}{2}
                   \dfrac{ \bra{k\!+\!1}\!\gamma^{\mu}|P_{1k}\!\ket{n} }
                         { \braket{k\!+\!1|n} } \\
      l_3^{\mu} =  \dfrac{1}{2}
                   \dfrac{ \bra{k\!+\!1}\!\gamma^{\mu}|P_{k+2,n-1}\!\ket{n} }
                         { \braket{k\!+\!1|n} } \\
      l_4^{\mu} = -\dfrac{1}{2}
                   \dfrac{ \bra{n}\!\gamma^{\mu}|P_{k+2,n-1}\!\ket{k\!+\!1} }
                         { \braket{n|k\!+\!1} }
\end{cases}
\label{2mekinematics} \end{equation}

      Here, we note that the solution conjugate to (\ref{2mekinematics})
corresponds to non-zero MHV vertices in massless corners,
which would leave only one minus helicity for two massive legs.
Therefore, such boxes vanish for the NMHV amplitude.
Moreover, both kinematic solutions mean that the massless corners
cannot have different helicities, which leaves us
with the only non-zero family of two-mass-easy NMHV boxes.

      To compute such coefficient from the double cut (\ref{genericcut}),
we only need to multiply it by $l_3^2$ and $l_4^2$,
which immediately cancels MHV-like poles
$\braket{k\!+\!1|l_2}$ and $ \braket{l_1|n}$,
and constrain loop momentum spinors to satisfy (\ref{2mekinematics}).
Doing the same operation to the master integral would produce
$-2(4\pi)^{\frac{d}{2}}i$, where the factor of 2
comes from the two on-shell solutions,
so we normalize the cut expression accordingly:
      \begin{equation} \begin{aligned}
             C_{\mathcal{N}=1 \text{ chiral}}^{\text{box,2me}} =
                  - \frac{1}{2(4\pi)^{\frac{d}{2}}i}
                    \frac{ 1 }
                         { \braket{12} \dots \braket{k\!-\!1|k}
                           \braket{k\!+\!1|k\!+\!2} \dots \braket{n\!-\!1|n} }
                    \frac{ \braket{l_1|m_3} [n|l_1]
                           [l_2|k\!+\!1] \braket{m_3|-\!l_2} }
                         { \braket{l_1|1} \braket{l_1|l_2}^2
                           \braket{k|l_2} } & \\ \times \!\!\!
                    \sum_{s=m_1+2}^{m_1-3} \, \sum_{t=s+2}^{m_1-1}
                    R_{m_1 st} D_{m_1 st; m_2(-l_1)} D_{m_1 st; m_2 l_2}
                    \big(
                      \!\braket{-l_2|m_3} D_{m_1 st; m_2(-l_1)}
                      + \braket{ l_1|m_3} D_{m_1 st; m_2  l_2 }
                    \big)^2 & .
      \label{4cut}
      \end{aligned} \end{equation}

      Completely in the spirit of Section \ref{nmhvbubbles},
we divide the double sum in (\ref{4cut})
into cases $\mathcal{A}$ through $\mathcal{E}$
with respect to the loop momentum dependence
and provide the contributions separately:
\begin{equation} \begin{aligned}
    & C_{\mathcal{N}=1 \text{ chiral}}^{\text{box,2me}}
        (m_1^-,m_2^-\!\in\!\{1,\dots,k\}, m_3^-\!\in\{k\!+\!1,\dots,n\}) \\
        & = \sum_{ \{s,t\} \in \mathcal{A} } C_{\mathcal{A}}^{\text{box},s,t}
          + \sum_{ \{s,t\} \in \mathcal{B} } C_{\mathcal{B}}^{\text{box},s,t}
 +\!\!\!\!\!\!\sum_{ \{s,t=-l_1\} \in \mathcal{C} }
                                  \!\!\!\!\! C_{\mathcal{C}}^{\text{box},s}
 +\!\!\!\!\!\!\sum_{ \{s,t=-l_1\} \in \mathcal{D} }
                                  \!\!\!\!\! C_{\mathcal{D}}^{\text{box},s}
          + \sum_{ \{s,t\} \in \mathcal{E} } C_{\mathcal{E}}^{\text{box},s,t}
          + \sum_{ \{s,t\} \in \mathcal{F} } C_{\mathcal{F}}^{\text{box},s,t} .
\label{CboxN1frame}
\end{aligned} \end{equation}

      Note that, apart from multiplication by
$ \braket{l_1|n} \! \braket{k\!+\!1|l_2} \! [n|l_1]  [l_2|k\!+\!1] $,
what follows is basically
the same double cut expressions that we used in Section \ref{nmhvbubbles}
to generate bubble coefficients.
The main difference is that now loop momentum spinors
are understood to satisfy (\ref{2mekinematics}), i.~e.
\begin{equation}
\begin{cases}
      \bra{l_1} =  \bra{n} , \hspace{50pt}
          |l_1] =  \dfrac{ |P_{1k}\!\ket{k\!+\!1} }
                         { \braket{n|k\!+\!1} } , \\
      \bra{l_2} =  \bra{k\!+\!1} , \hspace{36pt}
          |l_2] = -\dfrac{ |P_{1k}\!\ket{n} }
                         { \braket{k\!+\!1|n} } .
\end{cases}
\label{2mespinors} \end{equation}

      This time, we go through the cases in alphabetical order
and we find now need to write formulas for subcases separately,
except if they are due to minus-helicity gluons $m_1^+$ and $m_2^+$
becoming adjacent.

\subsection{Cases $\mathcal{A}$ and $\mathcal{B}$}
\label{boxcase:ab}

      As in Section \ref{case:ab},
cases $\mathcal{A}$ and $\mathcal{B}$ can be considered together,
so for $s \in \{m_1+2,\dots,k-2\}$ and $t \in \{m_2+1,\dots,k,l_2\}$
we have:
\begin{equation} \begin{aligned}
C_{\mathcal{AB}}^{\text{box},s,t}(B) \! = \! -
\frac{1}{2(4\pi)^{\frac{d}{2}}i}
\frac{ \braket{m_1|m_3}^2
       \braket{m_1|B|P_{s,t-1}|m_2}^4 }
     { \braket{12} \dots \braket{k\!-\!1|k} \!
       \braket{k\!+\!1|k\!+\!2} \dots \braket{n\!-\!1|n} }
\frac{ \braket{l_1|m_1} \! \braket{l_1|m_3} \!
       [n|l_1] [l_2|k\!+\!1] \!
       \braket{m_1|l_2} \! \braket{m_3|l_2} \! }
     { \braket{l_1|1} \! \braket{k|l_2} }
& \\ \times
\braket{s\!-\!1|s} \! \braket{t\!-\!1|t} \! /
\big(P_{s,t-1}^2 \! \braket{m_1|P_{m_1+1,t-1}|P_{s,t-1}|s\!-\!1} \!
       \braket{m_1|P_{m_1+1,t-1}|P_{s+1,t-1}|s} & \\ \times
       \braket{m_1|P_{m_1+1,s-1}|P_{s,t-2}|t\!-\!1} \!
       \braket{m_1|P_{m_1+1,s-1}|P_{s,t-1}|t} & \big) .
\label{BoxStandardAB}
\end{aligned} \end{equation}

\subsection{Case $\mathcal{C}$}
\label{boxcase:c}

If negative-helicity gluons $m_1^+$ and $m_2^+$ are non-adjacent,
case $\mathcal{C}$ produces the following terms.

\subsubsection{First $\mathcal{C}$-case contribution}
\label{boxsubcase:firstc}
      The first term comes from $s=m_1+2$:
\begin{equation} \begin{aligned}
C_{\mathcal{C}}^{\text{box},s=m_1+2} \! = \! -
\frac{1}{2(4\pi)^{\frac{d}{2}}i}
\frac{ \braket{m_1|m_1\!+\!1} \!
       \braket{m_1\!+\!1|m_1\!+\!2} }
     { \braket{12} \dots \braket{k\!-\!1|k} \!
       \braket{k\!+\!1|k\!+\!2} \dots \braket{n\!-\!1|n} }
\frac{ \braket{l_1|m_1} \! \braket{l_1|m_3} \! [n|l_1] [l_2|k\!+\!1] \!
       \braket{m_2|l_2} \! \braket{m_3|l_2} }
     { \braket{l_1|1} \! \braket{l_1|l_2} \! \braket{k|l_2} } &
\\ \times
\frac{ \big(\! \bra{m_1}\!P_{1,m_1}|m_1\!+\!1]
             - \braket{m_1|l_1}[l_1|m_1\!+\!1] \big)
       \big(\! \bra{m_2}\!P_{1,m_1}|m_1\!+\!1]
             - \braket{m_2|l_1}[l_1|m_1\!+\!1] \big) }
     { \big( P_{1,m_1}^2 - \bra{l_1}\!P_{1,m_1}|l_1] \big)
       \big( P_{1,m_1+1}^2 - \bra{l_1}\!P_{1,m_1+1}|l_1] \big)
       \bra{l_1}\!P_{1,m_1}|m_1\!+\!1] [m_1\!+\!1|P_{m_1+2,k}\!\ket{l_2} } &
\\ \times
\frac{ \Big(\!
       \braket{m_1|m_3} \braket{l_1|l_2}
       \big(\!\bra{m_2}\!P_{1,m_1}|m_1\!+\!1]
            - \braket{m_2|l_1}[l_1|m_1\!+\!1] \big)
     + \braket{m_1|m_2} \braket{l_1|m_3}\![m_1\!+\!1|P_{m_1+2,k}\!\ket{l_2} \!
       \Big)^2 }
     { \braket{m_1|P_{1,m_1-1}|P_{1,m_1+1}|m_1\!+\!2}
     - \braket{m_1|l_1}\![l_1|P_{1,m_1+1}\!\ket{m_1\!+\!2}
     - \bra{m_1}\!P_{1,m_1-1}|l_1]\!\braket{l_1|m_1\!+\!2} } & .
\label{BoxFirstC}
\end{aligned} \end{equation}

\subsubsection{Generic $\mathcal{C}$-case contribution}
\label{boxsubcase:standardc}
      Subsequent terms come from $s \in \{m_1\!+\!3,\dots,m_2\}$:
\begin{equation} \begin{aligned}
C_{\mathcal{C}}^{\text{box},s} \! = &
\frac{1}{2(4\pi)^{\frac{d}{2}}i}
\frac{ \braket{s\!-\!1|s} }
     { \braket{12} \dots \braket{k\!-\!1|k} \!
       \braket{k\!+\!1|k\!+\!2} \dots \braket{n\!-\!1|n} }
\frac{ \braket{l_1|m_1} \! \braket{l_1|m_3} \! [n|l_1] [l_2|k\!+\!1] \!
       \braket{m_2|l_2} \! \braket{m_3|l_2} }
     { \braket{l_1|1} \! \braket{l_1|l_2} \! \braket{k|l_2} }
\\ \times &
\frac{ 1 }
     { \big( P_{1,s-1}^2 - \bra{l_1}\!P_{1,s-1}|l_1] \big)
       \braket{l_1|P_{1,s-1}|P_{m_1+1,s-1}|m_1} \!
       \braket{m_1|P_{m_1+1,s-1}|P_{s,k}|l_2} }
\\ \times &
\big(\! \bra{m_1}\!P_{m_1+1,s-1}|l_1]\!\braket{l_1|m_2}
      - \braket{m_1|P_{m_1+1,s-1}|P_{1,s-1}|m_2} \!\big)
\\ \times &
\big(\! \bra{m_1}\!P_{m_1+1,s-1}|l_1]\!\braket{l_1|m_1}
      - \braket{m_1|P_{m_1+1,s-1}|P_{1,m_1-1}|m_1} \!\big)
\\ \times &
\big(\! \braket{l_1|l_2} \! \braket{m_2|m_3} \!
        \braket{m_1|P_{1,s-1}|P_{s,m_1-1}|m_1}
      + \braket{m_1|m_2} \! \braket{m_3|l_2} \!
        \braket{l_1|P_{1,s-1}|P_{s,m_1-1}|m_1}
      \\ & \hspace{184pt}
      - \braket{m_2|m_3} \! \braket{l_1|m_1} \!
        \braket{m_1|P_{m_1+1,s-1}|P_{1,k}|l_2} \!\big)^2
\\ / &
\Big(
\big(\! \braket{m_1|P_{1,m_1-1}|P_{1,s-2}|s\!-\!1}
      - \braket{m_1|l_1}\![l_1|P_{1,s-2}\!\ket{s\!-\!1}
      - \bra{m_1}\!P_{1,m_1-1}|l_1]\!\braket{l_1|s\!-\!1} \!\big)
      \\ & \hspace{41pt} \times \!
\big(\! \braket{m_1|P_{1,m_1-1}|P_{1,s-1}|s}
      - \braket{m_1|l_1}\![l_1|P_{1,s-1}\!\ket{s}
      - \bra{m_1}\!P_{1,m_1-1}|l_1]\!\braket{l_1|s} \!\big)
\Big) .
\label{BoxStandardC}
\end{aligned} \end{equation}

\subsection{Case $\mathcal{D}$}
\label{boxcase:d}

\subsubsection{Generic $\mathcal{D}$-case contribution}
\label{boxsubcase:standardd}
      For $s \in \{m_2\!+\!1,\dots,k\}$ case $\mathcal{D}$ generates
\begin{equation} \begin{aligned}
C_{\mathcal{D}}^{\text{box},s} \! = &
\frac{1}{2(4\pi)^{\frac{d}{2}}i}
\frac{ \braket{m_1|m_2}^4 \braket{s\!-\!1|s} }
     { \braket{12} \dots \braket{k\!-\!1|k} \!
       \braket{k\!+\!1|k\!+\!2} \dots \braket{n\!-\!1|n} }
\frac{ \braket{l_1|m_1} \! \braket{l_1|m_3} \! [n|l_1]
       [l_2|k\!+\!1] \! \braket{m_3|l_2} }
     { \braket{l_1|1} \! \braket{l_1|l_2} \! \braket{k|l_2} }
\\ \times &
\frac{ \big(\! \braket{m_1|l_1}\![l_1|P_{s,k}\!\ket{l_2}
             - \braket{m_1|P_{1,m_1-1}|P_{s,k}|l_2} \!\big) }
     { \braket{l_1|P_{1,s-1}|P_{m_1+1,s-1}|m_1} \!
       \braket{m_1|P_{m_1+1,s-1}|P_{s,k}|l_2} }
\\ \times &
\big(\! \braket{l_1|m_1} \! \braket{m_3|P_{m_1,s-1}|P_{s,k}|l_2}
      + \braket{m_3|m_1} \! \braket{l_1|P_{1,m_1-1}|P_{s,k}|l_2} \!\big)^2
\\ / &
\Big(
\big(\! \braket{m_1|P_{1,m_1-1}|P_{1,s-2}|s\!-\!1}
      - \braket{m_1|l_1}\![l_1|P_{1,s-2}\!\ket{s\!-\!1}
      - \bra{m_1}\!P_{1,m_1-1}|l_1]\!\braket{l_1|s\!-\!1} \!\big)
      \\ & \hspace{41pt} \times \!
\big(\! \braket{m_1|P_{1,m_1-1}|P_{1,s-1}|s}
      - \braket{m_1|l_1}\![l_1|P_{1,s-1}\!\ket{s}
      - \bra{m_1}\!P_{1,m_1-1}|l_1]\!\braket{l_1|s} \!\big)
\Big) .
\label{BoxStandardD}
\end{aligned} \end{equation}

\subsubsection{Special $\mathcal{D}$-case contribution
               for adjacent negative helicities}
\label{boxsubcase:adjacentd}
      However, if $m_2=m_1\!+\!1$, the first $\mathcal{D}$-case term
should be considered separately:
\begin{equation} \begin{aligned}
C&_{\mathcal{D}}^{\text{box},s=m_1+2=m_2+1} \! =
\frac{1}{2(4\pi)^{\frac{d}{2}}i}
\frac{ \braket{m_1|m_1\!+\!1} \! \braket{m_1\!+\!1|m_1\!+\!2} }
     { \braket{12} \dots \braket{k\!-\!1|k} \!
       \braket{k\!+\!1|k\!+\!2} \dots \braket{n\!-\!1|n} }
\\ \times &
\frac{ \braket{l_1|m_1} \! \braket{l_1|m_3} \! [n|l_1]
       [l_2|k\!+\!1] \! \braket{m_3|l_2} }
     { \braket{l_1|1} \! \braket{l_1|l_2} \! \braket{k|l_2} }
\frac{ \big(\! \braket{m_1|P_{1,m_1-1}|P_{m_1+2,k}|l_2}
             - \braket{m_1|l_1} \![l_1|P_{m_1+2,k}\!\ket{l_2} \!\big) }
     { \big( P_{1,m_1}^2 - \bra{l_1}\!P_{1,m_1}|l_1] \big)
       \bra{l_1}\!P_{1,m_1}|m_1\!+\!1]
       [m_1\!+\!1|P_{m_1+2,k}\!\ket{l_2} }
\\ \times &
\big(\! \braket{m_1|m_3}\!\braket{l_1|P_{1,m_1+1}|P_{m_1+2,k}|l_2}
      - \braket{l_1|m_3} \! \braket{m_1|m_1\!+\!1|P_{m_1+2,k}|l_2} \!\big)^2
\\ / &
\big(\! \braket{m_1|P_{1,m_1-1}|P_{1,m_1+1}|m_1\!+\!2}
      - \braket{m_1|l_1}\![l_1|P_{1,m_1+1}\!\ket{m_1\!+\!2}
      - \bra{m_1}\!P_{1,m_1-1}|l_1]\!\braket{l_1|m_1\!+\!2} \!\big)
\label{BoxAdjacentD}
\end{aligned} \end{equation}

\subsection{Case $\mathcal{E}$}
\label{boxcase:e}
      Case $\mathcal{E}$ generates
\begin{equation} \begin{aligned}
C_{\mathcal{E}}^{\text{box},s,t} \! = \! -
\frac{1}{2(4\pi)^{\frac{d}{2}}i}
\frac{ \braket{m_2|m_3}^2
       \braket{m_1|P_{t,m_1-1}|P_{m_1+1,s-1}|m_1}^4 }
     { \braket{12} \dots \braket{k\!-\!1|k}
       \braket{k\!+\!1|k\!+\!2} \dots \braket{n\!-\!1|n} }
\frac{ \braket{l_1|m_2} \! \braket{l_1|m_3} \!
       [n|l_1] [l_2|k\!+\!1] \!
       \braket{m_2|l_2} \! \braket{m_3|l_2} \! }
     { \braket{l_1|1} \! \braket{k|l_2} }
& \\ \times
\braket{s\!-\!1|s} \! \braket{t\!-\!1|t} \! /
\big( P_{t,s-1}^2 \! \braket{m_1|P_{t,m_1-1}|P_{t,s-2}|s\!-\!1} \!
       \braket{m_1|P_{t,m_1-1}|P_{t,s-1}|s} & \\
       \braket{m_1|P_{m_1+1,s-1}|P_{t,s-1}|t\!-\!1} \!
       \braket{m_1|P_{m_1+1,s-1}|P_{t+1,s-1}|t} & \big) ,
\label{BoxStandardE}
\end{aligned} \end{equation}
where $s \in \{m_1+2,\dots,m_2\}$ and $t \in \{2,\dots,m_1-1\}$,
whereas for the subcase where $t=1$,
one can use the same formula with $t\!-\!1=-l_1$.

\subsection{Case $\mathcal{F}$}
\label{boxcase:f}
      Finally, for $s \in \{m_2+1,\dots,k,l_2\}$, $t \in \{1,2,\dots,m_1-1\}$
we have
\begin{equation} \begin{aligned}
C_{\mathcal{F}}^{\text{box},s,t} \! = \! - &
\frac{1}{2(4\pi)^{\frac{d}{2}}i}
\frac{ \braket{m_1|m_2}^4
       \braket{m_1|P_{t,m_1-1}|P_{t,s-1}|m_3}^2 }
     { \braket{12} \dots \braket{k\!-\!1|k} \!
       \braket{k\!+\!1|k\!+\!2} \dots \braket{n\!-\!1|n} }
\frac{ \braket{l_1|m_3} \!
       [n|l_1] [l_2|k\!+\!1] \!
       \braket{m_3|l_2} \! }
     { \braket{l_1|1} \! \braket{k|l_2} }
\\ \times &
\braket{l_1|P_{t,s-1}|P_{t,m_1-1}|m_1} \!
\braket{m_1|P_{t,m_1-1}|P_{t,s-1}|l_2}
\\ \times &
\braket{s\!-\!1|s} \! \braket{t\!-\!1|t} \! /
\big( P_{t,s-1}^2 \! \braket{m_1|P_{t,m_1-1}|P_{t,s-2}|s\!-\!1} \!
       \braket{m_1|P_{t,m_1-1}|P_{t,s-1}|s} \\ & \hspace{77pt}
       \braket{m_1|P_{m_1+1,s-1}|P_{t,s-1}|t\!-\!1} \!
       \braket{m_1|P_{m_1+1,s-1}|P_{t+1,s-1}|t} \! \big) ,
\label{BoxStandardF}
\end{aligned} \end{equation}
where again we include the three subcases with $s=l_2$ and/or $t\!-\!1=-l_1$.

\section{Checks}
\label{checks}

      The first check we used to ensure the validity of our results
was verifying that the sum of all bubble coefficients (\ref{bubblesum})
coincides numerically with the tree amplitude,
as discussed in Section (\ref{singular}).
We ensured this for all distinct helicity configurations at 6, 7 and 8 points.

      As another strong and independent cross-check,
we compared our results with numerical data
kindly produced with the help of the powerful NGluon package
\cite{Badger:2010nx} by one of its authors.
To simulate the $\mathcal{N}=1$ chiral multiplet in the loop,
we had to add separate contributions from the fermion and the scalar loop.
Moreover, to remove the discrepancies
due to different implementation of spinor-helicity formalism,
we compared ratios of the master integral coefficients to the tree amplitude.
In this way, we witnessed agreement
for all types of coefficients
up to machine precision of 13 digits for 8-point amplitudes
and 12 digits for 17-point amplitudes.

      Producing numerical tests for a large number (such as 25) of external gluons
becomes more involved, as their kinematics gets more and more singular.
There are numerical instabilities at the level of coefficient/tree ratios
which we believe to come from the spurious poles in $R_{rst}$ \eqref{Rrst}.
They cancel in the sum over $s$ and $t$, but can contaminate the numerical accuracy.
In fact, this issue occurs for the tree amplitude itself.

\section{Discussion and outlook}
\label{discussion}

      We have studied one-loop NMHV amplitudes
in $\mathcal{N}=1$ super-Yang-Mills theory for any number of external gluons
and managed to find general analytic formulas
for all missing scalar integral coefficients:
\begin{itemize}
\item bubbles with arbitrary helicity assignment;
\item two-mass-easy and one-mass boxes with two minus-helicity gluons attached to one of the massive corners, but otherwise arbitrary.
\end{itemize}
We have also numerically verified the remaining all-$n$ formulas
calculated previously in \cite{Dunbar:2009ax}
which we provide in Appendices \ref{app:boxes} and \ref{app:triangles} for completeness.

      Our principal method was spinor integration
\cite{Britto:2005ha,Anastasiou:2006jv}.
It is a general one-loop method which combines mathematical elegance
with simplicity of computer implementation.
Even though we adapted it to the case of bubbles
with massless $\mathcal{N}=1$ chiral supermultiplet in the loop,
the method is general and can also be applied to
theories with massive particle content
\cite{Britto:2006fc}
and arbitrary loop-momentum power-counting
\cite{Anastasiou:2006gt}.

      For all our results, we performed numerical tests at 8 and 17 points
and found agreement with numerical data produced by other methods.

      Thus, NMHV amplitudes $\mathcal{N}=1$ SYM
add to the body of one-loop amplitudes known for all $n$.
Of course, such amplitudes were already numerically accessible
for phenomenological studies for multiplicities of order 20
\cite{Giele:2008bc,Badger:2010nx}.
Hopefully, new analytic results will prove useful
for the search of general mathematical structure of amplitudes,
such as recursion relations between separate coefficients
or their meaning in (momentum) twistor space \cite{Hodges:2009hk}.
Moreover, the formulas we provide
might not be the best possible way to write down the NMHV amplitudes.
We look forward to further studies that might uncover a simpler way to look at them,
such as rewriting them using more suitable variables or a better integral basis.

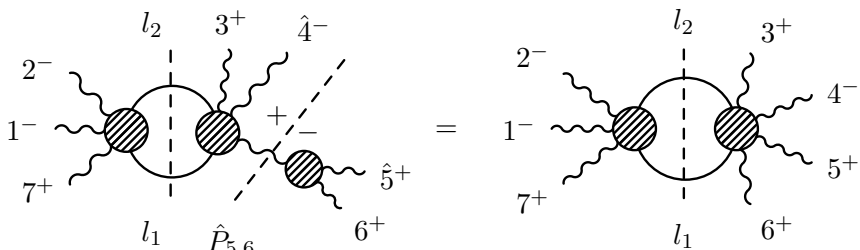
\begin{figure}[h]
      \centering
      \parbox{150pt}{ \begin{fmffile}{graph11}
      \fmfframe(10,10)(10,10){ \begin{fmfgraph*}(120,60)
            \fmflabel{$7^+$}{g7}
            \fmflabel{$l_1$}{bottom}
            \fmflabel{$1^-$}{g1}
            \fmflabel{$2^-$}{g2}
            \fmflabel{$3^+$}{g3}
            \fmflabel{$l_2$}{top}
            \fmflabel{$\hat{4}^-$}{g4}
            \fmflabel{$\hat{5}^+$}{g5}
            \fmflabel{$6^+$}{g6}
            \fmflabel{$\hat{P}_{5,6}$}{P56bottom}
            \fmfleft{,g7,,g1,,g2,}
            \fmfright{g6,,g5,,,,,,}
            \fmftop{,,,,top,,g3,,g4,,P56top,}
            \fmfbottom{,,,,bottom,,P56bottom,,,,,}
            \fmf{wiggly}{g7,vleft}
            \fmf{wiggly}{g1,vleft}
            \fmf{wiggly}{g2,vleft}
            \fmf{wiggly,tension=0.2}{g3,vright}
            \fmf{wiggly,tension=0.5}{g4,vright}
            \fmf{wiggly,tension=1.7}{v56,vright}
            \fmf{wiggly,tension=1.5}{g5,v56}
            \fmf{wiggly,tension=1.5}{g6,v56}
            \fmf{dashes}{top,bottom}
            \fmf{dashes,
                 label=$\begin{array}{c}\!+\\[-10pt]\qquad-\end{array}$,
                 lab.dist=-0.42h}{P56top,P56bottom}
            \fmf{plain,left,tension=1.0}{vleft,vright}
            \fmf{plain,right,tension=1.0}{vleft,vright}
            \fmfblob{0.27h}{vleft}
            \fmfblob{0.27h}{vright}
            \fmfblob{0.23h}{v56}
      \end{fmfgraph*} }
      \end{fmffile} }
       = \qquad
      \parbox{127pt}{ \begin{fmffile}{graph12}
      \fmfframe(10,10)(10,10){ \begin{fmfgraph*}(100,60)
            \fmflabel{$7^+$}{g7}
            \fmflabel{$l_1$}{bottom}
            \fmflabel{$1^-$}{g1}
            \fmflabel{$2^-$}{g2}
            \fmflabel{$3^+$}{g3}
            \fmflabel{$l_2$}{top}
            \fmflabel{$4^-$}{g4}
            \fmflabel{$5^+$}{g5}
            \fmflabel{$6^+$}{g6}
            \fmfleft{,g7,,g1,,g2,}
            \fmfright{,,,g5,,,,g4,,,}
            \fmftop{,,top,g3,}
            \fmfbottom{,,bottom,g6,}
            \fmf{wiggly}{g7,vleft}
            \fmf{wiggly,tension=0.75}{g1,vleft}
            \fmf{wiggly}{g2,vleft}
            \fmf{wiggly}{g3,vright}
            \fmf{wiggly,tension=1.17}{g4,vright}
            \fmf{wiggly,tension=1.17}{g5,vright}
            \fmf{wiggly}{g6,vright}
            \fmf{dashes}{top,bottom}
            \fmf{plain,left,tension=1.0}{vleft,vright}
            \fmf{plain,right,tension=1.0}{vleft,vright}
            \fmfblob{0.27h}{vleft}
            \fmfblob{0.27h}{vright}
      \end{fmfgraph*} }
      \end{fmffile} }
      \caption{Recursion for the $P_{7,2}$-channel cut for
               $ A_{\mathcal{N}=1 \text{ chiral}}^{\text{1-loop,NMHV}}
                   (1^-,2^-,3^+,4^-,5^+,6^+,7^+) $
               \label{recursion}}
      \end{figure}

      To illustrate one possible train of thought for further developments,
we have found several series of bubble coefficients that obey
simple BCFW recursion relations inherited from the tree amplitudes
which constitute the corresponding unitarity cuts.
As a simple example, one can easily verify that $P_{n,2}$-bubble coefficients
in amplitudes of the form $A(1^-,2^-,3^+,4^-,5^+,\dots,n^+)$
are simply related through the $[45\rangle$-shift for $n>7$.
However, recursion fails if one tries to derive the 7-point coefficient
from the 6-point one, even though the cuts still possess that relation,
see fig. \ref{recursion}.

      Of course, such a relation would also fail
even if we flip the helicity of the 3rd gluon.
That would produce the split-helicity case in which the recursion
is well understood and takes place only if one packs adjacent scalar bubble integrals
into two-mass triangles with a Feynman parameter in the numerator
and then works with coefficients of that modified basis \cite{Bern:2005hh}.
The problem is that, unlike the split-helicity case,
the NMHV integral basis consists not only from bubbles,
but also from three-mass triangles and various boxes,
and it is not understood how to repackage the full set
of one-loop integrals to make the recursion work.

      This brings about another example within the same NMHV amplitude family:
we witnessed the validity of the $[45\rangle$-shift relation
between three-mass triangles $(23,4567,81)$ and $(23,456,71)$,
but not between $(23,456,71)$ and $(23,45,61)$. 
For some reason, the recursion seems to work, but later than expected,
which leaves it unreliable for any predictive calculations.
However, it seems to be the perfect tool
to obtain better understanding of the underlying structure
of the NMHV amplitudes beyond the tree level.
For instance, impressive developments in $\mathcal{N}=4$ SYM
at the all-loop integrand level \cite{ArkaniHamed:2012nw}
also heavily rely on the BCFW construction implemented in super-twistor variables.
It then seems natural that the on-shell recursion might eventually prove helpful
to tame \emph{integrated} loop amplitudes as well, hopefully, 
for arbitrary configurations of negative and positive helicities.

\section*{Acknowledgments}

I would like to thank my supervisor Ruth Britto
for her guidance and encouragement,
as well as for useful comments on this manuscript.
I would also like to express my gratitude to Simon Badger and Edoardo Mirabella
who provided valuable numerical cross-checks of the presented analytic results.
I am also very grateful to Bo Feng, Gregory Korchemsky, David Kosower,
Piotr Tourkine and Yang Zhang for helpful discussions.

I would like to extend my appreciation to the Bethe Center for Theoretical Physics
of the University of Bonn for the hospitality at the initial stage of the project.
This work was partially supported by the Agence Nationale de la Recherche under grant number ANR-09-CEXC-009-01.

\appendix

\section{Sign conventions}
\label{app:signs}

\subsection{Momentum flipping in spinors}
\label{app:momentumflip}

      When one performs four-dimensional cuts, some loop momenta are put on shell
and spinor expressions for amplitudes are used to construct the cut integrand.
Once one chooses a direction for a loop momentum,
an amplitude on one side of the cut will depend on loop-momentum spinors
with that momentum with a minus sign.
In this paper, we deal with such spinors in the following way:
      \begin{eqnarray} \begin{aligned}
            \ket{-l} & = i \ket{l} , \\
            |\!-\!l] & = i |l] .
      \label{flippingmomentum}
	\end{aligned} \end{eqnarray}

\subsection{Spinor residues}
\label{app:residues}

      Simple spinor residues are defined as
      \begin{equation}
            \Res_{\lambda=\zeta} \frac{1}{\braket{\zeta|\la}}
                                 \frac{N(\la,\lb)}{D(\la,\lb)}
                  = \frac{N(\zeta,\tilde{\zeta})}
                         {D(\zeta,\tilde{\zeta})}
                  =-\Res_{\lambda=\zeta} \frac{1}{\braket{\la|\zeta}}
                                         \frac{N(\la,\lb)}{D(\la,\lb)} .
      \label{simplespinorpole}
	\end{equation}
      Multiple spinor poles can in principle be extracted using the following formula:
      \begin{equation}
            \Res_{\lambda=\zeta} \frac{1}{\braket{\zeta|\la}^k}
                                 \frac{N(\la,\lb)}{D(\la,\lb)}
                  = \frac{1}{\braket{\eta|\zeta}^{k-1}}
                    \bigg\{ \frac{1}{(k-1)!}
                            \frac{\mathrm{d}^{(k-1)}}{\mathrm{d}t^{(k-1)}}
                            \frac{N(\zeta-t\eta,\tilde{\zeta})}
                                 {D(\zeta-t\eta,\tilde{\zeta})}
                    \bigg\} \bigg|_{t=0} ,
      \label{multiplespinorpole}
	\end{equation}
where $\eta$ is an arbitrary auxiliary spinor not equal to the pole spinor $\zeta$.
However, in $\mathcal{N}=1$ SYM there are no multiple poles.

      Care must be taken when dealing with poles of the form $\bra{\la}\!K|q]$
because it is equivalent to a pole $\braket{\zeta|\la}$ with the following value of $\zeta$:
      \begin{eqnarray} \begin{aligned}
            \bra{\zeta} & = [q|K| , &
            \ket{\zeta} & = - |K|q] , \\
            [\tilde{\zeta}| & = - \bra{q}\!K| , &
            |\tilde{\zeta}] & = |K\!\ket{q} .
      \label{compositepole}
	\end{aligned} \end{eqnarray}

\section{MHV-constructible box coefficients}
\label{app:boxes}

      In Section \ref{nmhvboxes}, we computed the coefficients of two-mass-easy boxes
with two minus-helicity gluons on one of the massive legs.
The result is also if the opposite leg with one minus-helicity gluon becomes massless,
which gives a family of one-mass boxes with two pluses opposite to each other.

      In this section, we gather all-multiplicity formulas for all remaining boxes
that were calculated previously in \cite{Dunbar:2009ax}.
We checked them numerically using spinor integration formula (\ref{Cbox})
through 8 points.

\subsection{One-mass boxes}
\label{app:box1m}

      \begin{figure}[h]
      \centering
      \parbox{127pt}{ \begin{fmffile}{graph9}
      \fmfframe(10,10)(10,10){ \begin{fmfgraph*}(100,100)
            \fmflabel{$(p\!+\!2)^+$}{pp2}
            \fmflabel{$\dots$}{dots1}
            \fmflabel{$m_1^-$}{m1}
            \fmflabel{$\dots$}{dots2}
            \fmflabel{$(p\!-\!2)^+$}{pm2}
            \fmflabel{$(p\!-\!1)^-$}{pm1}
            \fmflabel{$p^+$}{p}
            \fmflabel{$(p\!+\!1)^-$}{pp1}
            \fmfleft{,,,dots2,,pm2,,,left,,,,,,,,}
            \fmfright{right}
            \fmftop{,pm1,,,,,,,,top,,,,,,,,p,}
            \fmfbottom{,m1,,,dots1,,pp2,,,bottom,,,,,,,,pp1,}
            \fmf{wiggly,tension=0}{pp2,v1}
            \fmf{wiggly,tension=1.0}{m1,v1}
            \fmf{wiggly,tension=0}{pm2,v1}
            \fmf{wiggly,tension=1.0}{pm1,v2}
            \fmf{wiggly,tension=1.0}{p,v3}
            \fmf{wiggly,tension=1.0}{pp1,v4}
            \fmf{dashes}{top,bottom}
            \fmf{dashes}{left,right}
            \fmf{plain,tension=0.50}{v1,v2}
            \fmf{plain,tension=0.50}{v2,v3}
            \fmf{plain,tension=0.50}{v3,v4}
            \fmf{plain,tension=0.50}{v4,v1}
            \fmfblob{0.17w}{v1}
            \fmfblob{0.17w}{v2}
            \fmfblob{0.17w}{v3}
            \fmfblob{0.17w}{v4}
      \end{fmfgraph*} }
      \end{fmffile} }
      \caption{MHV-constructible one-mass quadruple cut
               of $ A_{\mathcal{N}=1 \text{ chiral}}^{\text{1-loop,NMHV}}
                   (m_1^-\!\in\!\{p+2,\dots,p-2\},
                    (p-1)^-,(p+1)^-) $
               \label{box1m}}
      \end{figure}
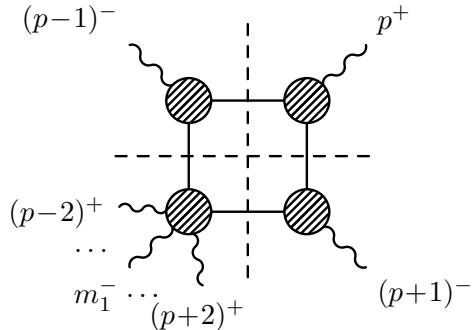

      First of all, there is another family of one-mass boxes
with two minuses opposite to each other, see fig. \ref{box1m}.
They are given by
\begin{equation} \begin{aligned}
      C^{\text{box,1m},-P_{p-1,p+1}}_{\mathcal{N}=1 \text{ chiral}}
      (\overset{+}{\dots},m_1^-,\overset{+}{\dots},
       (p\!-\!1)^-,p^+,(p\!+\!1)^-,\overset{+}{\dots}) =
    - \frac{1}{2(4\pi)^{\frac{d}{2}} i}
      \frac{ 1 }
           { \braket{p\!+\!2|p\!+\!3} \dots \braket{p\!-\!3|p\!-\!2} } &
      \\ \times
      \frac{ P_{p-1,p}^2 P_{p,p+1}^2  \bra{m_1}\!P_{p-1,p+1}|p\!-\!1]
             \bra{m_1}\!P_{p-1,p+1}|p] \bra{m_1}\!P_{p-1,p+1}|p\!+\!1] }
           { P_{p-1,p+1}^2 [p\!-\!1|p\!+\!1]^2
             \bra{p\!-\!2}\!P_{p-1,p+1}|p\!+\!1]
             \bra{p\!+\!2}\!P_{p-1,p+1}|p\!-\!1] } & .
\label{N1chBox1m}
\end{aligned} \end{equation}

\subsection{Two-mass-hard boxes}
\label{app:box2mh}

      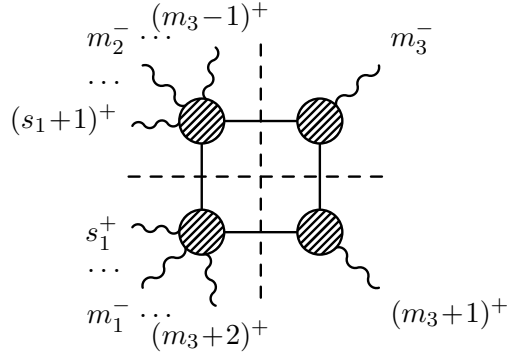
\begin{figure}[h]
      \centering
      \parbox{127pt}{ \begin{fmffile}{graph8}
      \fmfframe(10,10)(10,10){ \begin{fmfgraph*}(100,100)
            \fmflabel{$(m_3\!+\!2)^+$}{m32}
            \fmflabel{$\dots$}{dots1}
            \fmflabel{$m_1^-$}{m1}
            \fmflabel{$\dots$}{dots2}
            \fmflabel{$s_1^+$}{s1}
            \fmflabel{$(s_1\!+\!1)^+$}{s11}
            \fmflabel{$\dots$}{dots3}
            \fmflabel{$m_2^-$}{m2}
            \fmflabel{$\dots$}{dots4}
            \fmflabel{$(m_3\!-\!1)^+$}{m3m1}
            \fmflabel{$m_3^-$}{m3}
            \fmflabel{$(m_3\!+\!1)^+$}{m31}
            \fmfleft{,,,dots2,,s1,,,left,,,s11,,dots3,,,}
            \fmfright{right}
            \fmftop{,m2,,,dots4,,m3m1,,,top,,,,,,,,m3,}
            \fmfbottom{,m1,,,dots1,,m32,,,bottom,,,,,,,,m31,}
            \fmf{wiggly,tension=0}{m32,v1}
            \fmf{wiggly,tension=1.0}{m1,v1}
            \fmf{wiggly,tension=0}{s1,v1}
            \fmf{wiggly,tension=0}{s11,v2}
            \fmf{wiggly,tension=1.0}{m2,v2}
            \fmf{wiggly,tension=0}{m3m1,v2}
            \fmf{wiggly,tension=1.0}{m3,v3}
            \fmf{wiggly,tension=1.0}{m31,v4}
            \fmf{dashes}{top,bottom}
            \fmf{dashes}{left,right}
            \fmf{plain,tension=0.50}{v1,v2}
            \fmf{plain,tension=0.50}{v2,v3}
            \fmf{plain,tension=0.50}{v3,v4}
            \fmf{plain,tension=0.50}{v4,v1}
            \fmfblob{0.17w}{v1}
            \fmfblob{0.17w}{v2}
            \fmfblob{0.17w}{v3}
            \fmfblob{0.17w}{v4}
      \end{fmfgraph*} }
      \end{fmffile} }
      \caption{Two-mass-hard quadruple cut
               of $ A_{\mathcal{N}=1 \text{ chiral}}^{\text{1-loop,NMHV}}
                   (m_1^-\!\in\!\{m_3\!+\!2,\dots,s\},
                    m_2^-\!\in\!\{s\!+\!1,\dots,m_3\!-\!1\},m_3^-) $
               \label{box2mh}}
      \end{figure}

      Next, if a two-mass box has two massless legs adjacent to each other,
they must have different helicities.
This leaves two negative-helicity gluons for the other two massive legs,
which constitutes what is called a two-mass-hard box:
\begin{equation} \begin{aligned}
      & \!\!\!\!\!\!\!\!\!\!\!\!\!\!\,
      C^{\text{box,2mh},K_1,K_2}_{\mathcal{N}=1 \text{ chiral}}
      (\,\overset{+}{\dots},m_1^-,\overset{+}{\dots},m_2^-,
         \overset{+}{\dots},m_3^-,\overset{+}{\dots}\,) \\ = - &
      \frac{1}{2(4\pi)^{\frac{d}{2}} i}
      \frac{ 1 }
           { \braket{m_3\!+\!1|m_3\!+\!2}
             \braket{m_3\!+\!2|m_3\!+\!3} \dots \braket{s_1\!-\!1|s_1}
             \braket{s_1\!+\!1|s_1\!+\!2} \dots \braket{m_3\!-\!2|m_3\!-\!1} }
      \\ \times &
      \frac{ (K_2+p_{m_3})^2 P_{m_3,m_3+1}^2 \! \braket{m_1|m_3\!+\!1}
             \bra{m_1}\!K_2|m_3] \bra{m_2}\!K_2|m_3]
             \braket{m_3\!+\!1|K_1|K_2|m_2} }
           { K_2^2
             \bra{m_3\!+\!1}\!K_1|m_3] \bra{m_3\!+\!1}\!K_2|m_3]
             \bra{s_1}\!K_2|m_3] \bra{s_1\!+\!1}\!K_2|m_3]
             \braket{m_3\!+\!1|K_1|K_2|m_3\!-\!1} }
      \\ \times & \braket{m_1|K_2+p_{m_3}|K_2|m_2}^2 .
\label{N1chBox2mh}
\end{aligned} \end{equation}

      As explained in Section \ref{nmhvboxes},
if a two-mass box has two massless legs opposite to each other,
the only non-zero family of such two-mass-easy boxes
is the one computed in the main text.

\subsection{Three-mass boxes}
\label{app:box3m}

      \begin{figure}[h]
      \centering
      \parbox{127pt}{ \begin{fmffile}{graph7}
      \fmfframe(10,10)(10,10){ \begin{fmfgraph*}(100,100)
            \fmflabel{$(s_3\!+\!2)^+$}{s32}
            \fmflabel{$\dots$}{dots1}
            \fmflabel{$m_1^-$}{m1}
            \fmflabel{$\dots$}{dots2}
            \fmflabel{$s_1^+$}{s1}
            \fmflabel{$(s_1\!+\!1)^+$}{s11}
            \fmflabel{$\dots$}{dots3}
            \fmflabel{$m_2^-$}{m2}
            \fmflabel{$\dots$}{dots4}
            \fmflabel{$s_2^+$}{s2}
            \fmflabel{$(s_2\!+\!1)^+$}{s21}
            \fmflabel{$\dots$}{dots5}
            \fmflabel{$m_3^-$}{m3}
            \fmflabel{$\dots$}{dots6}
            \fmflabel{$s_3^+$}{s3}
            \fmflabel{$(s_3\!+\!1)^+$}{s31}
            \fmfleft{,,,dots2,,s1,,,left,,,s11,,dots3,,,}
            \fmfright{,,,,,,,,right,,,s3,,dots6,,,}
            \fmftop{,m2,,,dots4,,s2,,,top,,,s21,,dots5,,,m3,}
            \fmfbottom{,m1,,,dots1,,s32,,,bottom,,,,,,,,s31,}
            \fmf{wiggly,tension=0}{s32,v1}
            \fmf{wiggly,tension=1.0}{m1,v1}
            \fmf{wiggly,tension=0}{s1,v1}
            \fmf{wiggly,tension=0}{s11,v2}
            \fmf{wiggly,tension=1.0}{m2,v2}
            \fmf{wiggly,tension=0}{s2,v2}
            \fmf{wiggly,tension=0}{s21,v3}
            \fmf{wiggly,tension=1.0}{m3,v3}
            \fmf{wiggly,tension=0}{s3,v3}
            \fmf{wiggly,tension=1.0}{s31,v4}
            \fmf{dashes}{top,bottom}
            \fmf{dashes}{left,right}
            \fmf{plain,tension=0.50}{v1,v2}
            \fmf{plain,tension=0.50}{v2,v3}
            \fmf{plain,tension=0.50}{v3,v4}
            \fmf{plain,tension=0.50}{v4,v1}
            \fmfblob{0.17w}{v1}
            \fmfblob{0.17w}{v2}
            \fmfblob{0.17w}{v3}
            \fmfblob{0.17w}{v4}
      \end{fmfgraph*} }
      \end{fmffile} }
      \caption{Three-mass quadruple cut
               of $ A_{\mathcal{N}=1 \text{ chiral}}^{\text{1-loop,NMHV}}
                   (m_1^-\!\in\!\{s_3\!+\!2,\dots,s_1\},
                    m_2^-\!\in\!\{s_1\!+\!1,\dots,s_2\},
                    m_3^-\!\in\!\{s_2\!+\!1,\dots,s_3\}) $
               \label{box3m}}
      \end{figure}
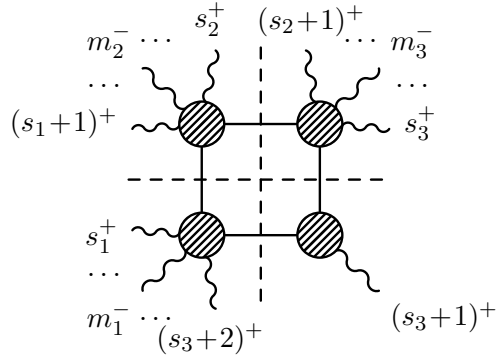

      Finally, if a three-mass box must have all three negative helicities
attributed to the massive legs. Its coefficient is given by
\begin{equation} \begin{aligned}
      C^{\text{box,3m},s_1,s_2,s_3}_{\mathcal{N}=1 \text{ chiral}}
        (\,\overset{+}{\dots},m_1^-,\overset{+}{\dots},m_2^-,
         \overset{+}{\dots},m_3^-,\overset{+}{\dots}\,) =
    - \frac{1}{2(4\pi)^{\frac{d}{2}} i}
      \frac{ \braket{m_1|s_3\!+\!1} \! \braket{m_3|s_3\!+\!1} }
           { K_2^2 \prod_{i \neq s_1,s_2}^n \! \braket{i|i\!+\!1} }
      \hspace{88pt} & \\ \times
      \frac{ \braket{m_1|K_2|K_3|s_3\!+\!1} \! \braket{m_2|K_2|K_3|s_3\!+\!1} \!
             \braket{m_2|K_2|K_1|s_3\!+\!1} \! \braket{m_3|K_2|K_1|s_3\!+\!1} \!
             [s_3\!+\!1|K_1|K_2|K_3\!\ket{s_3\!+\!1} }
           { \braket{s_3\!+\!1|K_1|K_2|s_3\!+\!1}^2
             \braket{s_1|K_2|K_3|s_3\!+\!1} \! \braket{s_1\!+\!1|K_2|K_3|s_3\!+\!1} \!
             \braket{s_2|K_2|K_1|s_3\!+\!1} \! \braket{s_2\!+\!1|K_2|K_1|s_3\!+\!1} }
      & \\ \times
      \big(\!\braket{m_1|m_2} \! \braket{m_3|K_2|K_1|s_3\!+\!1}
           + \braket{m_3|m_2} \! \braket{m_1|K_2|K_3|s_3\!+\!1}\!\big)^2 & .
\label{N1chBox3m}
\end{aligned} \end{equation}

\section{Three-mass triangle coefficients}
\label{app:triangles}

      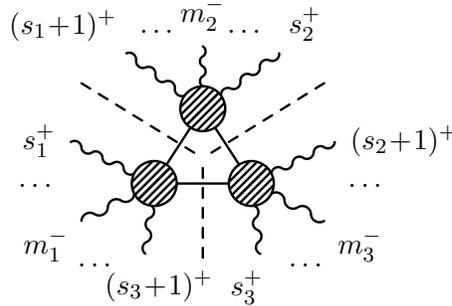
\begin{figure}[h]
      \centering
      \parbox{127pt}{ \begin{fmffile}{graph10}
      \fmfframe(10,10)(10,10){ \begin{fmfgraph*}(100,80)
            \fmflabel{$(s_3\!+\!1)^+$}{s31lab}
            \fmflabel{$\dots$}{dots1}
            \fmflabel{$m_1^-$}{m1}
            \fmflabel{$\dots$}{dots2}
            \fmflabel{$s_1^+$}{s1}
            \fmflabel{$(s_1\!+\!1)^+$}{s11}
            \fmflabel{$\dots$}{dots3}
            \fmflabel{$m_2^-$}{m2}
            \fmflabel{$\dots$}{dots4}
            \fmflabel{$s_2^+$}{s2}
            \fmflabel{$(s_2\!+\!1)^+$}{s21}
            \fmflabel{$\dots$}{dots5}
            \fmflabel{$m_3^-$}{m3}
            \fmflabel{$\dots$}{dots6}
            \fmflabel{$s_3^+$}{s3lab}
            \fmfleft{,,m1,,dots2,,s1,,,left,,}
            \fmfright{,,m3,,dots5,,s21,,,right,,}
            \fmftop{,,,s11,,dots3,,m2,,dots4,,s2,,,}
            \fmfbottom{,,,dots1,s31,s31lab,,bottom,,s3lab,s3,dots6,,,}
            \fmf{wiggly,tension=0}{s31,v1}
            \fmf{wiggly,tension=1.0}{m1,v1}
            \fmf{wiggly,tension=0}{s1,v1}
            \fmf{wiggly,tension=0}{s11,v2}
            \fmf{wiggly,tension=1.2}{m2,v2}
            \fmf{wiggly,tension=0}{s2,v2}
            \fmf{wiggly,tension=0}{s21,v3}
            \fmf{wiggly,tension=1.0}{m3,v3}
            \fmf{wiggly,tension=0}{s3,v3}
            \fmf{dashes}{left,center}
            \fmf{dashes}{right,center}
            \fmf{dashes,tension=1.5}{bottom,center}
            \fmf{plain,tension=0.50}{v1,v2}
            \fmf{plain,tension=0.50}{v2,v3}
            \fmf{plain,tension=0.50}{v3,v1}
            \fmfblob{0.17w}{v1}
            \fmfblob{0.17w}{v2}
            \fmfblob{0.17w}{v3}
      \end{fmfgraph*} }
      \end{fmffile} }
      \caption{Three-mass triple cut
               of $ A_{\mathcal{N}=1 \text{ chiral}}^{\text{1-loop,NMHV}}
                   (m_1^-\!\in\!\{s_3\!+\!1,\dots,s_1\},
                    m_2^-\!\in\!\{s_1\!+\!1,\dots,s_2\},
                    m_3^-\!\in\!\{s_2\!+\!1,\dots,s_3\}) $
               \label{triangle3m}}
      \end{figure}

      Three-mass triangle integral coefficients can be found
not only from double cuts, but from triple cuts as well,
which, in contrast with the spinor integration method,
can produce explicitly rational expressions \cite{BjerrumBohr:2007vu,Dunbar:2009ax}.
The coefficient of the $\mathcal{N}\!=\!1$ chiral NMHV three-mass triangle
constrained by the cut in fig. \ref{triangle3m}
is given by the following all-$n$ formula:
\begin{equation} \begin{aligned}
      C^{\text{tri},s_1,s_2,s_3}_{\mathcal{N}=1 \text{ chiral}}
        (\,\overset{+}{\dots},m_1^-,\overset{+}{\dots},m_2^-,
         \overset{+}{\dots},m_3^-,\overset{+}{\dots}\,) & \\ =
      - \frac{1}{(4\pi)^{\frac{d}{2}} i}
        \frac{1}{ K_2^2 \prod_{i \neq s_1,s_2,s_3}^n \! \braket{i|i\!+\!1} }
        \sum_{i=1}^{6} & \frac{ \prod_{j=1}^5 \braket{c_j d_i} }
                              { \prod_{j \neq i}^6 \braket{d_j d_i} }
            J_1^0(d_i;m_1,m_3,X;{K_1,K_2,K_3}) ,
\label{N1chTri3m}
\end{aligned} \end{equation}
where we label the triangle by three indices $s_1$, $s_2$, $s_3$ such that
each minus-helicity leg $m_i^-$ belongs to a different triangle leg
with massive momentum $ K_i = P_{(s_{i-1}+1),s_i} $.
The auxiliary function is defined as
      \begin{equation}
            J_1^0(d;a,b,c;{K_1,K_2,K_3}) = -
                \frac{ \braket{d|[K_1,K_2]|b} \! \braket{a|[K_1,K_2]|c} }
                     { 2 \braket{d|K_1|K_2|d} \Delta_3(K_1,K_2,K_3) }
                \!+\!
                \frac{ \braket{d b} \! \braket{a c} }
                     { 2 \braket{d|K_1|K_2|d} }
                \!-\!
                \frac{ \braket{d b} \! \braket{d c} \! \braket{a|[K_1,K_2]|d} }
                     { 2 \braket{d|K_1|K_2|d}^2 } ,
      \label{J0} \end{equation}
with the standard notation for
      \begin{equation}
            \Delta_3(K_1,K_2,K_3) = - K_1^4 - K_2^4 - K_3^4
                                  + 2 K_1^2 K_2^2 + 2 K_2^2 K_3^2 + 2 K_3^2 K_1^2 .
      \label{Delta3}
	\end{equation}
In (\ref{N1chTri3m}), the spinors $d_i$ and $c_j$ go through the following values:
      \begin{equation}
            \{ \ket{d_i} \}_{i=1}^6 = \left\{
                                          \ket{s_3}, \ket{s_3\!+\!1},
                                          |K_3|K_2\ket{s_1}, |K_3|K_2\ket{s_1\!+\!1},
                                          |K_1|K_2\ket{s_2}, |K_1|K_2\ket{s_2\!+\!1}
                                      \right\} ,
      \label{dspinors}
      \end{equation}
      \begin{equation}
            \{ \ket{c_j} \}_{j=1}^5 = \left\{
                                          |K_3|K_2\ket{m_1}, |K_3|K_2\ket{m_2},
                                          |K_1|K_2\ket{m_2}, |K_1|K_2\ket{m_3},
                                          \ket{X}
                                      \right\} ,
      \label{cspinors}
      \end{equation}
where
      \begin{equation}
            \ket{X} = \ket{m_3} \braket{m_1|K_1|K_2|m_2}
                    - \ket{m_1} \braket{m_2|K_2|K_3|m_3} .
      \label{Xspinor}
      \end{equation}
      
      We checked expressions that we found using
spinor integration formula (\ref{Ctri})
against these formulas and found numerical agreement through 8 points,
up to machine precision.

\section{Two-mass and one-mass triangle-related momenta}
\label{app:simplifications}

      In Section \ref{trianglepoles}, we mentioned that massive poles
$P^i_\pm$ are defined through square roots of momentum invariants.
We also explained that such poles are related to three-mass triple cuts
and, consequently, three-mass triangles.
Indeed, one can easily see that they would naturally appear
in our triangle coefficient formula (\ref{Ctri}).
Of course, that formula is applicable to two-mass and one-mass triangles.
Such triangles are obtained from a double cut
by cutting a propagator adjacent to the cut.
Such coefficients do not have even superficial irrationality
because for corresponding $P^i_\pm$ the square roots can be taken immediately.
For a $P_{1k}$-channel cut, these momenta are:
\begin{align*}
&     \begin{cases}
            P^k_+ & = p_k - \frac{2 p_k P_{1k}}{P_{1k}^2} P_{1k} \\
            P^k_- & = p_k
      \end{cases} & \hspace{50pt}
&     \begin{cases}
            P^{k+2}_+ & = -p_{k+1} + \frac{2 p_{k+1} P_{1k}}{P_{1k}^2} P_{1k} \\
            P^{k+2}_- & = -p_{k+1}
      \end{cases} \\
&     \begin{cases}
            P^2_+ & = -p_1 \\
            P^2_- & = -p_1 + \frac{2 p_1 P_{1k}}{P_{1k}^2} P_{1k}
      \end{cases} & \hspace{50pt}
&     \begin{cases}
            P^n_+ & = p_n \\
            P^n_- & = p_n - \frac{2 p_n P_{1k}}{P_{1k}^2} P_{1k}
      \end{cases}
\end{align*}
In fact, corresponding denominators can be factorized in a simpler way,
for example,
      \begin{equation}
            \braket{\la|Q_k|P_{1k}|\la} = \braket{\la|k|P_{1k}|\la}
                                        = \braket{\la|k} [k|P_{1k}\!\ket{\la} ,
      \label{Qk}
	\end{equation}
and the two residues can be taken without introducing $P^k_\pm$.

\section{Simplified bubble formulas for subcases of $\mathcal{C}$}
\label{app:subcases}

      For some of the bubble coefficient contributions given in the main text,
massive poles can become massless in helicity configurations
in which either $m_1$ or $m_2$ turn out to be adjacent to the cut.
As we explain in Appendix \ref{app:simplifications},
such poles can be taken more easily
without introducing new massless momenta $P^i_\pm$.
In this section, we provide simplified versions of such contributions
for case $\mathcal{C}$.
In principle, such a configuration can also occur for case $\mathcal{D}$
if $m_2=m_1\!+\!1$,
but this reduces to the case already considered in Section \ref{simplebubbles}.

      More general formulas (\ref{PolesStandardC}) and (\ref{PolesFirstC}) usually
remain valid as well for the subcases that follow, which can be verified numerically.
However, they are likely to lead to numerical instabilities
for some particular helicity configurations,
so we prefer to use the simplified versions.

\subsection{Massless $\mathcal{C}$-case contribution for $m_2=k$}
\label{app:masslessc}

      If $m_2=k$, poles $P^{m_2}_\pm$ that come
from propagator $(l_1-P_{1,m_2-1})^2 = (l_2+p_k)^2$ become massless,
so the last $\mathcal{C}$-case contribution can be computed
using simplified formulas:
\begin{equation} \begin{aligned}
R_{\mathcal{C}}^{s=m_2=k} =
\frac{P_{1k}^2}{(4\pi)^{\frac{d}{2}}i}
\frac{1}{ \braket{12} \dots \braket{k\!-\!2|k\!-\!1}
          \braket{k\!+\!1|k\!+\!2} \dots \braket{n\!-\!1|n}
          \bra{m_1}\!P_{m_1+1,k-1}|k] } \hspace{50pt} \\ \times
\bigg\{
\frac{ F_{\mathcal{C}}^{s=m_2=k}(\la_1,\lb_1) }
     { \braket{1|k\!+\!1} \braket{1|n} }
+
\frac{ F_{\mathcal{C}}^{s=m_2=k}(\la_{k+1},\lb_{k+1}) }
     { \braket{k\!+\!1|1} \braket{k\!+\!1|n} }
+
\frac{ F_{\mathcal{C}}^{s=m_2=k}(\la_n,\lb_n) }
     { \braket{n|1} \braket{n|k\!+\!1} } \hspace{108pt} \\
+
\frac{ \bra{m_1}\!P_{1k}|k]^2 \bra{m_3}\!P_{1k}|k]^2
       \bra{k}\!P_{1k}|q] }
     { P_{1k}^4 P_{1,k-1}^2
       \bra{1}\!P_{1k}|k] \bra{k\!+\!1}\!P_{1k}|k] \bra{n}\!P_{1k}|k]
       \bra{k\!-\!1}\!P_{1k}|k] \bra{k}\!P_{1k}|k]
       \bra{m_1}\!P_{m_1+1,k-1}|k] [k|q] } & \\ \times
 \big( \bra{m_1}\!P_{1k}|k] \braket{m_1|P_{m_1+1,k-1}|P_{1k}|m_3}
     + \bra{m_3}\!P_{1k}|k] \braket{m_1|P_{1,m_1-1}|P_{m_1+1,k-1}|m_1}
 \big)^2 \\
-
\frac{ \bra{m_1}\!P_{1k}|q]^2 \bra{m_3}\!P_{1k}|q]^2
       \bra{m_1}\!P_{m_1+1,k-1}|q]^2 }
     { P_{1k}^4 \bra{1}\!P_{1k}|q] \bra{k\!+\!1}\!P_{1k}|q] \bra{n}\!P_{1k}|q]
       \bra{m_1}\!P_{m_1+1,k}|q]
       \bra{m_1}\!P_{m_1+1,k-1}|P_{1,k-1}|P_{1k}|q] [k|q] } & \\ \times
\frac{ \big( \bra{m_1}\!P_{1k}|q] \braket{m_1|P_{m_1+1,k-1}|P_{1k}|m_3}
           + \bra{m_3}\!P_{1k}|q] \braket{m_1|P_{1,m_1-1}|P_{m_1+1,k-1}|m_1}
       \big)^2 }
     { \bra{m_1}\!P_{1k}|q] \bra{k\!-\!1}\!P_{1,k-2}|q]
     - \bra{k\!-\!1}\!P_{1k}|q] \bra{m_1}\!P_{1,m_1-1}|q] } &
\bigg\} ,
\label{PolesStandardSimpleC}
\end{aligned} \end{equation}
where 
\begin{equation} \begin{aligned}
F_{\mathcal{C}}^{s=m_2=k}(\la,\lb) =
\frac{ \braket{\la|m_1}^2 \braket{\la|m_3}^2
       \braket{\la|P_{1k}|P_{m_1+1,k-1}|m_1}^2 [\lb|q] }
     { \braket{\la|P_{1,k-1}|P_{m_1+1,k-1}|m_1}
       \braket{\la|P_{1k}|P_{m_1+1,k}|m_1}
       \bra{\la}\!P_{1k}|\lb] \bra{\la}\!P_{1k}|k] \bra{\la}\!P_{1k}|q] } & \\ \times
\frac{ \big( \braket{\la|m_1} \braket{m_1|P_{m_1+1,k-1}|P_{1k}|m_3}
           + \braket{\la|m_3} \braket{m_1|P_{1,m_1-1}|P_{m_1+1,k-1}|m_1}
       \big)^2 }
     { \braket{\la|m_1} \braket{\la|P_{1k}|P_{1,k-2}|k\!-\!1}
     - \braket{\la|k\!-\!1} \braket{\la|P_{1k}|P_{1,m_1-1}|m_1} } & .
\label{FsimpleC}
\end{aligned} \end{equation}

\subsection{First $\mathcal{C}$-case contribution for $m_1=k-2$ and $m_2=k$}
\label{app:simple2c}

      The previous simplified subcase is not valid for $s=m_2=k$,
if the general form of the contribution contains two massive poles,
i.~e. when $s=m_2=k=m_1+2$. The last pole becomes massless,
but the first one remains massive:
\begin{equation} \begin{aligned}
R_{\mathcal{C}}^{s=k} \! = \! -
\frac{P_{1k}^2}{(4\pi)^{\frac{d}{2}}i}
\frac{1}{ \braket{12} \dots \braket{k\!-\!3|k\!-\!2} \! [k\!-\!1|k] \!
          \braket{k\!+\!1|k\!+\!2} \dots \braket{n\!-\!1|n} }
\bigg\{
\frac{ F_{\mathcal{C}}^{s=k}(\la_1,\lb_1) }
     { \braket{1|k\!+\!1} \! \braket{1|n} }
\! + \!
\frac{ F_{\mathcal{C}}^{s=k}(\la_{k+1},\lb_{k+1}) }
     { \braket{k\!+\!1|1} \! \braket{k\!+\!1|n} } & \\
+
\frac{ F_{\mathcal{C}}^{s=k}(\la_n,\lb_n) }
     { \braket{n|1} \! \braket{n|k\!+\!1} }
\! - \!
\frac{ [k\!-\!1|k] }
     { P_{1k}^4 P_{1,k-1}^2 \! \braket{k\!-\!2|k\!-\!1} }
\frac{ \bra{k\!-\!2}\!P_{1k}|k]^2 \bra{m_3}\!P_{1k}|k]^2
       \braket{k\!-\!2|P_{1,k-1}|P_{1k}|m_3}^2 \bra{k}\!P_{1k}|q] }
     { \bra{1}\!P_{1k}|k] \bra{k\!+\!1}\!P_{1k}|k] \bra{n}\!P_{1k}|k]
       \bra{k\!-\!1}\!P_{1,k-2}|k] \bra{k}\!P_{1k}|k] [k|q] } & \\
+
\frac{ \bra{k\!-\!2}\!P_{1k}|q]^2 \bra{m_3}\!P_{1k}|q]^2
       [k\!-\!1|q]^2
 \big( P_{1k}^2 \braket{k\!-\!2|m_3} [k\!-\!1|q]
     + \bra{m_3}\!P_{1k}|q] \bra{k\!-\!2}k|k\!-\!1]
 \big)^2 }
     { P_{1k}^4 \bra{1}\!P_{1k}|q] \bra{k\!+\!1}\!P_{1k}|q] \bra{n}\!P_{1k}|q]
       [q|P_{1,k-2}|P_{k-1,k}|q] [k\!-\!1|P_{1,k-2}|P_{1k}|q]
       \bra{k\!-\!2}\!P_{k-1,k}|q] [k|q] } & \\
+ \, \tilde{M}_{\mathcal{C}}^{s=k}(\la^{k-1}_+,\lb^{k-1}_+)
   + \tilde{M}_{\mathcal{C}}^{s=k}(\la^{k-1}_-,\lb^{k-1}_-) &
\label{PolesFirstSimple2C}
\bigg\} ,
\end{aligned} \end{equation}
where 
\begin{equation}
F_{\mathcal{C}}^{s=k}(\la,\lb) =
\frac{ \braket{\la|k\!-\!2}^2 \braket{\la|m_3}^2
       \bra{\la}\!P_{1k}|k\!-\!1]^2 [\lb|q]
\big(\!\braket{k\!-\!2|m_3} \bra{\la}\!P_{1k}|k\!-\!1]
     - \braket{\la|m_3} \bra{k\!-\!2}\!k|k\!-\!1]
\big)^2 }
     { \braket{\la|P_{1,k-2}|P_{k-1,k}|\la}
       \bra{\la}\!P_{1k}|\lb] \bra{\la}\!P_{1k}|k] \bra{\la}\!P_{1k}|q]
       \bra{\la}\!P_{1,k-2}|k\!-\!1] \braket{\la|P_{1k}|P_{k-1,k}|k\!-\!2} } ,
\label{Fsimple2C}
\end{equation}
and the first massive pole gives
\begin{equation} \begin{aligned}
\tilde{M}_{\mathcal{C}}^{s=k}(\la,\lb) = & -
\frac{1}{4( (P_{1,k-2} \cdot P_{k-1,k})^2
             - P_{1,k-2}^2 P_{k-1,k}^2) )} \\ \times &
\frac{ \braket{\la|k\!-\!2}^2 \braket{\la|m_3}^2
       \bra{\la}\!P_{1k}|k\!-\!1]^2 [\lb|q]
\big(\!\braket{k\!-\!2|m_3} \bra{\la}\!P_{1k}|k\!-\!1]
     - \braket{\la|m_3} \bra{k\!-\!2}\!k|k\!-\!1] \big)^2 }
     { \braket{\la|1} \braket{\la|k\!+\!1} \braket{\la|n}
       \bra{\la}\!P_{1k}|k] \bra{\la}\!P_{1k}|q]
       \bra{\la}\!P_{1,k-2}|k\!-\!1] \braket{\la|P_{1k}|P_{k-1,k}|k\!-\!2} } .
\label{MassivePolesFirstSimple2}
\end{aligned} \end{equation}

\subsection{First $\mathcal{C}$-case contribution for $m_1=1$}
\label{app:simple1c}

      Finally, there might be another reason
for the contribution with two massive poles to get simplified ---
the first negative-helicity gluon can become adjacent to the cut: $m_1=1$.
In that case, the first massive pole becomes massless,
while the second one stays massive:
\begin{equation} \begin{aligned}
R_{\mathcal{C}}^{s=3} \! = \! -
\frac{P_{1k}^2}{(4\pi)^{\frac{d}{2}}i}
\frac{1}{ [12] \braket{34} \dots \braket{k\!-\!1|k} \!
               \braket{k\!+\!1|k\!+\!2} \dots \braket{n\!-\!1|n} }
\bigg\{
\frac{ F_{\mathcal{C}}^{s=3}(\la_k,\lb_k) }
     { \braket{k|k\!+\!1} \! \braket{k|n} }
\! + \!
\frac{ F_{\mathcal{C}}^{s=3}(\la_{k+1},\lb_{k+1}) }
     { \braket{k\!+\!1|k} \! \braket{k\!+\!1|n} } & \\
+
\frac{ F_{\mathcal{C}}^{s=3}(\la_n,\lb_n) }
     { \braket{n|k} \! \braket{n|k\!+\!1} }
\! + \!
\frac{ [12] }
     { P_{1k}^4 P_{2,k}^2 \! \braket{23}}
\frac{ \bra{m_2}\!P_{1k}|1]^2 \bra{m_3}\!P_{1k}|1]^2
       \braket{m_2|P_{2,k}|P_{1k}|m_3}^2 \bra{1}\!P_{1k}|q] }
     { \bra{k}\!P_{1k}|1] \bra{k\!+\!1}\!P_{1k}|1] \bra{n}\!P_{1k}|1]
       \bra{2}\!P_{1k}|1] \bra{1}\!P_{1k}|1] [1|q] } & \\
+
\frac{ \bra{m_2}\!P_{1k}|q]^2 \bra{m_3}\!P_{1k}|q]^2 [2|q]^2
       \big( P_{1k}^2 \braket{m_2|m_3} [2|q]
          + \bra{m_3}\!P_{1k}|q] \bra{m_2}\!1|2])^2
       \big) }
     { P_{1k}^4 \bra{k}\!P_{1k}|q]
       \bra{k\!+\!1}\!P_{1k}|q] \bra{n}\!P_{1k}|q]
       [q|P_{1,2}|P_{3,k}|q]
       [2|P_{3,k}|P_{1k}|q]
       \bra{3}\!P_{1,2}|q] [1|q] } & \\
+ \; M_{\mathcal{C}}^{s=3}(\la^3_+,\lb^3_+)
   + M_{\mathcal{C}}^{s=3}(\la^3_-,\lb^3_-) &
\bigg\} ,
\label{PolesFirstSimple1C}
\end{aligned} \end{equation}
where 
\begin{equation}
F_{\mathcal{C}}^{s=3}(\la,\lb) =
\frac{ \braket{\la|m_2}^2 \braket{\la|m_3}^2
       \bra{\la}\!P_{1k}|2]^2 [\lb|q]
       \big(\!\braket{m_2|m_3} \bra{\la}\!P_{1k}|2]
            - \braket{\la|m_3} \bra{m_2}\!1|2] 
       \big)^2 }
     { \braket{\la|P_{1,2}|P_{3,k}|\la}
       \bra{\la}\!P_{1k}|\lb] \bra{\la}\!P_{1k}|1] \bra{\la}\!P_{1k}|q]
       \bra{\la}\!P_{3,k}|2] \braket{\la|P_{1k}|P_{1,2}|3} } ,
\label{Fsimple1C}
\end{equation}
and
\begin{equation} \begin{aligned}
M_{\mathcal{C}}^{s=3}(\la,\lb) =
\frac{-\braket{\la|m_2}^2 \! \braket{\la|m_3}^2 \!
       \bra{\la}\!P_{1k}|2]^2 [\lb|q] 
\big(\!\braket{m_2|m_3} \bra{\la}\!P_{1k}|2]
     - \braket{\la|m_3} \bra{m_2}\!1|2] \big)^2 }
     { 4( (P_{1,2} \cdot P_{3,k})^2 - P_{1,2}^2 P_{3,k}^2) \!
       \braket{\la|k} \! \braket{\la|k\!+\!1} \! \braket{\la|n} \!
       \bra{\la}\!P_{1k}|1] \! \bra{\la}\!P_{1k}|q] \!
       \bra{\la}\!P_{3,k}|2] \! \braket{\la|P_{1k}|P_{1,2}|3} }
\label{MassivePoleFirstSimple1} .
\end{aligned} \end{equation}

\section{Mathematica implementation}
\label{app:implementation}

      We distribute the Mathematica file N1chiralAll.nb along with this paper,
in which all final formulas are collected.
They are intended to be used along the package S@M package \cite{Maitre:2007jq}.
The main end-user functions are
\begin{description}
\item{BubSimplest[$k\_Integer$, {$plus\_$, $minus\_$}, $p\_List$, $K\_$]}:
      $\overline{\text{MHV}}$-MHV bubble (\ref{simplestbubbledpw});
\item{BubSimple[$k\_Integer$, $m\_Integer$, $p\_List$, $P\_List$]}:
      NMHV bubble (\ref{simplebubble})
      with minus-helicity gluons $k^-$ and $(k\!-\!1)^-$
      adjacent to the $P_{1k}$-channel cut;
\item{BubStandard[$k\_Integer$, $m1\_Integer$, $m2\_Integer$, $m3\_Integer$,
                  $p\_List$, $P\_List$, $Ptri\_List$, $q\_$]}:
      most generic NMHV bubble (\ref{CbubN1frame}) with $m_2 \neq m_1\!+\!1$;
\item{BubAdjacent[$k\_Integer$, $m1\_Integer$, $m3\_Integer$,
                  $p\_List$, $P\_List$, $Ptri\_List$, $q\_$]}:
      generic NMHV bubble (\ref{CbubN1frame}) with $m_2 = m_1\!+\!1$;
\item{BoxStandard[$k\_Integer$, $m1\_Integer$, $m2\_Integer$, $m3\_Integer$,
                  $p\_List$, $P\_List$, $l1\_$, $l2\_$]}:
      two-mass-easy or one-mass box (\ref{CboxN1frame}) for $m_2 \neq m_1\!+\!1$;
\item{BoxAdjacent[$k\_Integer$, $m1\_Integer$, $m3\_Integer$,
                  $p\_List$, $P\_List$, $l1\_$, $l2\_$]}:
      two-mass-easy or one-mass box (\ref{CboxN1frame}) for $m_2 \neq m_1\!+\!1$;
\item{Box1m[$minus\_Integer$, $plus\_Integer$, $p\_List$, $P\_List$]}:
      one-mass box (\ref{N1chBox1m});
\item{Box2mh[$s1\_Integer$, $m1\_Integer$, $m2\_Integer$, $m3\_Integer$,
             $p\_List$, $P\_List$]}:
      two-mass-hard box(\ref{N1chBox2mh});
\item{Box3m[$s1\_Integer$, $s2\_Integer$, $s3\_Integer$, 
            $m1\_Integer$, $m2\_Integer$, $m3\_Integer$, $p\_List$, $P\_List$]}:
      three-mass box(\ref{N1chBox3m});
\item{Tri3m[$s1\_Integer$, $s2\_Integer$, $s3\_Integer$,
            $m1\_Integer$, $m2\_Integer$, $m3\_Integer$, $p\_List$, $P\_List$]}:
      three-mass triangle (\ref{N1chTri3m}).
\end{description}
Most of these functions have a subset of the following arguments:
\begin{description}
\item{$p\_List$}
is the vector of external gluon momenta $\{p_1,p_2,\dots,p_n\}$.
\item{$P\_List$}
is the array of external momentum sums $\{P_{i,j}\}_{i,j=1}^n$.
\item{$Ptri\_List$}
is the array of massless momentum solutions $\{\{P^i_+,P^i_-\}\}_{i=m_1+1}^{m_2}$
for massive poles, \\ \emph{excluding} those adjacent to the cut ($i=2,k$).
\item{$k\_Integer$}
is the number of particles in the cut channel,
i.~e. the \emph{cut momentum is defined as} $P[[1,k]]$. \\
If one wishes to deal with another channel,
one should relabel input arrays $p$ and $P$ accordingly.
\item{$m_1\_Integer, m_2\_Integer, m_3\_Integer$}
indicate positions of the minus-helicity gluons inside $p$.
In our conventions $ m_1,m_2 \in \{1,\dots,k\} $
and $ m_3 \in \{k\!+\!1,\dots,n\} $.
\item{$s\_Integer,t\_Integer$}
indicate which contribution is being evaluated,
i.~e. from which $\mathcal{R}_{m_1 s t}$ invariant it comes from.
Since we consider the non-zero cases in which
labels $s$ and $t$ become equal to $l_2$ and $-l_1$ separately,
those arguments are understood to go through
only external gluon labels: \\
$ s \in \{m_1\!+\!2,\dots,k\} $;
$ t \in \{m_2\!+\!1,\dots,k\} \cup \{2,\dots,m_1\!-\!1\} $.
\item{$q\_$}
defines the arbitrary reference spinor $|q]$.
\item{$l\_$}
is the free argument corresponding to loop momentum variables $\la,\lb$
and is present in auxiliary functions, such as $F_{\mathcal{AB}}(\la,\lb)$.
\item{$l1\_,l2\_$}
are the $l_1,l_2$ arguments in two-mass-easy box functions,
intended to be replaced by the quadruple cut spinor solutions (\ref{2mespinors}).
The corresponding replacement rule can be generated using function
ReplaceLoopSpinors[$k\_Integer$, $p\_List$, $P\_List$].
\end{description}

To illustrate how to use that functionality,
we attach the sample calculation file Ammppmppp.nb
in which we calculate all non-zero bubbles coefficients of
$ A^{\text{1-loop}}_{\mathcal{N}=1 \text{ chiral}}
 (1^-,2^-,3^+,4^+,5^-,6^+,7^+,8^+) $,
as well as some triangles and boxes.


\bibliographystyle{JHEP}
\bibliography{references}

\end{document}